\documentclass[a4paper,10pt]{article}
\usepackage{jheppub,color,graphicx,subfigure,dsfont,slashed,multirow,lscape,%
  hyperref}
\usepackage[utf8]{inputenc}
\usepackage[compat=1.1.0]{tikz-feynman}	
\usepackage[section]{placeins}
\usepackage[makeroom]{cancel}
\usepackage{listings,upquote,booktabs,pgffor,multirow,caption,amssymb,slashed,amsmath}
\graphicspath{{Figures/}}

\renewcommand{\arraystretch}{1.2}

\def\ie{{\it i.e.}}
\def\eg{{\it e.g.}}

\newcommand{\bfl}{{\sc\small MoGRe}}
\newcommand{\collier}{{\sc\small Collier}}
\newcommand{\cuttools}{{\sc\small CutTools}}
\newcommand{\fastjet}{{\sc\small FastJet}}
\newcommand{\fa}{{\sc\small FeynArts}}
\newcommand{\fr}{{\sc\small FeynRules}}
\newcommand{\iregi}{{\sc\small Iregi}}
\newcommand{\lhapdf}{{\sc\small LHAPDF6}}
\newcommand{\madfks}{{\sc\small MadFKS}}
\newcommand{\madgolem}{{\sc\small MadGolem}}
\newcommand{\madloop}{{\sc\small MadLoop}}
\newcommand{\madspin}{{\sc\small MadSpin}}
\newcommand{\mados}{{\sc\small MadSTR}}
\newcommand{\madwidth}{{\sc\small MadWidth}}
\newcommand{\mg}{{\sc\small MG5\_aMC}}
\newcommand{\mglong}{{\sc\small MadGraph5}\_a{\sc MC@NLO}}
\newcommand{\mthmtc}{{\sc\small Mathematica}}
\newcommand{\ninja}{{\sc\small Ninja}}
\newcommand{\nloct}{{\sc\small NLOCT}}
\newcommand{\openloops}{{\sc\small OpenLoops}}
\newcommand{\prospino}{{\sc\small Prospino2}}
\newcommand{\pythia}{{\sc\small Pythia8.2}}
\newcommand{\herwig}{{\sc\small Herwig}}
\newcommand{\resummino}{{\sc\small Resummino}}

\newcommand{\be}{\begin{equation}}
\newcommand{\ee}{\end{equation}}
\def\bsp#1\esp{\begin{split}#1\end{split}}
\def\bpm{\begin{pmatrix}}
\def\epm{\end{pmatrix}}
\newcommand\sss{\scriptscriptstyle}
\def\abs#1{\left|#1\right|}

\newcommand\as{\alpha_{\sss S}}
\newcommand{\msbar} {\overline{\rm MS}}
\newcommand{\drbar} {\overline{\rm DR}}
\newcommand\DR{{\rm DR}}
\newcommand\DRpI{{\rm DR}+{\rm I}}
\newcommand\DS{{\rm DS}}
\newcommand\boost{\mathbb B}
\newcommand{\epsbar}{\bar\epsilon}
\newcommand{\gau}[1][]{\ensuremath{\tilde{\chi}^{#1}}}	
\newcommand{\anti}[1]{#1^\star}
\newcommand{\neu}[1][]{\ensuremath{\gau[0]_{#1}}}	
\newcommand\aem{\alpha}
\newcommand\amp{{\cal A}}
\newcommand\bb{\bar{b}}
\newcommand\bk{\bar{k}}
\newcommand\bkb{\bar{k}_\beta}
\newcommand\bkd{\bar{k}_\delta}
\newcommand\bkg{\bar{k}_\gamma}
\newcommand\bkrec{\bk_{\rm rec}}
\newcommand\bx{\bar{x}}
\newcommand\bt{\bar{t}}
\newcommand\Gb{\Gamma_\beta}
\newcommand\kb{k_\beta}
\newcommand\kd{k_\delta}
\newcommand\kbt{k_\beta^2}
\newcommand\kg{k_\gamma}
\newcommand\krec{k_{\rm rec}}
\newcommand\mb{m_\beta}
\newcommand\mbt{m_\beta^2}
\newcommand\md{m_\delta}
\newcommand\mga{m_\gamma}
\newcommand\mrec{m_{\rm rec}}
\newcommand\mrect{m_{\rm rec}^2}
\newcommand\nbeta{\cancel{\beta}}
\newcommand\proj{{\mathbb P}}
\newcommand\sg{\tilde{g}}
\newcommand\sn{\tilde{\chi}}
\newcommand\sq{\tilde{q}}
\newcommand\vbkrec{\vec{\bar{k}}_{\rm rec}}
\newcommand\vkb{\vec{k}_\beta}
\newcommand\vkrec{\vec{k}_{\rm rec}}

\newcommand{\panti}[1]{#1^{(\star)}}

\chardef\MyArticleWithColor=\pdfcolorstackinit page direct{0 g}

\title{Automated simulations beyond the Standard Model: supersymmetry}

\author[a]{Stefano~Frixione}
\author[b,c]{\!\!, Benjamin~Fuks}
\author[d]{\!\!, Valentin~Hirschi}
\author[e]{\!\!, Kentarou Mawatari}
\author[b]{\!\!, Hua-Sheng~Shao}
\author[f]{\!\!, Marthijn~P.~A.~Sunder}
\author[g]{\! and Marco~Zaro}

\emailAdd{stefano.frixione@cern.ch}
\emailAdd{fuks@lpthe.jussieu.fr}
\emailAdd{hirschva@itp.phys.ethz.ch}
\emailAdd{kentarou.mawatari@het.phys.sci.osaka-u.ac.jp}
\emailAdd{huasheng.shao@lpthe.jussieu.fr}
\emailAdd{mpasunder@uni-muenster.de}
\emailAdd{m.zaro@nikhef.nl}

\affiliation[a]{INFN, Sezione di Genova, Via Dodecaneso 33, I-16146, Genoa,
  Italy}
\affiliation[b]{Laboratoire de Physique Th\'eorique et Hautes Energies (LPTHE), UMR 7589, Sorbonne Universit\'e et CNRS, 4 place Jussieu, 75252 Paris Cedex 05, France}
\affiliation[c]{Institut Universitaire de France, 103 boulevard Saint-Michel,  75005 Paris, France}
\affiliation[d]{Institute for Theoretical Physics, ETH Z\"urich, 8093 Z\"urich,
    Switzerland}
\affiliation[e]{Department of Physics, Osaka University, Toyonaka, Osaka 560-0043, Japan}
\affiliation[f]{Institut f\"ur Theoretische Physik, Westf\"alische Wilhelms-Universit\"at
 M\"unster, Wilhelm-Klemm-Stra\ss{}e 9, D-48149 M\"unster, Germany}
\affiliation[g]{Nikhef, Theory Group, Science Park 105, 1098 XG, Amsterdam, The Netherlands}

\abstract{The \mglong\ framework aims to automate all types
of leading- and next-to-leading-order-accurate simulations for any user-defined
model that stems from a renormalisable Lagrangian. In this paper,
we present all of the key ingredients of such models in the context
of supersymmetric theories. In order to do so, we extend the
\fr\ package by giving it the possibility of dealing
with different renormalisation options that are relevant to supersymmetric
models. We also show how to deal with the problem posed by the
presence of narrow resonances, thus generalising the so-called
on-shell subtraction approaches. We extensively compare our total rate
results with those of both \prospino\ and \resummino, and
present illustrative applications relevant to the 13~TeV LHC, both
at the total-rate and differential levels.
The computer programs that we have used to obtain the predictions
presented here are all publicly available.}

\begin{document}
\preprint{OU-HET 1021, NIKHEF/2019-017}

\maketitle
\flushbottom

\section{Introduction}\label{sec:intro}
After more than fifty years since its proposal, the Standard Model (SM) has 
been proven to be an extremely successful theory of Nature: its predictions 
agree well with the vast majority of the data collected so far, sometimes 
at an astonishing level of precision. Despite its success, however,
the SM leaves some deep questions unanswered, and suffers from various 
conceptual issues and limitations; thus, it is widely understood as a 
low-energy effective theory that is supposed to emerge from a suitable
UV-complete theory. Among the candidates for the latter, weak-scale 
supersymmetry (SUSY) constitutes one of the best-motivated options from
a theoretical viewpoint. Naturally extending the Poincar\'e algebra
by linking the fermionic and bosonic degrees of freedom of the 
theory~\cite{Golfand:1971iw,Volkov:1973ix,Wess:1973kz, Wess:1974tw,
Wess:1974jb,Salam:1974yz,Salam:1974jj,Ferrara:1974ac,Ferrara:1974pu}, 
SUSY and in particular its minimal incarnation, the so-called Minimal 
Supersymmetric Standard Model (MSSM)~\cite{Nilles:1983ge,Haber:1984rc}, 
addresses several of the shortcomings of the SM. 
For example, SUSY can tackle some aspects of the hierarchy problem
inherent to the SM by stabilising all scalar masses relatively to quantum
corrections~\cite{Witten:1981nf}, it can ensure the unification of the three
SM gauge couplings at high energies~\cite{Ibanez:1981yh,Dimopoulos:1981yj,
Ellis:1990wk}, and many SUSY realisations include a candidate explaining the
presence of dark matter in the Universe~\cite{Goldberg:1983nd,Ellis:1983ew}.

The existence of a superpartner with a mass equal to the one of
its SM counterpart is experimentally excluded, so SUSY 
must be broken. For theoretical and phenomenological reasons, this
breaking must be soft and is expected to shift the SUSY particle masses in the
TeV regime. Thus, also owing to the solid theoretical motivations of
supersymmetric theories, the quest for SUSY particles still
plays a major role in the searches carried out by the LHC experiments,
and features prominently in the strategies that are currently being laid 
out for future-collider projects. Both the ATLAS and CMS collaborations 
have begun to release full LHC run~2 results; the persistent absence 
of beyond-the-SM (BSM) signals implies that the space of parameters of 
SUSY theories is constrained in an increasingly severe manner. For example,
squarks and gluinos are bounded to have masses well above 
1--2 TeV~\cite{ATLAS:2018yhd,ATLAS:2019mge,CMS:2019twi,CMS:2019see},
whilst the bounds on the electroweak superpartners now reach several 
hundreds of GeV~\cite{ATLAS:2019cfv,CMS:2019hos}. 
However, such experimental limits have
been extracted either in the framework of specific MSSM benchmark scenarios, 
or in the context of simplified models of new physics that are inspired 
by the MSSM, and that feature only a small number of new particles and new
interactions with respect to those of the SM~\cite{Alwall:2008ag,Alves:2011wf}. 
Therefore, these limits can be evaded both in various non-minimal 
supersymmetric theories and with specific, fine-tuned, parameter 
configurations of the MSSM.

Experimental analyses for SUSY searches are currently mostly based on Monte
Carlo simulations of signals in which tree-level matrix elements of different partonic 
multiplicities are consistently combined and interfaced with parton-shower
(PS) Monte Carlos; the absence of double counting between matrix elements 
and parton showers is guaranteed by the use of a merging prescription --
CKKW~\cite{Catani:2001cc,Krauss:2002up}, MLM~\cite{Mangano:2006rw},
and CKKW-L~\cite{Lonnblad:2001iq,Lavesson:2005xu} 
(see also refs.~\cite{Hoeche:2009rj,Hamilton:2009ne,Lonnblad:2011xx}).
For any given jet multiplicity, such predictions are therefore
leading-order (LO) plus leading-logarithm (LL) accurate. They
are typically further improved by normalising them, at the fully
inclusive level, to the best available (in the sense of perturbative
information) total cross sections for the production of the 
relevant SUSY particles.
Typically, these cross sections combine fixed-order predictions at the 
next-to-leading-order (NLO) accuracy in $\as$ (which we shall call 
``NLO QCD'' henceforth, understanding that quarks, gluons, squark, and 
gluinos run in the loops), with calculations that carry out the resummation 
of threshold logarithms at the next-to-leading-logarithmic (NLL) accuracy.
In some cases, the results include some of the next-to-next-to-leading-order 
(NNLO) and next-to-next-to-leading-logarithmic (NNLL) contributions
as well.

As the LHC moves steadily towards a precision-physics phase,
BSM search strategies, in view of their null results thus far,
must evolve too. Among other things, finding elusive signals requires
improving the control over systematics. At the theoretical level,
this demands a better use of higher-order results than the simple
rescaling of tree-level merged simulations, without losing the realistic
description of multi-jet final states provided by parton showers. Fortunately, 
solutions originally devised for SM physics, which have been thoroughly 
and successfully tested since the start of the LHC operations, can be
applied to BSM scenarios as well, pretty much as their tree-level
counterparts. More specifically, NLO computations can be matched to parton 
showers (NLO+PS henceforth) by means of the MC@NLO~\cite{Frixione:2002ik}
and {\sc\small Powheg}~\cite{Nason:2004rx} methods (or of any of their variants and
less-used alternatives~\cite{Nagy:2005aa,Bauer:2006mk,Nagy:2007ty,
Giele:2007di,Bauer:2008qh,Hoeche:2011fd,Hamilton:2013fea,Jadach:2015mza})
for any underlying SUSY process. Likewise, the extensions of tree-level 
merging techniques to NLO~\cite{Lavesson:2008ah,Hamilton:2010wh,Hoche:2010kg,
Giele:2011cb,Alioli:2011nr,Hoeche:2012yf,Frederix:2012ps,Platzer:2012bs,
Alioli:2012fc,Lonnblad:2012ix,Hamilton:2012rf,Alioli:2013hqa} work equally 
well in the SM and BSM theories. These facts have stimulated recent 
theoretical work (which we shall briefly review in section~\ref{sec:sim}), 
that is characterised by being essentially on a process-by-process 
basis and in simplified models.

The main goal of this paper is to render systematically feasible,
for any user-defined process and in the fully-fledged MSSM, fixed-order, 
matched, and merged simulations that include NLO QCD effects. In order
to do this, we shall rely on automated techniques, which are by now
extremely well established; in particular, we shall work in the
\mglong\ framework~\cite{Alwall:2014hca} (\mg\ henceforth). We remind
the reader that \mg\ is a meta-code, which uses hard-wired information
on Quantum Field Theories (such as general rules of Feynman diagrammatics,
the structures of matrix elements, phase spaces, and cross sections,
and so forth), and external information equivalent to the Lagrangian
of the theory one is interested in (these are a set of rules, called
a {\em UFO model}~\cite{Degrande:2011ua}, obtained automatically from the 
Lagrangian by means of codes such as \fr~\cite{Alloul:2013bka}, possibly used
jointly with \nloct~\cite{Degrande:2014vpa}), to construct on the fly a computer
code specific to the production process that one is interested to simulate.

Thanks to the significant amount of development and validation activity
performed on \mg\ in the past few years, most of the work for the current
paper has gone into the construction of a UFO model for the MSSM
(see section~\ref{sec:model}). A notable exception, at the level of
the meta-code, is the following.
In the context of theories with rich particle spectra, that
include relatively narrow resonances, the computation of 
contributions beyond the LO might give results 
spoiling the ``convergence'' of the perturbative series.
This happens when real-emission corrections include partonic sub-processes
that feature those resonances in $s$-channels, whose integration
over the phase space is either divergent (if the resonance
propagator is not Dyson-summed), or grows like an inverse power
of the resonance width (and is therefore numerically dominant).
A familiar example in the SM is that of $tW^-$ production,
for which a subset of real corrections features a $tW^-\bar{b}$
final state, to which diagrams with an $s$-channel $\bar{t}$ quark
contribute. We point out that the lack of convergence of the
perturbative series in these cases is perfectly justified, since
one is trying to compute higher-order corrections to the cross
section of a process which is simply ill-defined. In the SM example 
just mentioned, $tW^-$ production does not exist as
such: its definition requires the simultaneous presence of a
particle (the top) and of one of the decay products of its
antiparticle (the $W^-$).
In spite of this, one can try and give an operative meaning to
these ill-defined processes: they are conceptually useful, since
they correspond to an intuitive physical picture which is easy to
understand, and they can constitute, within reasonable approximations,
valuable perturbative tools. In this sense, BSM theories in general,
and SUSY in particular, provide one with many possibilities to
test different strategies. In this paper we shall discuss the
definitions of such strategies, that we collectively call 
{\em Simplified Treatments of Resonances} (STR henceforth), and
their implementations in \mg\ in a way suited
to NLO+PS simulations. Thus, STR encompass the procedures known
as Diagram Removal (DR) and Diagram Subtraction (DS), and generalise
the so-called On-Shell Subtractions (OSS) schemes.

This paper is organised as follows. In section~\ref{sec:model} we briefly
review the status of simulations in SUSY theories, and the basic 
characteristics of \mg\ and of the models it uses for BSM physics.
We also introduce the theoretical framework of the MSSM, and its
various renormalisation options. We validate our implementation against
\prospino\ and \resummino\ in section~\ref{sec:validate}. The total cross
sections of several SUSY benchmark processes in simplified scenarios are
presented in section~\ref{sec:rates}. The general algorithms for the treatments 
of resonances in perturbative computations, and their implementations in
\mg, are discussed in section~\ref{sec:OS}. A case study of jets plus missing
energy at NLO+PS accuracy at the 13~TeV LHC is considered in section~\ref{sec:pheno},
in the context of a non-simplified benchmark point. We draw our conclusions
in section~\ref{sec:conclusion}. Some details about one- and two-point
loop integrals, the \bfl\ package (the acronym standing for
\textit{\underline{Mo}re \underline{G}eneral \underline{Re}normalisation} in \fr), and
the settings relevant to the decoupling 
mass limit in \mg\ can be found in the appendices \ref{app:PV}, \ref{app:bfl},
and \ref{app:decoupling}, respectively.


\section{Theoretical framework}\label{sec:model}
\subsection{The Minimal Supersymmetric Standard Model} \label{sec:mssm}
\subsubsection{Field content and supersymmetric Lagrangian}
The MSSM, the simplest SUSY
extension of the SM, results from the direct
supersymmetrisation of the SM~\cite{Nilles:1983ge,Haber:1984rc}. The gauge
symmetry group of the theory is the SM one, and it relies on the three vector
supermultiplets $V_B$, $V_W$ and $V_G$,
\be
  V_B \equiv \big(B_\mu, \widetilde B^\alpha\big)\ , \qquad
  V_W \equiv \big(W_\mu, \widetilde W^\alpha\big)\ , \qquad
  V_G \equiv \big(G_\mu, \widetilde G^\alpha\big)\ ,
\ee
that are respectively connected to the hypercharge $U(1)_Y$, weak $SU(2)_L$ and strong $SU(3)_c$ symmetries.
All the SM gauge bosons $B_\mu$, $W_\mu$ and $G_\mu$ are
supplemented by their fermionic gaugino superpartner $\widetilde B$,
$\widetilde W$ and $\widetilde G$, all fields lying in the adjoint
representations of their respective
gauge group. The matter sector of the theory includes three generations of five
chiral supermultiplets $L_L$, $E_R$, $Q_L$, $U_R$ and $D_R$,
\be\bsp
  & L_L({{\bf 1}}, {{\bf 2}}, -1/2) \equiv
     \Bigg(\ell_L = \bpm \nu_L \\ e_L \epm, \
     \widetilde \ell_L = \bpm \widetilde \nu_L\\ \widetilde e_L\epm\Bigg)
      \ , \qquad
    E_R({{\bf 1}},{{\bf 1}}, 1) =
      \big( e_R^c, \widetilde e_R^\dag\big) \ , \\
  & Q_L({ {\bf 3}}, {{\bf 2}}, 1/6) \equiv
     \Bigg(q_L = \bpm u_L \\ d_L \epm, \
     \widetilde q_L = \bpm \widetilde u_L\\ \widetilde d_L\epm\Bigg) \ , \\
  & U_R({{\bf \bar 3}},{{\bf 1}}, -2/3) =
      \big( u_R^c, \widetilde u_R^\dag\big) \ , \qquad
    D_R({{\bf \bar 3}},{{\bf 1}},  1/3) =
      \big( d_R^c, \widetilde d_R^\dag\big)\ ,
\esp\ee
where the SM left-handed ($q_L$) and right-handed ($u_R$ and $d_R$)
quarks, as well as left-handed ($\ell_L$) and right-handed ($e_R$) leptons, are
complemented by scalar counterparts, the left-handed ($\widetilde q_L$) and
right-handed ($\widetilde u_R$ and $\widetilde d_R$) squarks and left-handed
($\widetilde\ell_L$) and right-handed ($\widetilde e_R$) sleptons. We have moreover
indicated, in the equation above, the representation of the various
supermultiplets under the MSSM gauge group
$SU(3)_c\times SU(2)_L\times U(1)_Y$.
The MSSM Higgs sector is constituted of two chiral supermultiplets $H_D$ and
$H_U$,
\be\bsp
 &H_D({{\bf 1}}, {{\bf 2}}, -1/2) \equiv
   \Bigg( H_d = \bpm H_d^0 \\ H_d^- \epm,\
    \widetilde H_d = \bpm \widetilde H_d^0 \\ \widetilde H_d^- \epm\Bigg)\ , \\
 &H_U({{\bf 1}}, {{\bf 2}}, 1/2) \equiv
   \Bigg( H_u = \bpm H_u^+ \\ H_u^0 \epm,\
    \widetilde H_u = \bpm \widetilde H_u^+ \\ \widetilde H_u^0\epm \Bigg)\ ,
\esp\ee
which allows for the cancellation of chiral anomalies and to generate masses for
all up-type and down-type particles.

All kinetic and gauge-interaction Lagrangian terms are fixed by gauge and
SUSY invariance, and can be casted very compactly within the superspace
formalism~\cite{Salam:1974yz,Salam:1974jj,Ferrara:1974ac},
\be\bsp
   {\cal L}_{\rm kin} = &\ \sum_\Phi
   \Bigg[
       \Phi^\dag \Big(e^{-2 g_y V_B} e^{-2 g_w V_W} e^{-2 g_s V_G}\Big)
         \Phi \Bigg]_{\theta\cdot\theta \bar\theta\cdot\bar\theta}
    \\ &\qquad + \Bigg\{ \Bigg[
    \frac14 W_B^\alpha W_{B\alpha} +
    \frac{1}{16 g_w^2} W_{Wk}^\alpha W^k_{W\alpha}  + 
    \frac{1}{16 g_s^2} W_{Ga}^\alpha W^a_{G\alpha} 
   \Bigg]_{\theta\cdot\theta} + {\rm h.c.} \Bigg\}\ ,
\esp\ee
where the notation $[\ .\ ]_X$ indicates that after expanding the superfield
inside the bracket in terms of the Grassmanian variables $\theta$ and
$\bar\theta$, only the $X$-component is retained. The first line of the
Lagrangian refers to the chiral sector of the theory and includes a sum upon the
superfields associated with all the previously-introduced chiral
supermultiplets. The vector superfields appearing in the exponents are
considered contracted with the relevant representation matrices, and
$g_y$, $g_w$ and $g_s$ denote the hypercharge, weak and strong coupling
constants. The last line of this Lagrangian describes the gauge sector, and
involves squares of the superfield strength tensors associated with the three
gauge subgroups, the summed spin indices $\alpha$ being indicated explicitly.

Assuming $R$-parity conservation to avoid the presence of baryon- and
lepton-number violating interactions, which will challenge the experimental
observations~\cite{Farrar:1978xj}, the superpotential interactions include
Yukawa
couplings, generating in particular the SM quark and lepton masses, as well as
the off-diagonal Higgs mass-mixing $\mu$ term. The corresponding Lagrangian in
superspace is written as
\be
 {\cal L}_{\rm superW} = \Big[W_{\rm MSSM}(\Phi)\Big]_{\theta\cdot\theta}
    + {\rm h.c.}\ ,
\ee
where the superpotential $W_{\rm MSSM}(\Phi)$ is given, in the flavour space and
with all flavour indices understood for clarity, by
\be
  W_{\rm MSSM}(\Phi) = U_R \ {\bf y^u}\ Q_L H_U - D_R \ {\bf y^d}\ Q_L H_D +
    E_R \ {\bf y^e}\ L_L H_D  + \mu H_U H_D  \ .
\label{eq:superW}\ee
In this expression, the matrices ${\bf y^u}$, ${\bf y^d}$ and ${\bf y^e}$ are
the usual $3\times3$ Yukawa matrices.

The expansion of the above Lagrangian in terms of the component fields relies on
standard techniques detailed, \eg, in refs.~\cite{Wess:1992cp,Fuks:2014xpa}, and
allows for the extraction of the SUSY-conserving part of the MSSM Lagrangian. The
results read
\be\label{eq:lag}\bsp
 & {\cal L}^{\rm (SUSY)}_{\rm MSSM} = \sum_k \bigg[
    - \frac14 V_k^{\mu \nu} V^k_{\mu \nu}
    + \frac{i}{2}\bigg(\widetilde V^k \sigma^\mu D_\mu \overline{\widetilde V}_k
    - D_\mu \widetilde V^k\sigma^\mu \overline{\widetilde V}_k\bigg) \bigg]\\
  & \ + \sum_i \bigg[
      D_\mu \phi_i^\dag D^\mu \phi^i
    + \frac{i}{2} \Big(\psi^i \sigma^\mu D_\mu\bar\psi_i - D_\mu \psi^i
       \sigma^\mu  \bar\psi_i\Big)
    + i\sqrt{2} \sum_k \Big(g_k \overline{\widetilde V}^k \!\cdot\!
      \bar\psi_i T_k \phi^i
    + {\rm h.c.} \Big)\bigg]\\
  &\ -\frac12 \sum_{i,j}\Bigg[
    \frac{\partial^2 W_{\rm MSSM}(\phi)}{\partial\phi^i\partial\phi^j}
       \psi^i \cdot \psi^j + {\rm h.c.}\Bigg]\\ &\
   - \sum_i{\frac{\partial W_{\rm MSSM}(\phi)}{\partial \phi^i}
     \frac{\partial W_{\rm MSSM}^\star(\phi^\dag)}{\partial \phi^\dag_i}}
   - \frac12 \sum_k\Bigg[\bigg(g_k \sum_j\Big[\phi_j^\dag T^k \phi^j\Big]\bigg)
      \bigg(g_k\sum_i\Big[\phi_i^\dag T_k \phi^i\Big]\bigg)\Bigg]  \ ,
\esp \ee
where all fermionic fields are two-component left-handed ($\psi$ and
$\widetilde V$) or right-handed ($\bar\psi$ and $\overline{\widetilde V}$)
spinors, the dot products are invariant products in spin space,
$\sigma_\mu= (1,\sigma^i)$ consists of one of the possible four-vectors
built upon the Pauli matrices and we used the usual gauge field strength tensor
$V_{\mu\nu}$ and covariant derivatives $D_\mu$ taken in the appropriate
representation of the gauge group. The summation over $k$ refers to the three
MSSM gauge groups (the generic notation $T_k$ and $g_k$ being used for the gauge-group 
representation matrices and coupling constants), whereas we use the
indices $i$ and $j$ and the generic
$(\phi, \psi)$ notation for the sum over the scalar and fermionic component of
the model supermutiplets. In addition, the gaugino-scalar-fermion interactions
are included in the last term of the second line, the
third line is constituted of the Yukawa interactions deduced from the
superpotential, and the scalar potential is shown in the last line of the above
Lagrangian.

\subsubsection{Soft supersymmetry breaking and particle mixings}
As for any phenomenologically realistic SUSY theory, the MSSM exhibits
soft SUSY breaking to introduce a mass splitting between a SM particle
and its superpartner. We supplement the Lagrangian of eq.~\eqref{eq:lag} by all
possible soft terms breaking SUSY explicitly~\cite{Girardello:1981wz},
\be\bsp 
   {\cal L}_{\rm MSSM}^{\rm (soft)} =
   &\ \frac12 \Big[ 
     M_1 \widetilde B \!\cdot\! \widetilde B + 
     M_2 \widetilde W \!\cdot\! \widetilde W + 
     M_3 \widetilde G \!\cdot\! \widetilde G + 
     {\rm h.c.} \Big]
      - \tilde q_L^\dag    {\bf m^2_{\tilde Q}} \tilde q_L
      - \tilde u_R         {\bf m^2_{\tilde U}} \tilde u_R^\dag
   \\ &\
      - \tilde d_R         {\bf m^2_{\tilde D}} \tilde d_R^\dag
      - \tilde \ell_L^\dag {\bf m^2_{\tilde L}} \tilde \ell_L
      - \tilde e_R         {\bf m^2_{\tilde E}} \tilde e_R^\dag
      - m_{H_u}^2 H_u^\dag H_u
      - m_{H_d}^2 H_d^\dag H_d\\
   &\ - \Big[
             \tilde u_R^\dag {\bf T^u}\tilde q_L \!\cdot\! H_u 
           - \tilde d_R^\dag {\bf T^d}\tilde q_L \!\cdot\! H_d 
           - \tilde e_R^\dag {\bf T^e}\tilde \ell_L \!\cdot\! H_d
           + b H_u \!\cdot\! H_d  + {\rm h.c.}\Big]\ .
\esp\label{eq:lsoft}\ee
The soft SUSY-breaking Lagrangian first includes mass terms for the
gauginos, the mass parameters of the $U(1)_Y$, $SU(2)_L$ and $SU(3)_c$
gauginos being denoted by $M_1$, $M_2$ and $M_3$. The following seven terms consist
of mass terms for all scalar fields, the parameters ${\bf m_{\tilde Q}}$,
${\bf m_{\tilde L}}$, ${\bf m_{\tilde U}}$, ${\bf m_{\tilde D}}$ and
${\bf m_{\tilde E}}$ being $3\times3$ Hermitian matrices in the flavour space and
$m_{H_u}$ and $m_{H_d}$ are the two Higgs mass parameters. In addition, bilinear
and trilinear soft multiscalar interactions can be deduced from the form of the
superpotential, the corresponding coupling strengths being organised in the
three $3\times 3$ ${\bf T^u}$, ${\bf T^d}$ and ${\bf T^e}$ matrices (in the flavour
space) and the complex number $b$ related to the soft SUSY-breaking
Higgs mixing term.

At the minimum of the scalar potential, the neutral components of both Higgs
doublets $H_u^0$ and $H_d^0$ get non-vanishing vacuum expectation values and
the electroweak symmetry is broken. As a result, the electroweak vector bosons
$B_\mu$ and $W_\mu^3$ mix into the massless photon $A_\mu$ and massive
$Z_\mu$ states,
\be
  \bpm  A_\mu \\ Z_\mu \epm = \bpm \cos\theta_w & \sin\theta_w\\ -\sin\theta_w
    & \cos\theta_w \epm \bpm B_\mu \\ W_\mu^3 \epm \ ,
\ee
where we have introduced the electroweak mixing angle $\theta_w$. As in the SM,
the charged weak boson physical states are defined by diagonalising the third
generator of $SU(2)$ in the adjoint representation, which gives rise to
\be
  W_\mu^\pm = \frac{1}{\sqrt{2}} (W_\mu^1 \mp i W^2_\mu) \ .
\ee
The eight degrees of freedom included within the two Higgs doublets give rise to
three Goldstone bosons $G^\pm$ and $G^0$ that become the longitudinal modes of
the weak gauge bosons and five physical Higgs bosons $h^0$, $H^0$, $A^0$ and $H^\pm$
defined by diagonalising the Higgs sector,
\be\bsp
  H_u^0 = \frac{v_u}{\sqrt{2}} + \cos\alpha\ h^0 + \sin\alpha\ H^0 +
    i \cos\beta\ A^0 + i \sin\beta\ G^0\ , \quad &
  H_u^+ = \cos\beta\ H^+ +\sin\beta\ G^+\ , \\
  H_d^0 = \frac{v_d}{\sqrt{2}} - \sin\alpha\ h^0 + \cos\alpha\ H^0 +
    i \sin\beta\ A^0 - i \cos\beta\ G^0\ , \quad &
  H_d^- = \sin\beta\ H^- - \cos\beta\ G^-\ ,
\esp\ee
where we have introduced the Higgs mixing angles $\alpha$
and $\beta$, the tangent of the latter being given by the ratio of the vacuum
expectation values of the neutral Higgs gauge eigenstates, $\tan\beta=v_u/v_d$.

Mixings in the fermionic electroweak sector are also induced by the breaking of
the electroweak symmetry. In the neutral sector, the gaugino and Higgsino gauge
eigenstates mix into the physical neutralino eigenstates ${\chi}^0_i$
(with $i=1, 2, 3, 4$), whereas in the charged sector, they
mix into the two chargino states ${\chi}^\pm_i$ (with $i=1, 2$),
\be
  \bpm {\chi}^0_1 \\ {\chi}^0_2 \\ {\chi}^0_3 \\ {\chi}^0_4 \epm = N
  \bpm i \tilde B \\ i \tilde W^3 \\ \tilde H_d^0 \\  \tilde H_u^0\epm \ ,\qquad
  \bpm {\chi}^+_1 \\ {\chi}^+_2 \epm = V \bpm i \tilde W^+ \\ \tilde H^+_u \epm
  \quad \text{and}\quad
  \bpm {\chi}^-_1 \\ {\chi}^-_2 \epm = U \bpm i \tilde W^- \\ \tilde H^-_d \epm \ .
\ee
In those relations, the mixing matrices $N$, $U$ and $V$ are unitary and allow
to diagonalise the neutral and charged electroweakino mass matrices.

As in the SM, the diagonalisation of the quark sector requires four unitary
matrices $V_u$, $V_d$, $U_u$ and $U_d$ and the one of the lepton sector relies
on three unitary rotation matrices $V_e$, $V_\nu$ and $U_e$ as we have omitted
the right-handed neutrino (super)fields in the model definition. This leads to
the following field redefinitions in the flavour space,
\be\bsp
  d_L   \to V_d &d_L\ , \quad
  d_R^c \to d_R^c U_d^\dag\ ,\quad
  u_L   \to V_u u_L \quad \text{and} \quad
  u_R^c \to u_R^c U_u^\dag\ ,\\
  & e_L   \to V_e e_L \ , \quad 
  e_R^c \to e_R^c U_e^\dag\quad\text{and} \quad 
  \nu_L \to V_\nu \nu_L\ ,
\esp\ee
and we follow the traditional approach of casting these rotations through a
redefinition of the left-handed down-type quark field only via the CKM matrix
$V_{\rm CKM}$,
\be
  d_L \to V_{\rm CKM} d_L = V_u^\dag V_d d_L \ .
\label{eq:sckm1}\ee
In agreement with SUSY, these field redefinitions are promoted to the
superfield level so that one must consider an extra rotation acting on the basis
of left-handed down-type squarks similar to the one of eq.~\eqref{eq:sckm1},
\be
  \tilde d_L \to V_{\rm CKM} \tilde d_L\ .
\label{eq:sckm2}\ee
The two field redefinitions of eq.~\eqref{eq:sckm1} and eq.~\eqref{eq:sckm2}
define the so-called super-CKM basis~\cite{Hall:1985dx} in which the resulting
$6\times 6$ squark mass matrices are non-diagonal. Following the SUSY Les
Houches Accord (SLHA) conventions~\cite{Skands:2003cj}, the superpotential and
soft parameters are redefined according to
\be\bsp
  & {\bf y^u} \to {\bf \hat y^u} = U_u^\dag {\bf y^u} V_u  \ ,\qquad
  {\bf y^d} \to {\bf \hat y^d} V_d^\dag V_u = U_d^\dag {\bf y^d} V_u \ ,
  \qquad
  {\bf y^e} \to {\bf \hat y^e} = U_e^\dag {\bf y^e} V_e  \ ,\\
  & {\bf T^u} \to {\bf \hat T^u} = U_u^\dag {\bf T^u} V_u\ , \qquad
  {\bf T^d} \to {\bf \hat T^d V_{\rm CKM}^\dag} = U_d^\dag {\bf T^d} V_u \ ,
  \qquad
  {\bf T^e} \to  {\bf \hat T^e} = U_e^\dag {\bf T^e} V_e \ , \\
  & {\bf m^2_{\tilde Q}} \to V_{\rm CKM} {\bf \hat m^2_{\tilde Q}}
    V_{\rm CKM}^\dag \ ,\qquad
  {\bf m^2_{\tilde U}} \to {\bf \hat m^2_{\tilde U}} =
    U_u^\dag {\bf m^2_{\tilde U}} U_u \ , \qquad
  {\bf m^2_{\tilde D}} \to {\bf \hat m^2_{\tilde D}} =
     U_d^\dag {\bf m^2_{\tilde D}} U_d \ ,\\
  & {\bf m^2_{\tilde L}} \to {\bf \hat m^2_{\tilde L}} =
     V_e^\dag {\bf m^2_{\tilde L}} V_e \ ,  \qquad
  {\bf m^2_{\tilde E}} \to {\bf \hat m^2_{\tilde E}} =
     U_e^\dag {\bf m^2_{\tilde E}} U_e \ ,
\esp\ee
where the hatted quantities refer to the new free parameters of the theory. The
matrices ${\bf \hat y^u}$, ${\bf \hat y^d}$ and ${\bf \hat y^e}$ are diagonal
and real $3 \times 3$ matrices in the flavour space, whereas all the other matrices
are in principle possibly flavour-violating and $CP$-violating. We however assume
a constrained realisation of the MSSM in which organising principles of the soft
terms forbid any source of flavour and $CP$ violation on top of those inherent to
the CKM matrix. In this case, all subsequent flavour-violating effects in the
squark sector are small, and each squark flavour turns out to be aligned with the
associated quark flavour.

Following the SLHA conventions and the standard MSSM literature, the
${\bf \hat T^f}$ matrices are decomposed as
\be\label{eq:TtoA}
  {\bf \hat T^f} = {\bf A^f} {\bf \hat y^f}\ ,
\ee
where the overall strength of the trilinear scalar interactions for a (s)fermion
species $f$ is embedded in
the three $3\times 3$ ${\bf A^f}$ (diagonal and real) matrices in the flavour
space. As a consequence and for not
too extreme values of the coupling strengths ${\bf A^f}$, typical MSSM scenarios
only exhibit a flavour-conserving mixing of the third generation sfermions, any
other mixing being subdominant and negligible. Such a mixing is modelled through
the stop ($\theta_{\tilde t}$), sbottom ($\theta_{\tilde b}$) and stau
($\theta_{\tilde \tau}$) mixing angles, and the stop ($\tilde t_1$,
$\tilde t_2$), sbottom ($\tilde b_1$, $\tilde b_2$) and stau ($\tilde\tau_1$,
$\tilde \tau_2$) mass-ordered physical states are related to the corresponding
gauge eigenstates as
\be\bsp
  \bpm \tilde t_1\\ \tilde t_2\epm \!=\!
    S_{\tilde t} \bpm \tilde t_L\\ \tilde t_R \epm \!=\!
    \bpm \phantom{-}\cos\theta_{\tilde t} & \sin\theta_{\tilde t}\\
         -\sin\theta_{\tilde t} & \cos\theta_{\tilde t} \epm
    \! \bpm \tilde t_L\\ \tilde t_R \epm \ ,\ \
  \bpm \tilde b_1\\ \tilde b_2\epm \!=\!
    S_{\tilde b} \bpm \tilde b_L\\ \tilde b_R \epm & \!=\!
    \bpm \phantom{-}\cos\theta_{\tilde b} & \sin\theta_{\tilde b}\\
         -\sin\theta_{\tilde b} & \cos\theta_{\tilde b} \epm
    \! \bpm \tilde b_L\\ \tilde b_R \epm \ ,\\
  \bpm \tilde \tau_1\\ \tilde \tau_2\epm =
    S_{\tilde \tau} \bpm \tilde \tau_L\\ \tilde \tau_R \epm =
    \bpm \phantom{-}\cos\theta_{\tilde \tau} & \sin\theta_{\tilde \tau}\\
         -\sin\theta_{\tilde \tau} & \cos\theta_{\tilde \tau} \epm &
    \bpm \tilde \tau_L\\ \tilde \tau_R \epm \ .
\esp\label{eq:3rdgenmix}\ee

\subsubsection{Two-component and four-component fermions}
All the two-component fermionic fields introduced so far are finally combined to
form Dirac and Majorana spinors $\Psi$, that are the fermion representations
supported at the level of the Monte Carlo event generators which we plan
to use for our phenomenological study. The SM, gluino,
chargino and neutralino four-component fermions are defined by
\be\bsp
  &\
    \Psi_u = \bpm u_L \\ \bar u_R^c\epm \ , \quad 
    \Psi_d = \bpm d_L \\ \bar d_R^c\epm \ , \quad 
    \Psi_e = \bpm e_L \\ \bar e_R^c\epm \ , \quad 
    \Psi_\nu = \bpm \nu_L \\ 0\epm \ , \\
  &\qquad \qquad
    \Psi_{\chi^0} = \bpm \chi^0 \\ \bar \chi^0 \epm \ , \quad 
    \Psi_{\chi^\pm} = \bpm \chi^\pm \\ \bar \chi^\mp \epm \ , \quad 
    \Psi_{\tilde g} = \bpm i \tilde g \\ -i \overline {\tilde g}\epm \ .
\esp\ee 
Although the tree-level form of all mixing matrices introduced so far can be
easily calculated from the Lagrangians of eq.~\eqref{eq:lag} and
eq.~\eqref{eq:lsoft}, loop correction effects are important. One-loop and known
two-loop contributions are hence in general included in all available MSSM
spectrum generators~\cite{Chankowski:1992er,Dabelstein:1994hb,Pierce:1996zz,
Goodsell:2014bna}.

\subsection{Renormalisation}
\subsubsection{Generalities}\label{sec:general}
Focusing in the following on NLO calculations in $\as$, we rotate the Lagrangians
of eqs.~\eqref{eq:lag} and \eqref{eq:lsoft} to the mass basis and omit from the
discussion any term that is irrelevant with respect to the strong interaction,
\be\bsp
  & {\cal L}_{\rm MSSM}^{\rm (QCD)} = {\cal L}_{\rm SM}^{\rm (QCD)}
    + \sum_{\tilde q_k} \Big[ D_\mu \tilde q_k^\dag D^\mu \tilde q_k -
          m_{\tilde q_k}^2 \tilde q_k^\dag \tilde q_k \Big]
    + \frac{i}{2} {\bar{\Psi}_{\tilde g}}\slashed{D}\Psi_{\tilde g} - \frac12m_{\tilde g} {\bar{\Psi}_{\tilde g}}\Psi_{\tilde g}\\
  &\quad
   - \frac{g_s^2}{2} \bigg\{ \sum_q\Big[
     (S_{\tilde q})_{j2}(S_{\tilde q})_{2i}^\ast\ \tilde q_j^\dag T\tilde q_i -
     (S_{\tilde q})_{j1}(S_{\tilde q})_{1i}^\ast\ \tilde q_j^\dag T\tilde q_i
       \Big]\bigg\}^2\\
  &\quad
    + \sqrt{2} g_s \sum_q
       \Big[ -(S_{\tilde q})_{j1}\ \tilde q_j^\dag\ T \big({\bar{\Psi}_{\tilde g}}P_L\Psi_q\big)
          + \big({\bar{\Psi}_q} P_L \Psi_{\tilde g}\big) T (S_{\tilde q})_{2j}^\ast \tilde q_j 
          + {\rm h.c.} \Big] + \ldots
\esp\label{eq:lsqcd}\ee
In this expression, ${\cal L}_{\rm SM}^{\rm (QCD)}$ denotes the QCD part of the
SM Lagrangian involving quarks and gluons, the sum over $\tilde q_k$ refers
to a sum over all twelve squark mass-eigenstates ($\tilde u_L$, $\tilde c_L$,
$\tilde t_1$, $\tilde u_R$, $\tilde c_R$, $\tilde t_2$, $\tilde d_L$,
$\tilde s_L$, $\tilde b_1$, $\tilde d_R$, $\tilde s_R$, $\tilde b_2$) of masses
$m_{\tilde q_k}$ and the sums over $q$ refer to sums over all six quark
flavours. In the former sums, the mixing matrices associated with the first and
second generation squarks are taken as $2\times 2$ identity matrices, so that
the first and second generation mass-eigenstates $\tilde q_1$ and $\tilde q_2$
are subsequently identified with the left-handed and right-handed squarks
$\tilde q_L$ and $\tilde q_R$, respectively. In addition, $m_{\tilde g}$ stands
for the gluino mass, $T$ for the fundamental representation matrices of
$SU(3)$, $P_{L,R}$ for the left-handed and right-handed chirality projectors,
$g_s=\sqrt{4\pi\as}$ for the strong coupling constant and the covariant
derivatives are restricted to their QCD component.

Ultraviolet divergences appearing at the one-loop level are absorbed into the
counterterms generated by the renormalisation of the above Lagrangian. Following
the usual procedure, all bare bosonic fields $\Phi$ and fermionic fields $\Psi$
are replaced by their renormalised counterparts,
\be\label{eq:renofield}
  \Phi \to \big[ 1 + \frac12 \delta Z_\Phi\big] \Phi\qquad\text{and}\qquad
  \Psi \to \big[ 1 + \frac12 \delta Z^L_\Psi P_L +
      \frac12 \delta Z^R_\Psi P_R \big]\Psi\ ,
\ee
with the exception of third generation squarks for which matrix renormalisation
is in order as they mix,
\be
  \bpm \tilde q_1\\ \tilde q_2\epm \to
  \bpm \tilde q_1\\ \tilde q_2\epm + \frac12 \bpm
   \delta Z_{\tilde q,11} & \delta Z_{\tilde q, 12} \\
   \delta Z_{\tilde q,21} & \delta Z_{\tilde q, 22} \epm
  \bpm \tilde q_1\\ \tilde q_2\epm \qquad\text{with}\qquad
   \tilde q = \tilde b, \tilde t \ .
\label{eq:stopsbotmix}\ee
Although the structure of the gluino-squark-quark interactions (the last line of
eq.~\eqref{eq:lsqcd}) could induce the mixing of any squark flavour at the
one-loop level, those effects are proportional to the corresponding quark
masses. Considering $n_{lf}=4$ flavours of massless quarks, the first two
generations are kept non-mixing so that gauge and mass eigenstates are
equivalent. In addition, the bare parameters of the MSSM Lagrangian, generically
denoted by $y$ (for couplings) and $m$ (for masses), are renormalised as
\be
 y \to y + \delta y \qquad\text{and}\qquad m\to m + \delta m\ .
\label{eq:renoprm}\ee

In this work, we calculate the various renormalisation constants appearing in
the renormalisation procedure of the Lagrangian of eq.~\eqref{eq:lsqcd} in the
on-shell (OS) scheme where the input parameters are physical observables such as
the physical particle masses. There is however no unique definition of such a
scheme in SUSY by virtue of existing interrelations between various
mass and coupling parameters, which will be addressed in
section~\ref{sec:susy-onshell}. Fermion self-energy corrections $\Sigma(p)$ are
decomposed in terms of independent Lorentz structures,
\be
  \Sigma(p) = -i \Big[
    \Sigma^V_L(p^2) \slashed{p} P_L + \Sigma^V_R(p^2) \slashed{p} P_R
      + \Sigma^S_L(p^2) P_L + \Sigma^S_R(p^2) P_R \Big]\ ,
\ee
from which the OS fermionic wave-function renormalisation constants
$\delta Z_f^{L,R}$ and mass renormalisation constant $\delta m_f$ can be
deduced. Imposing that the renormalised mass is the pole of the propagator and
that the residue of the propagator pole equals one, we get
\be\bsp
  \delta Z_f^{L,R} =&\ \Sigma^V_{L,R}(m_f^2)
    + m_f^2 \Big[\Sigma^{\prime V}_L(m_f^2) + \Sigma^{\prime V}_R(m_f^2)\Big]
    + m_f \Big[\Sigma^{\prime S}_L(m_f^2) +\Sigma^{\prime S}_R(m_f^2)\Big]\ ,\\
  \delta m_f = &\ -\frac12 \Big[\Sigma^V_L(m_f^2) + \Sigma^V_R(m_f^2)\Big]
      - \frac12\Big[\Sigma^S_L(m_f^2) +\Sigma^S_R(m_f^2)\Big] \ ,
\esp\ee
where the prime denotes a derivative with respect to $p^2$. Gauge-boson
self-energy corrections $D_{\mu\nu}(p)$ are reduced, in the case of the gluon
(that is the only relevant gauge boson as long as only $\as$ corrections are
concerned), to their purely transverse component,
\be
  D_{\mu\nu}(p) = i \Big[\eta^{\mu\nu} - \frac{p_\mu p_\nu}{p^2}\Big]
     D^T(p^2)\ .
\ee
The wave-function renormalisation constant $\delta Z_g$ is obtained after
imposing OS renormalisation conditions, and reads
\be
  \delta Z_g = D^{\prime T}(0) \ .
\ee
Finally, as above-mentioned, the scalar quark sector of the theory relies on
matrix renormalisation, any off-diagonal element being vanishing in the case of
the first and second generation non-mixing squarks. We rewrite the scalar
self-energies $\Pi_{ij}(p)$ as
\be
  \Pi_{ij}(p) = -i \Pi^S_{ij}(p^2) \ ,
\ee
where the indices $i$ and $j$ are either 1 or 2. This allows for the derivation
of the diagonal ($\delta Z_{ii}$) and non-diagonal ($\delta Z_{ij}$)
wave-function renormalisation constants,
\be
  \delta Z_{ii} = \Pi^{\prime S}_{ii}(m_i^2)
  \qquad\text{and}\qquad
  \delta Z_{ij} = \delta Z_{ji} =
     \frac{\Pi^S_{ij}(m_j^2) - \Pi^S_{ij}(m_i^2)}{m_j^2-m_i^2} \ ,
\ee
where $m_i^2$ indicates the squared mass of the $i^{\rm th}$ eigenstate, as well
as of the mass renormalisation constant,
\be
  \delta m_i^2 = - \Pi^S_{ii}(m_i^2)\ .
\ee
We do not address the complex-mass renormalisation scheme in this paper, 
so that we kept implicit that only the real-part of the self-energies is considered in the above expressions.
The case of complex-mass renormalisation conditions is especially delicate in the case of NLO computations 
within SUSY theories because their mass spectrum is arbitrary to a large extend thus making it necessary to
implement the most general analytic continuation of the two-point functions~\cite{Frederix:2018nkq}.

\subsubsection{The Standard Model sector}
Beginning with the SM sector, the wave-function renormalisation
constants $\delta Z_q^{L,R}$ of the massless quarks $q=u$, $d$, $s$, $c$ and
$\delta Z_Q^{L,R}$ of the massive bottom and top quarks $Q=b$, $t$, as well as
the one of the gluon $\delta Z_g$ are given by
\be\bsp
  &\ \delta Z_g = -\frac{g_s^2}{24 \pi^2} \Bigg[
   \sum_{Q=b,t}\bigg\{B_0\big(0, m_Q^2, m_Q^2\big)
    + 2 m_Q^2 B_0^\prime\big(0, m_Q^2, m_Q^2\big)
      -\frac13\bigg\}
    + n_c\bigg\{ B_0\big(0, m_{\tilde g}^2, m_{\tilde g}^2\big) \\ &\qquad
      + 2 m_{\tilde g}^2 B_0^\prime\big(0, m_{\tilde g}^2, m_{\tilde g}^2\big)
      - \frac13\bigg\}
    +\sum_{\tilde q_k}\bigg\{
            \frac14 B_0\big(0, m_{\tilde q_k}^2, m_{\tilde q_k}^2\big)
        - m_{\tilde q}^2
            B_0^\prime\big(0, m_{\tilde q_k}^2, m_{\tilde q_k}^2\big)
       +\frac16 \bigg\} \Bigg]\ , \\
 &\ \delta Z_q^{L,R} = \frac{g_s^2 C_F}{8 \pi^2}\
      B_1\big(0; m_{\tilde g}^2, m_{\tilde q_{L,R}}^2\big) \ , \\
 &\ \delta Z_Q^{L,R} = \frac{g_s^2 C_F}{16 \pi^2} \bigg[
        1 
      + 2 B_1\big(m_Q^2; m_Q^2, 0\big)
      + 4 m_Q^2 B_1^\prime\big(m_Q^2; m_Q^2, 0\big)
      + 8 m_Q^2 B_0^\prime\big(m_Q^2; m_Q^2,0\big)
  \\ &\qquad
      + 2 \sum_{i=1}^2 B_1\big(m_Q^2; m_{\tilde g}^2, m_{\tilde Q_i}^2\big)
         \big|(S_{\tilde Q})_{j1,2}\big|^2
      + 2m_Q^2\sum_{i=1}^2
         B_1^\prime\big(m_Q^2; m_{\tilde g}^2, m_{\tilde Q_i}^2\big)
   \\ &\qquad
      + 4m_Q^2\sum_{i=1}^2
       B_0^\prime\big(m_Q^2; m_{\tilde g}^2, m_{\tilde Q_i}^2\big)
        (S_{\tilde Q})_{j1}^\ast (S_{\tilde Q})_{j2}
      \bigg] \ ,
\esp\ee
where the $B_0$ and $B_1$ functions and their derivatives are the real part of the
usual two-point Passarino-Veltman integrals~\cite{Passarino:1978jh} collected in
appendix~\ref{app:PV}. Moreover, $n_c=3$ and $C_F=(n_c^2-1)/(2 n_c)$ stand
respectively for the number of colours and for the quadratic Casimir invariant
connected with the fundamental representation of $SU(3)$, and the sum upon
$\tilde q_k$ refers to a sum over all squark states. In addition, the
bottom and top mass OS renormalisation constants $\delta m_Q$ (with
$Q=b$, $t$) are given by
\be\bsp
   \delta m_Q = &\  -\frac{g_s^2 C_F m_Q}{16 \pi^2} \bigg[
      - 1 + 4 B_0\big(m_Q^2; m_Q^2, 0\big)
      + 2 B_1\big(m_Q^2; m_Q^2, 0\big)
      + \sum_{i=1}^2 B_1\big(m_Q^2; m_{\tilde g}^2, m_{\tilde Q_i}^2\big)
   \\ &\qquad
      + \sum_{i=1}^2 (S_{\tilde Q})_{j1}^\ast (S_{\tilde Q})_{j2}
             B_0\big(m_Q^2; m_{\tilde g}^2, m_{\tilde Q_i}^2\big) \bigg] \ ,
\esp\ee

\subsubsection{Gluino renormalisation}
Gluino renormalisation in the OS scheme is standard, and the corresponding
wave-function renormalisation constant $\delta Z_{\tilde g} \equiv
\delta Z_{\tilde g}^L = \delta Z_{\tilde g}^R$ (as the gluino is a Majorana
fermion) and mass renormalisation constant $\delta m_{\tilde g}$ read
\be\bsp
  \delta Z_{\tilde g} = &\ \frac{g_s^2}{16\pi^2} \Bigg[
    n_c + 2 n_c B_1(m_{\tilde g}^2; m_{\tilde g}^2, 0)
      + 8 n_c m_{\tilde g}^2 B_0^\prime(m_{\tilde g}^2; m_{\tilde g}^2, 0)
      + 4 n_c m_{\tilde g}^2 B_1^\prime(m_{\tilde g}^2; m_{\tilde g}^2, 0)
  \\&\
      + \sum_{\tilde q_k}\bigg\{
          B_1(m_{\tilde g}^2; m_q^2, m_{\tilde q_k}^2) +
          2 m_{\tilde g}^2 B_1^\prime(m_{\tilde g}^2; m_q^2, m_{\tilde q_k}^2)
        \bigg\}
     \Bigg]\ , \\
  \delta m_{\tilde g} = &\ \frac{g_s^2 m_{\tilde g}}{16\pi^2} \bigg[
         n_c - 4 n_c B_0(m_{\tilde g}^2; m_{\tilde g}^2, 0)
       - 2 n_c B_1(m_{\tilde g}^2; m_{\tilde g}^2, 0)
       - \sum_{\tilde q_k} B_1(m_{\tilde g}^2; m_q^2, m_{\tilde q}^2)
    \bigg]\ .
\esp\ee

\subsubsection{On-shell squark renormalisation}
\label{sec:susy-onshell}

{\bf The naive on-shell scheme}\\
Using the standard OS formulas as presented in section~\ref{sec:general} for
deriving the wave-function renormalisation constants of the first and second
generation squarks $\delta Z_{\tilde q}$ and third generation squarks
$\delta Z_{\tilde Q}$, we obtain
\be\bsp
  & \delta Z_{\tilde q} = \frac{g_s^2 C_F}{8\pi^2} \bigg[
       B_0(m_{\tilde q}^2; m_{\tilde q}^2, 0)
     - B_0(m_{\tilde q}^2; m_{\tilde g}^2, 0)
     + 2 m_{\tilde q}^2 B_0^\prime(m_{\tilde q}^2; m_{\tilde q}^2, 0)
     + \Big(m_{\tilde g}^2 - m_{\tilde q}^2\Big)
        B_0^\prime(m_{\tilde q}^2; m_{\tilde g}^2, 0)
   \bigg] \ , \\
  & (\delta Z_{\tilde Q})_{ii} = \frac{g_s^2 C_F}{8\pi^2} \bigg[
       B_0(m_{\tilde Q_i}^2; m_{\tilde Q_i}^2, 0)
     - B_0(m_{\tilde Q_i}^2; m_{\tilde g}^2, m_Q^2)
     + 2 m_{\tilde Q_i}^2 B_0^\prime(m_{\tilde Q_i}^2; m_{\tilde Q_i}^2, 0)
  \\ &\hspace{3cm}
     + \Big(m_{\tilde g}^2 + m_Q^2 - m_{\tilde Q_i}^2 - 4 m_{\tilde g} m_Q
        \Re\big\{(S_{\tilde Q})_{i1}^\ast (S_{\tilde Q})_{i2}\big\}\Big)
        B_0^\prime(m_{\tilde Q_i}^2; m_{\tilde g}^2, m_Q^2)
   \bigg] \ , \\
  & (\delta Z_{\tilde Q})_{ij} = -\frac{g_s^2 C_F m_{\tilde g} m_Q}
     {4\pi^2} 
     \Big( (S_{\tilde Q})_{i1}^\ast (S_{\tilde Q})_{j2} +
       (S_{\tilde Q})_{i2}^\ast (S_{\tilde Q})_{j1} \Big)
       \frac{B_0(m_{\tilde Q_j}^2; m_Q^2, m_{\tilde g}^2) - 
          B_0(m_{\tilde Q_i}^2; m_Q^2, m_{\tilde g}^2)}
          {m_{\tilde Q_j}^2-m_{\tilde Q_i}^2}\ .
\esp\ee
Similarly, the corresponding mass renormalisation constants read
\be\label{eq:naivemasses}\bsp
  & \delta m_{\tilde q}^2 = \frac{g_s^2 C_F}{8\pi^2} \bigg[
      A_0(m_{\tilde q}^2) - A_0(m_{\tilde g}^2)
    - 2 m_{\tilde q}^2 B_0(m_{\tilde q}^2; m_{\tilde q}^2, 0)
    + (m_{\tilde q}^2 - m_{\tilde g}^2) B_0(m_{\tilde q}^2; m_{\tilde g}^2, 0)
   \bigg] \ ,\\
  & \delta m_{\tilde Q_i}^2 = \frac{g_s^2 C_F}{8\pi^2} \bigg[
    \Big(m_{\tilde Q_i}^2\!-\!m_{\tilde g}^2\!-\!m_Q^2\!+\!4 m_{\tilde g} m_Q
      \Re\big\{(S_{\tilde Q_i})_{i1}^\ast (S_{\tilde Q})_{i2}\big\}\Big)
       B_0(m_{\tilde Q}^2; m_{\tilde g}^2, m_Q^2)
      + \frac{A_0(m_{\tilde Q_i}^2)}{2}
    \\ &\hspace{0.5cm}
      - A_0(m_{\tilde g}^2) - A_0(m_Q^2)
      - 2 m_{\tilde Q_i}^2 B_0(m_{\tilde Q_i}^2; m_{\tilde Q_i}^2, 0)
      + \frac12 \sum_{k=1,2} \Big\{ \big|(S_{\tilde Q})_{ik}\big|^2
         A_0(m_{\tilde Q_k}^2) \Big\}
    \bigg] \ ,
\esp\ee
and the renormalisation of the top and bottom squark mixing angles is related to
the one of their respective wave functions~\cite{Eberl:1996np},
\be
  \delta\theta_{\tilde t} = \frac14 \Big[
     \delta Z_{\tilde t,12} - \delta Z_{\tilde t,21}^\ast\Big]
    \qquad\qquad
    \text{and}\qquad\qquad
  \delta\theta_{\tilde b} = \frac14 \Big[
     \delta Z_{\tilde b,12} - \delta Z_{\tilde b,21}^\ast\Big] \ .
\label{eq:renomix}\ee

This scheme however breaks weak interaction gauge invariance, as the physical
squark masses are not allowed to be taken all independent. Left-handed up-type
and down-type squarks of a given generation are indeed connected by $SU(2)_L$ so
that they can consequently not be renormalised independently. Such a scheme is
however useful and valid for many phenomenological applications relying on
simplified models inspired by the MSSM in which only a few particles and a
subset of all MSSM Lagrangian terms are supplemented to the SM, as
for instance in the work of refs.~\cite{Degrande:2014sta,Degrande:2015vaa} or
for the numerical results presented in the following sections of this paper. In
the latter case, the relations between the physical squark masses are ignored as
the relevant terms are not present in the simplified model Lagrangian, so that
all fields can be renormalised independently. This approach however breaks down
as soon as one considers an entire generation of squarks and wants to retain
$SU(2)_L$ gauge invariance as embedded in the MSSM.

In the rest of this subsection, we
additionally present two of the most popular SUSY OS schemes, that will
not be considered in our numerical simulation but that could easily be implemented
in our framework as will be shown in section~\ref{sec:bfl_example}. Whilst the
differences between all the possible schemes are formally of higher order, the
corresponding higher-order contributions could be potentially large in some
parts of the parameter space (for instance, when $\tan\beta$ is large).
Moreover, the different schemes necessitate different sets of input parameters,
which becomes relevant for comparing their respective predictions.\\

\noindent {\bf The `$m_b$ on-shell' scheme}\\
As above-mentioned, gauge invariance under weak interactions implies that the
masses of the left-handed up-type and down-type squarks of a given generation
are connected to a unique bare soft mass parameter ${\bf\hat m^2_{\tilde Q}}$.
The tree-level squared mass matrices ${\cal M}_{\tilde q_u}^2$ and
${\cal M}_{\tilde q_d}^2$ associated with the up-type and down-type squarks of a
given generation are indeed given, in the $(\tilde q_L, \tilde q_R)$ basis, by
\be\label{eq:sqmasses}\bsp
  {\cal M}_{\tilde q_u}^2 = &\ \bpm
    \big({\bf \hat m^2_{\tilde Q}}\big)_{nn} \! + \! m_{q_u}^2 \! +\!
      m_Z^2c_{2\beta}\big[\frac12 \!-\! \frac23 s_w^2\big] &
        m_{q_u} \Big(\big({\bf A^u}\big)_{nn} \!-\! \mu\cot\beta\Big)\\
     m_{q_u} \Big(\big({\bf A^u}\big)_{nn} \!-\! \mu\cot\beta\Big) &
      \big({\bf \hat m^2_{\tilde U}}\big)_{nn} \! +\! m_{q_u}^2 \! +\!
      \frac23 m_Z^2c_{2\beta} s_w^2\\
  \epm \ ,\\[.3cm]
  {\cal M}_{\tilde q_d}^2 = &\ \bpm
    \big({\bf \hat m^2_{\tilde Q}}\big)_{nn} \! +\! m_{q_d}^2 \! -\!
      m_Z^2 c_{2\beta}\big[ \frac12 \!-\! \frac13 s_w^2\big] & m_{q_d}
      \Big(\big({\bf A^d}\big)_{nn} \!-\! \mu \tan\beta\Big)\\
     m_{q_d} \Big(\big({\bf A^d}\big)_{nn} \!-\! \mu \tan\beta\Big) &
      \big({\bf \hat m^2_{\tilde D}}\big)_{nn} \! +\! m_{q_d}^2 \!-\! \frac13
      m_Z^2c_{2\beta} s_w^2\\
  \epm \ ,
\esp\ee
where $m_Z$, $c_{2\beta}$ and $s_w$ stand for the mass of the $Z$-boson,
$\cos2\beta$ and $\sin\theta_w$. We have furthermore indicated by a subscript
$n$ the relevant generation index, and $m_{q_u}$ and $m_{q_d}$ are the masses of
the corresponding up-type and down-type quarks $q_u$ and $q_d$. While for the
first and second generations the off-diagonal terms vanish and those two
matrices are diagonal, they must be further diagonalised for third generation
squarks with the help of the two rotation matrices $S_{\tilde t}$ and
$S_{\tilde b}$,
\be
  \bpm m_{\tilde t_1}^2 & 0 \\ 0 & m_{\tilde t_2}^2\epm =
    S_{\tilde t}\ {\cal M}_{\tilde t}^2\ S_{\tilde t}^\dag
  \qquad\text{and}\qquad
  \bpm m_{\tilde b_1}^2 & 0 \\ 0 & m_{\tilde b_2}^2\epm = 
    S_{\tilde b}\ {\cal M}_{\tilde b}^2\ S_{\tilde b}^\dag \ .
\label{eq:sqmassdiag}\ee
Consequently, one of the four mass parameters associated with the first or the
second generation of squarks is a dependent parameter and cannot be renormalised
by imposing naive OS renormalisation conditions. Similarly, care must be taken
with the stop/sbottom sector where we have six quark and squark masses
($m_b$, $m_t$, $m_{\tilde t_1}$, $m_{\tilde t_2}$, $m_{\tilde b_1}$ and
$m_{\tilde b_2}$), two mixing angles ($\theta_{\tilde t}$ and
$\theta_{\tilde b}$) as well as two soft trilinear interaction strengths ($A_t
\equiv ({\bf A^u})_{33}$ and $A_b\equiv({\bf A^d})_{33}$). All these parameters
are related and thus receive one-loop $\as$ corrections in a connected
manner~\cite{Yamada:1996jf,Bartl:1997yd,Bartl:1998xp,Eberl:1999he}.

In the so-called `$m_b$ on-shell' scheme, the renormalisation of the up-type and
down-type squark sectors is performed simultaneously~\cite{Hollik:2003jj,
Heinemeyer:2004xw,Hollik:2008yi}. We consider the masses of the left-handed down
and strange squarks as well as the one of the heaviest bottom squark as
dependent parameters,
\be\bsp
  & m_{\tilde d_L}^2 = m_{\tilde u_L}^2 - m_Z^2 c_{2\beta} c_w^2 \ ,
  \qquad\qquad
  m_{\tilde s_L}^2 = m_{\tilde c_L}^2 - m_Z^2 c_{2\beta} c_w^2 \ ,\\
  & m_{\tilde b_2}^2 = \frac{1}{\big|(S_{\tilde b})_{12}\big|^2} \bigg[
   \big|(S_{\tilde t})_{11}\big|^2 m_{\tilde t_1}^2 +
   \big|(S_{\tilde t})_{21}\big|^2 m_{\tilde t_2}^2 -
   \big|(S_{\tilde b})_{11}\big|^2 m_{\tilde b_1}^2 + m_b^2 - m_t^2
   - m_Z^2c_{2\beta} c^2_w \bigg]\ ,
\esp\label{eq:su2l}\ee
with $c_w\equiv\cos\theta_w$, so that the corresponding counterterms are given
by
\be\bsp
  & \delta m_{\tilde d_L}^2 =  \delta m_{\tilde u_L}^2\ , \qquad\qquad
  \delta m_{\tilde s_L}^2 =  \delta m_{\tilde c_L}^2\ , \\
 &  \delta m_{\tilde b_2}^2 = \frac{1}{s_{\tilde b}^2} \bigg[
   c_{\tilde t}^2 \delta m_{\tilde t_1}^2 +
   s_{\tilde t}^2 \delta m_{\tilde t_2}^2 -
   c_{\tilde b}^2 \delta m_{\tilde b_1}^2 + 2 m_b \delta m_b - 2 m_t \delta m_t
   - s_{2\tilde t} \Delta m_{\tilde t}^2  \delta\theta_{\tilde t}
   + s_{2\tilde b} \Delta m_{\tilde b}^2  \delta\theta_{\tilde b}
  \bigg] \ ,
\esp\label{eq:sq_renorm}\ee
where $\Delta m^2_{\tilde t} = m_{\tilde t_1}^2 - m_{\tilde t_2}^2$, $\Delta
m_{\tilde b}^2 = m_{\tilde b_1}^2 - m_{\tilde b_2}^2$, $s_{\tilde t, \tilde b}
\equiv \sin\theta_{\tilde t, \tilde b}$, $s_{2\tilde t, 2\tilde b} \equiv \sin 
2\theta_{\tilde t, \tilde b}$ and $c_{\tilde t,\tilde b}\equiv\cos\theta_{
\tilde t,\tilde b}$. We have explicitly introduced in those expressions the
dependence on the mixing angles whose renormalisation constants
$\delta\theta_{\tilde t}$ and $\delta\theta_{\tilde b}$ are given by
eq.~\eqref{eq:renomix}.

As a result, the renormalised masses of the left-handed down and strange squarks
and of the heaviest bottom squarks are shifted with respect to their pole masses
$m_{\tilde d_L, {\rm pole}}^2$, $m_{\tilde s_L, {\rm pole}}^2$ and
$m_{\tilde b_2, {\rm pole}}^2$~\cite{Bartl:1997yd},
\be\bsp &
  m_{\tilde d_L, {\rm pole}}^2 = m_{\tilde d_L}^2 + \delta m_{\tilde d_L}^2
    - \delta m_{\tilde d_L, {\rm pole}}^2 \ , \qquad
  m_{\tilde s_L, {\rm pole}}^2 = m_{\tilde s_L}^2 + \delta m_{\tilde s_L}^2
    - \delta m_{\tilde s_L, {\rm pole}}^2 \ , \\
  &\hspace*{4cm}
   m_{\tilde b_2, {\rm pole}}^2 = m_{\tilde b_2}^2 + \delta m_{\tilde b_2}^2
    - \delta m_{\tilde b_2, {\rm pole}}^2 \ ,
\esp\ee
where $\delta m_{\tilde d_L, {\rm pole}}$, $\delta m_{\tilde s_L, {\rm pole}}$
and $\delta m_{\tilde b_2, {\rm pole}}$ stand for the naive OS renormalisation
constants of eq.~\eqref{eq:naivemasses} and $\delta m_{\tilde d_L}^2$,
$\delta m_{\tilde s_L}^2$ and $\delta m_{\tilde b_2}^2$ are the `$m_b$ on-shell'
counterterms of eq.~\eqref{eq:sq_renorm}. The tree-level masses are moreover
given by eq.~\eqref{eq:su2l}. These UV-finite shifts must in particular
be accounted for when an entire MSSM spectrum is used, as typical MSSM spectrum
generators solely output pole squark masses.

By virtue of eq.~\eqref{eq:sqmasses} and eq.~\eqref{eq:sqmassdiag}, the
strengths of the soft trilinear squark-Higgs interactions $A_t$ and $A_b$ also
receive one-loop corrections in $\as$ through their connection with the
corresponding squark mixing angles,
\be
  s_{2\tilde t} = \frac{2 m_t \big(\mu\cot\beta - A_t\big)}
    {\Delta m_{\tilde t}^2}
  \qquad\qquad\text{and}\qquad\qquad
  s_{2\tilde b} = \frac{2 m_b \big(\mu\tan\beta - A_b\big)}
    {\Delta m_{\tilde b}^2} \ .
\ee
The corresponding counterterms are given by
\be\bsp
   \delta A_q = \frac{1}{m_q}\bigg[
     \frac12 \big(\delta m_{\tilde q_1}^2 - \delta m_{\tilde q_2}^2\big)
       s_{2\tilde q} +
     \Delta m_{\tilde q}^2 c_{2\tilde q} \delta \theta_{\tilde q} -
     \frac{1}{2 m_q} \Delta m_{\tilde q}^2 s_{2\tilde q} \delta m_q
     \bigg] \qquad\text{for}\qquad q=b, t\ ,
\esp\label{eq:dAt}\ee
as both the $\mu$ parameter and $\tan\beta$ do not receive $\as$ corrections at one loop.\\

\noindent{\bf The `$A_b$/$\theta_{\tilde b}$ on-shell' scheme}\\
As a consequence of eq.~\eqref{eq:dAt}, two of the three counterterms
$\delta m_q$, $\delta A_q$ and $\delta\theta_{\tilde q}$ are independent. There are thus
various options for fixing the renormalisation conditions, that all lead to
slight differences in the predictions. In the `$m_b$ OS' scheme, the two
$\delta A_q$ renormalisation constants are derived from the other counterterms.
This is however known to yield potentially-unacceptably large threshold
corrections to the bottom-quark pole mass due to the $\delta A_b$ counterterm
when $\tan\beta$ is substantial~\cite{Brignole:2002bz,Heinemeyer:2004xw,
Degrassi:2010eu,Heinemeyer:2010mm}. Whilst a
fully $\drbar$ renormalisation of the bottom sector ($\delta m_b$, $\delta A_b$
and $\delta\theta_{\tilde b}$) would avoid the problem, this is also known not
to make manifest the decoupling of heavy particles.

We therefore present here another commonly-used scheme in which the $A_b$
parameter is renormalised in the OS scheme via a kinematic condition on the
coupling of the pseudoscalar Higgs boson $A^0$ to a $\tilde b_1 \tilde b_2$
pair. This approach relies on the proportionality of the $A^0 \tilde b_1\tilde
b_2$ coupling to the product of the bottom Yukawa coupling (or the bottom mass)
and the bottom trilinear coupling, so that shifts in one quantity
can always be reabsorbed in the other one. In practice, we calculate the
one-loop corrections to the above-mentioned vertex with appropriately chosen
external momenta and include suitable wave-function corrections to avoid any
infrared divergence,
\be
  \delta A_b = \frac{\Big[A_b s_\beta + \mu c_\beta\Big]\Big[
     - \Delta m_{\tilde b}^2\ s_{2\beta}\
         {\cal F}(m_{\tilde b_1}^2, m_{\tilde b_2}^2)
     + \big(\delta m_{\tilde b_1}^2 - \delta m_{\tilde b_2}^2 \big) s_{2\beta}
     + 2 \Delta m_{\tilde b}^2\ c_{2\beta}\ \delta\theta_{\tilde b} \Big]}
   {2 m_b\ \mu\ c_\beta +
      s_\beta \big[2 A_b m_b - \Delta m_{\tilde b}^2\ s_{2\beta}\big]} \ .
\label{eq:dAb}\ee
In the above expression, the ${\cal F}$ function originates from the one-loop
corrections to the $A^0 \tilde b_1 \tilde b_2$ vertex,
\be\bsp&
  {\cal F}(m_1^2,m_2^2) = -\frac{g_s^2 C_F}{8 \pi^2}\Bigg[
   \frac{- m_{\tilde g}}
      {A_b + \mu\cot\beta} \Big[B_0^{\rm fin.}(m_1^2; m_b^2, m_{\tilde g}^2)
      + B_0^{\rm fin.}(m_2^2; m_b^2, m_{\tilde g}^2) \Big]
  \\ &\hspace{1cm}
    + \frac{m_1^2}{m_1^2-m_2^2} \bigg( 4 + 2\log\frac{\mu_R^2}{m_1^2}
       - \frac{m_1^2 - m_{\tilde g}^2 - m_b^2}{m_1^2}
        B_0^{\rm fin.}(m_1^2; m_b^2, m_{\tilde g}^2) \bigg)
  \\ &\hspace{1cm}
    + \frac{m_2^2}{m_2^2-m_1^2} \bigg( 4 + 2\log\frac{\mu_R^2}{m_2^2}
       - \frac{m_2^2 - m_{\tilde g}^2 - m_b^2}{m_2^2}
        B_0^{\rm fin.}(m_2^2; m_b^2, m_{\tilde g}^2) \bigg)
   \Bigg] \ ,
\esp\ee
where only the finite pieces of the loop integrals are retained (\ie\ all
pieces independent of $1/\epsbar$ in the conventions of appendix~\ref{app:PV})
and $\mu_R$
stands for the renormalisation/regularisation scale. As a consequence, the
bottom mass counterterm is now a dependent parameter,
\be
  \delta m_b =  \frac{2 m_b}{\tan\beta}\ \delta\theta_{\tilde b} -
     \frac{2 m_b^2}{s_{2\beta}\ \Delta m_{\tilde b}^2} \ \delta A_b + 
     \frac{m_b}{\Delta m_{\tilde b}^2}
         \big(\delta m_{\tilde b_1}^2 - \delta m_{\tilde b_2}^2 \big)  \ .
\label{eq:dmb}\ee

\subsubsection{Renormalisation of the strong coupling}
\label{sec:renoas}
Our calculations require that the running of the strong coupling constant
$\as$ originates solely from the contributions of the gluons and $n_{lf}$
flavours of light quarks. We therefore renormalise the strong coupling by
subtracting, at zero-momentum transfer, all massive particle contributions
and $\msbar$ contributions of all massless particles from the gluon
self-energy~\cite{Collins:1978wz,Bardeen:1978yd,Marciano:1983pj}.
They are then absorbed in the renormalisation constant of
the strong coupling $\delta \as$ with $n_{lf}=4$,
\be\bsp
  \frac{\delta\as}{\as} = &
      \frac{\as}{2\pi\bar\epsilon}
        \bigg[\frac{n_{lf}}{3} \!-\! \frac{11 n_c}{6}\bigg]
    + \frac{\as}{6\pi}
        \bigg[\frac{1}{\bar\epsilon} \!-\! \log\frac{m_b^2}{\mu_R^2}\bigg]
    + \frac{\as}{6\pi}
        \bigg[\frac{1}{\bar\epsilon} \!-\! \log\frac{m_t^2}{\mu_R^2}\bigg]
    + \frac{\as n_c}{6\pi} \bigg[
        \frac{1}{\bar\epsilon} \!-\! \log\frac{m_{\tilde g}^2}{\mu_R^2}\bigg]\\
  &\qquad
    + \frac{\as}{24\pi}\sum_{\tilde q}\bigg[
        \frac{1}{\bar\epsilon} \!-\! \log\frac{m_{\tilde q}^2}{\mu_R^2}\bigg]\ ,
\esp\label{eq:das}\ee
where the sum in the last term includes all twelve squark species. The
UV-divergent part of the renormalisation constant has been written
explicitly in terms of the quantity
\be
  \frac{1}{\epsbar}=\frac{1}{\epsilon} - \gamma_E + \log{4\pi} \ ,
\ee
where $\gamma_E$ is the Euler-Mascheroni constant and $\epsilon$ is related to
the number of space-time dimensions $D=4-2\epsilon$.

The above renormalisation procedure however leads to a violation of
SUSY as it introduces a mismatch between the strong coupling $g_s$ and
the Yukawa interaction $\hat g_s$ of a gluino with a squark and a quark. While
these two couplings are equal at tree-level, as shown by the last term of the
second line of eq.~\eqref{eq:lag}, the equality is destroyed by the difference
in the number of fermionic gluino degrees of freedom and bosonic gluon degrees
of freedom. This artificial breaking of SUSY is compensated by finite
counterterms restoring SUSY invariance.

As we impose that the definition of the strong coupling $g_s$ is the SM one due to the decoupling theorem, only the quark-squark-gluino vertices and quartic squark interactions
have to be shifted~\cite{Martin:1993yx}. The SUSY restoring counterterm
Lagrangian $ {\cal L}^{\rm (SCT)}_{\rm MSSM}$ is then given, in the gauge eigenbasis, by
\be\bsp
  {\cal L}^{\rm (SCT)}_{\rm MSSM} = &
   \frac{g_s^2}{2}\frac{\alpha_s}{4 \pi}
      \Big[ \tilde q_R^\dag\{T_a,T_b\}\tilde q_R +
         \tilde q_L^\dag\{T_a,T_b\}\tilde q_L\Big]
      \Big[ \tilde q_R^\dag\{T^a,T^b\}\tilde q_R +
         \tilde q_L^\dag\{T^a,T^b\}\tilde q_L\Big]
\\ &\quad
   - \frac{g_s^2}{2}\frac{\alpha_s}{4\pi}
     \Big[\tilde q_R^\dag T_a \tilde q_R - \tilde q_L^\dag T_a \tilde q_L\Big]
     \Big[\tilde q_R^\dag T^a \tilde q_R - \tilde q_L^\dag T^a \tilde q_L\Big]
\\ &\quad
   + \sqrt{2} g_s\frac{\alpha_s}{3 \pi}
      \Big[ - \tilde q_L^\dag T_a \big(\bar\Psi_{\tilde g}^a P_L \Psi_q \big)
         + \big(\bar\Psi_q P_L \Psi_{\tilde g}^a \big) T_a \tilde q_R 
         + {\rm h.c.} \Big] \ ,
\esp\ee
where adjoint colour indices have been included and a sum over (s)quark flavours is
understood for clarity.

\subsection{Technical details on the model implementation in \fr}
\label{sec:bfl_example}
In order to calculate SUSY particle-production (total and
differential) rates at colliders and to simulate MSSM signals by matching
fixed-order results at the NLO accuracy with parton showers, we rely on the \mg\
framework~\cite{Alwall:2014hca}. Our methodology is based on the joint usage of
the \fr~\cite{Alloul:2013bka}, \nloct~\cite{Degrande:2014vpa} and
\fa~\cite{Hahn:2000kx} packages to automatically produce a UFO
model~\cite{Degrande:2011ua} that can be used by \mg. However, there are
substantial
differences with respect to the procedure that has been followed for stop pair
production~\cite{Degrande:2015vaa}, in the SUSY QCD case~\cite{Degrande:2014sta}
and for slepton production~\cite{Fuks:2019iaj}, as a consequence of the
non-trivial renormalisation procedure for the mixing angle and the trilinear
scalar couplings detailed in section~\ref{sec:susy-onshell}.
After having implemented the model described in section~\ref{sec:mssm} and its
tree-level Lagrangian in terms of superfields, we make use of the superspace
module of \fr~\cite{Duhr:2011se} to re-express the MSSM Lagrangian in terms of
the model physical degrees of freedom and four-component fermions. The
renormalisation is then performed with the \bfl\ package, that is introduced in
appendix~\ref{app:bfl} and that is necessary for a flexible definition of the
renormalisation scheme.

We firstly impose that all external parameters insensitive to QCD
corrections are kept unrenormalised. We hence enforce vanishing
renormalisation constants for all electroweak inputs (the Fermi constant $G_F$,
the inverse of the electromagnetic coupling at the $Z$ pole $1/\alpha$
and the $Z$-boson mass $m_Z$), the parameters of the Higgs sector ($\tan\beta$,
the $\alpha$ angle and the $\mu$ parameter), the slepton trilinear couplings
($A_e$, $A_\mu$ and $A_\tau$), the soft masses associated with the electroweak
particles and the electroweakino mixing matrices ($U$, $V$ and $N$) that are
external parameters in the SLHA conventions~\cite{Skands:2003cj}. Moreover, the
first and second generation squark trilinear couplings ($A_u$, $A_d$, $A_c$ and
$A_s$) are irrelevant as multiplied by a vanishing quark mass and will thus not
be renormalised. These constraints are imposed by using the
\verb+MoGRe`DefineUnrenormalizedParameter+ function introduced in
appendix~\ref{app:BFL_simp}.

Secondly, the quark mass dependence of the (remaining) trilinear squark-Higgs
couplings (${\bf \hat T^u}$ and ${\bf \hat T^d}$) as well as the one of the
fermion Yukawa couplings must be made explicit to guarantee the correct
functioning of \nloct. This is achieved by making use of the
\verb+RemovingInternalCst+ method introduced in appendix~\ref{sec:bfl:initmain}.
The same method is finally also used to replace all occurrences of the $g_s$
renormalisation constant in terms of the $\as$ one.

Next, we indicate to the code that fields that are insensitive
to the strong interaction at the one-loop level (the electroweak gauge and Higgs
bosons) do not need to be renormalised. This is achieved by making use of the
\verb+MoGRe`DefineUnrenormalizedField+ method detailed in
appendix~\ref{app:BFL_simp}. Whilst other purely electroweak fields such as
electroweakinos or
(s)leptons have in principle to be analogously tagged as unrenormalisable
objects, they do not appear in any QCD vertex so that they will be automatically
discarded by the code. We finally impose that all field wave-function
renormalisation constants are real (via the \verb+MoGRe`RealFieldRenormalization+
method presented in appendix~\ref{app:BFL_simp}).

In practice, the \bfl\ package is initialised as
\begin{lstlisting}
 SetOptions[MoGRe$\$$Renormalize, Exclude4Scalars->True,
  FlavorMixing -> {{st1,st2}, {sb1,sb2}}, CouplingOrders->{QCD}];
\end{lstlisting}
which indicates to the code that the stop and sbottom fields mix and that the
renormalisation of the four-scalar interactions can be ignored. While strictly
speaking, four-scalar interactions cannot be ignored, restrictions omitting them
are useful phenomenologically as these vertices rarely appear at tree-level.
The constraints
above-mentioned are then implemented as follows,
\begin{verbatim}
 MoGRe`DefineUnrenormalizedParameters[{
    Gf, aEWM1, MZ, MUH, alp, tb,
    Mx1, Mx2, mHu2, mHd2, meL, mmuL, mtauL, meR, mmuR, mtauR,
    Sequence@@Flatten[Table[{ae[i, i]}, {i, 1, 3}]],
    Sequence@@Flatten[Table[{au[i, i], ad[i, i]}, {i, 1, 2}]],
    Sequence@@Flatten[Table[{VV[i, j], UU[i, j]}, {i, 1, 2}, {j, 1, 2}]],
    Sequence@@Flatten[Table[{NN[i, j]}, {i, 1, 4}, {j, 1, 4}]]
 }];
 MoGRe`DeclareUnrenormalizedFields[W, A, Z];
 MoGRe`RealFieldRenormalization[] ;
 MoGRe`RemovingInternalCst[#] & /@ {gs,
   Sequence@@Flatten[Table[{yu[i,i], yd[i,i], tu[i,i], td[i,i]}, {i,1,3}]]};
\end{verbatim}

The exact details of the renormalisation scheme must then be specified, as shown
in appendix~\ref{sec:scheme}. Focusing on the naive OS scheme, the stop and
sbottom mixing matrices are renormalised on the basis of eq.~\eqref{eq:renomix},
which is implemented as
\begin{verbatim}
MoGRe`AddRenormalizationCondition[FR$delta[{Rtop[1, 1]}, {}], 1/4 Rtop[2, 1]*
   (FR$deltaZ[{st1, st2}, {{}}] - Conjugate[FR$deltaZ[{st2, st1}, {{}}]])];
MoGRe`AddRenormalizationCondition[FR$delta[{Rtop[1, 2]}, {}], 1/4 Rtop[1, 1]*
   (FR$deltaZ[{st1, st2}, {{}}] - Conjugate[FR$deltaZ[{st2, st1}, {{}}]])];
MoGRe`AddRenormalizationCondition[FR$delta[{Rtop[2, 1]}, {}], 1/4 Rtop[1, 1]*
   (FR$deltaZ[{st1, st2}, {{}}] - Conjugate[FR$deltaZ[{st2, st1}, {{}}]])];
MoGRe`AddRenormalizationCondition[FR$delta[{Rtop[2, 2]}, {}], 1/4 Rtop[2, 1]*
   (FR$deltaZ[{st1, st2}, {{}}] - Conjugate[FR$deltaZ[{st2, st1}, {{}}]])];
MoGRe`AddRenormalizationCondition[FR$delta[{Rbot[1, 1]}, {}], 1/4 Rbot[2, 1]*
   (FR$deltaZ[{sb1, sb2}, {{}}] - Conjugate[FR$deltaZ[{sb2, sb1}, {{}}]])];
MoGRe`AddRenormalizationCondition[FR$delta[{Rbot[1, 2]}, {}], 1/4 Rbot[1, 1]*i
   (FR$deltaZ[{sb1, sb2}, {{}}] - Conjugate[FR$deltaZ[{sb2, sb1}, {{}}]])];
MoGRe`AddRenormalizationCondition[FR$delta[{Rbot[2, 1]}, {}], 1/4 Rbot[1, 1]*
   (FR$deltaZ[{sb1, sb2}, {{}}] - Conjugate[FR$deltaZ[{sb2, sb1}, {{}}]])];
MoGRe`AddRenormalizationCondition[FR$delta[{Rbot[2, 2]}, {}], 1/4 Rbot[2, 1]*
   (FR$deltaZ[{sb1, sb2}, {{}}] - Conjugate[FR$deltaZ[{sb2, sb1}, {{}}]])];
\end{verbatim}
the stop and sbottom mixing matrices $S_{\tilde t}$ and $S_{\tilde b}$
being available as the parameters \verb+Rtop+ and \verb+Rbot+ in the \fr\
implementation. The renormalisation of the stop and sbottom sector is finalised
by imposing the way in which the $A_t$ and $A_b$ parameters are
renormalised, according to eq.~\eqref{eq:dAt},
\begin{verbatim}
  MoGRe`AddRenormalizationCondition[FR$delta[{Au[3,3]}, {}], 1/MT * (
    (Mst1*FR$delta[{Mst1}, {}] - Mst2*FR$delta[{Mst2}, {}]) +
    1/MT*(Mst1^2-Mst2^2)*Rtop[1,1]*Rtop[1,2]*FR$delta[{MT},{}] +
    (Mst1^2-Mst2^2)*(Rtop[1,1]*FR$delta[{Rtop[1,2]},{}] +
       Rtop[1,2]*FR$delta[{Rtop[1,1]},{}])
  )];
  MoGRe`AddRenormalizationCondition[FR$delta[{Ad[3,3]}, {}], 1/MB * (
    (Msb1*FR$delta[{Msb1}, {}] - Msb2*FR$delta[{Msb2}, {}]) +
    1/MB*(Msb1^2-Msb2^2)*Rbot[1,1]*Rbot[1,2]*FR$delta[{MB},{}] +
    (Msb1^2-Msb2^2)*(Rbot[1,1]*FR$delta[{Rbot[1,2]},{}] +
       Rbot[1,2]*FR$delta[{Rbot[1,1]},{}])
  )];
\end{verbatim}

We subsequently make use of \nloct\ to generate the ultraviolet
counterterms and $R_2$ Feynman rules necessary to obtain, from the UFO
interface of \fr, an NLO UFO model for the MSSM. Since we induce an artificial
breaking of supersymmetry by the mismatch of the two gluino and ($D-2$) gluon
degrees of freedom, one needs to add to the model Lagrangian a set of finite
counterterms allowing to restore supersymmetry when dealing with one-loop
calculations. Enforcing the definition of $g_s$ to be the SM one,
quark-squark-gluino and four-scalar interactions are the only interactions that
need to be shifted. Those shifts are given by the following counterterm
Lagrangian~\cite{Martin:1993yx},
\be\bsp
  {\cal L}_{\rm shift} =
   &\ \sqrt{2} g_s\frac{\alpha_s}{3 \pi}
      \Big[ - \tilde q_L^\dag T_a \big(\bar\Psi_{\tilde g}^a P_L \Psi_q \big)
       + \big(\bar\Psi_q P_L \Psi_{\tilde g}^a \big) T_a \tilde q_R 
       + {\rm h.c.} \Big] \\
   &\ - \frac{g_s^2}{2}\frac{\alpha_s}{4\pi}
     \Big[\tilde q_R^\dag T_a \tilde q_R - \tilde q_L^\dag T_a \tilde q_L\Big]
     \Big[\tilde q_R^\dag T^a \tilde q_R - \tilde q_L^\dag T^a \tilde q_L\Big] \\
   &\ + \frac{g_s^2}{2}\frac{\alpha_s}{4 \pi}
        \Big[ \sq_R^\dag\{T_a,T_b\}\sq_R + \sq_L^\dag\{T_a,T_b\}\sq_L\Big]
        \Big[ \sq_R^\dag\{T^a,T^b\}\sq_R + \sq_L^\dag\{T^a,T^b\}\sq_L\Big]\ ,
\esp\ee
where ${\cal L}_{\rm shift}$ is written in the gauge eigenbasis. Those
counterterms are appropriately included in the MSSM UFO in two
steps~\cite{Degrande:2015vaa}. We first evaluate the associated Feynman rules
with \fr\ and then provide the resulting set of rules to the UFO interface by
means of the {\tt UVLoopCounterterms} option of the {\tt WriteUFO} method. In
contrast to the previous approaches, the resulting model can be used, within
\mg, beyond the simplified model context.

\subsection{Simulations and cross section calculations in SUSY\label{sec:sim}}

As was discussed in section~\ref{sec:intro}, it is common practice to
normalise the results of tree-level merged simulations by means of
higher-order inclusive cross sections. NLO+NLL total production
rates are known for light-flavour squarks~\cite{Beenakker:1994an,
Beenakker:1996ch,Bozzi:2005sy,Kulesza:2008jb,Kulesza:2009kq,Beenakker:2009ha,
Beenakker:2011fu,Kauth:2011vg,Falgari:2012hx}, third-generation 
squarks~\cite{Beenakker:1997ut,Beenakker:2010nq,Beneke:2010da}, 
gluinos~\cite{Beenakker:1995fp,Beenakker:1996ch,Kulesza:2008jb,Kulesza:2009kq,
Beenakker:2009ha,Beenakker:2011fu,Kauth:2011vg,Falgari:2012hx},
electroweakinos~\cite{Beenakker:1999xh,Debove:2009ia,Debove:2010kf,
Debove:2011xj,Fuks:2012qx}, sleptons~\cite{Beenakker:1999xh,
Bozzi:2006fw,Bozzi:2007qr,Bozzi:2007tea,Fuks:2013lya}, and for several 
mixed channels involving one strong and one electroweak 
superpartner~\cite{Berger:2000iu,Fuks:2016vdc}. 
In addition, NLO QCD corrections have been computed including superparticle
decays for squark 
pair-production~\cite{Hollik:2012rc,Hollik:2013xwa}\footnote{Both the $2\to2$
matrix element describing the production process and the $1\to2$ matrix 
elements related to the decay processes are NLO-QCD accurate, the different 
contributions being assumed to factorise as it is the case in the 
narrow-width approximation.}, while approximate NNLO
threshold contributions~\cite{Langenfeld:2009eg,Langenfeld:2010vu,
Langenfeld:2012ti}, (electro)weak (EW)
corrections~\cite{Hollik:2007wf,Beccaria:2008mi,Hollik:2008yi,Hollik:2008vm,
Mirabella:2009ap,Arhrib:2009sb,Germer:2010vn,Germer:2011an,Germer:2014jpa,
Hollik:2015lha}, and resummation effects at the NNLL 
level~\cite{Beenakker:2011sf,Pfoh:2013edr,Broggio:2013uba,
Beenakker:2013mva,Beenakker:2014sma,Beenakker:2016gmf,Beneke:2016kvz} 
have been considered for squark and gluino
production. Moreover, effects originating from $R$-parity 
violation~\cite{Alves:2002tj,XiaoPeng:2012dp} and non-minimal 
flavour-violation~\cite{Bozzi:2007me,Fuks:2008ab,Fuks:2011dg} have also 
been explored.
All of these results have been included in several publicly available 
computer programs, which are restricted to the evaluation of total rates.
\prospino~\cite{Beenakker:1996ch,Beenakker:1997ut,Beenakker:1999xh} and
\madgolem~\cite{Binoth:2011xi,GoncalvesNetto:2012yt,Goncalves:2014axa}
give predictions that are NLO-QCD accurate, while  
{\sc\small NLL-fast}~\cite{Beenakker:2015rna}, 
{\sc\small NNLL-fast}~\cite{Beenakker:2016lwe} and 
{\sc\small Resummino}~\cite{Fuks:2013vua} also resum threshold logarithms.
All codes, with the exception of \madgolem\ and \resummino, assume 
mass-degenerate squark spectra.

Theoretical predictions at the differential level, that include both
NLO and PS effects, are more recent and, so far, tackled on a 
process-by-process basis. The production of pairs of 
squarks~\cite{Gavin:2013kga,Gavin:2014yga,Degrande:2014sta,Degrande:2015vaa}, 
gluinos~\cite{Degrande:2015vaa}, electroweakinos~\cite{Baglio:2016rjx},
and sleptons~\cite{Jager:2012hd,Jager:2014aua,Fuks:2019iaj} have all
been considered in the past few years. Some of these computations have 
been carried out with \mg. We recall here that \mg\ makes 
use of the FKS method~\cite{Frixione:1995ms,Frixione:1997np} 
(automated in the module \madfks~\cite{Frederix:2009yq,
Frederix:2016rdc}) for dealing with IR singularities. The computations 
of one-loop amplitudes are carried out by switching dynamically between 
two integral-reduction techniques, OPP~\cite{Ossola:2006us} or 
Laurent-series expansion~\cite{Mastrolia:2012bu}, and 
tensor-integral reduction~\cite{Passarino:1978jh,Davydychev:1991va,
Denner:2005nn}. These have been automated in the 
module \madloop~\cite{Hirschi:2011pa,Alwall:2014hca}, 
which in turn exploits \cuttools~\cite{Ossola:2007ax}, 
\ninja~\cite{Peraro:2014cba,Hirschi:2016mdz}, \iregi~\cite{iregi}, or 
\collier~\cite{Denner:2016kdg}, together with an in-house implementation 
of the \openloops\ optimisation~\cite{Cascioli:2011va}. Finally, in the case 
of matching with PS, the MC@NLO formalism~\cite{Frixione:2002ik} 
is employed, whereas NLO multi-jet mergings rely either on
FxFx~\cite{Frederix:2012ps} or UNLOPS~\cite{Lonnblad:2012ix}.

We point out that the original \mg\ paper~\cite{Alwall:2014hca} 
had the goal of including as many information on Quantum Field Theories
as possible in the meta-code, so as to allow it to simulate both SM and BSM 
processes by using the inputs in the form of UFO models constructed by codes 
such as \fr\ or {\sc\small Sarah}~\cite{Staub:2012pb}. Recently, the program 
has been upgraded, and can for instance now handle mixed-coupling scenarios, 
in particular QCD+EW simultaneous corrections~\cite{Frederix:2018nkq}. 
However, the most general BSM calculations beyond LO feature a number of 
non-trivial characteristics that are absent in the SM. While 
fermion-flow-violating interaction vertices and non-renormalisable 
operators (which were not available at the time of the first 
release~\cite{Alwall:2014hca}) can now be handled, coloured-sextet particles 
and the renormalisation-group running of new couplings are {\em not} 
yet included in \mg. Thus, we stress again that the current
work will be limited to considering NLO QCD corrections to SUSY theories,
whereby only quarks, gluons, squark, and gluinos can run in the loops.
Moreover, real-emission contributions only consider additional massless SM
particles in the final state, given that massive particle contributions are
finite (and can thus be computed independently) and often numerically
subleading (see the analogous discussion of ref.~\cite{Frederix:2018nkq} that
addresses Heavy Boson Radiation (HBR) in the context of the computation of NLO
electroweak corrections).

We conclude this section by listing the UFO models that can presently
be used for BSM simulations. These include simplified models, 
in which the SM is extended by colour-triplet and octet 
scalar particles~\cite{Degrande:2014sta,Cacciapaglia:2018rqf}, both
gluinos and squarks~\cite{Degrande:2015vaa} or sleptons~\cite{Fuks:2019iaj}, as
well as by vector-like
quarks~\cite{Fuks:2016ftf,Cacciapaglia:2018qep}, a heavy top-philic
scalar~\cite{BuarqueFranzosi:2017jrj} or a spin-2 particle~\cite{Das:2016pbk}.
In the latter spin-2 case, new physics have also been previously explored in a
semi-automated
framework (in the sense where the virtual matrix elements are provided
externally) based on \mg~\cite{Frederix:2012dp, Frederix:2013lga, Das:2014tva}.
Various BSM setups in which the Higgs sector differs from the SM one have 
been released, such as the two-Higgs-doublet model~\cite{Degrande:2015vpa,
Degrande:2016hyf}, the Georgi-Machacek model~\cite{Degrande:2015xnm}, the 
Higgs characterisation model~\cite{Artoisenet:2013puc,Maltoni:2013sma,
Demartin:2014fia,Demartin:2015uha,Demartin:2016axk}, and the SM effective 
field theory including dimension-six operators~\cite{Degrande:2016dqg}.
Higher-dimension operators either affecting the sector of the top 
quark~\cite{Degrande:2014tta,Durieux:2014xla,Bylund:2016phk,Maltoni:2016yxb,
Zhang:2016omx,Franzosi:2015osa}, dijet production~\cite{Hirschi:2018etq}, 
or $Z$-boson production~\cite{Degrande:2013kka} can be added as well. 
Moreover, the model library also allows for NLO+PS calculations in BSM 
models involving TeV-scale neutrinos~\cite{Degrande:2016aje}, a left-right 
symmetry~\cite{Mattelaer:2016ynf}, as well as extra neutral and charged 
gauge bosons~\cite{Fuks:2017vtl}. Finally, dark matter simplified models 
in which the dark matter particle is produced in $s$-channels are also 
available~\cite{Mattelaer:2015haa,Backovic:2015soa,Neubert:2015fka,
Arina:2016cqj,Kraml:2017atm,Bell:2016ekl,Bell:2017rgi,Afik:2018rxl}. 

As was stressed in section~\ref{sec:intro}, part of the present paper is 
devoted to creating a UFO model of the MSSM, which is still missing in
an unrestricted framework (\ie\ when going beyond the simplified-model 
approach). The lifting of such a restriction has also to do with the
treatment of resonant contributions, also addressed here through the
STR procedures.


\section{Validation}
\label{sec:validate}
Fixed-order NLO-QCD predictions for the total rates of specific two-to-two
processes in the MSSM are currently available from three different standalone
tools, namely \prospino~\cite{Beenakker:1996ch,Beenakker:1997ut,
Beenakker:1999xh}, \resummino~\cite{Fuks:2013vua}, and 
\madgolem~\cite{Binoth:2011xi,GoncalvesNetto:2012yt,Goncalves:2014axa}. 
A partial comparison of the results obtained with \mg\ and \madgolem\ 
has already been performed for coloured-scalar 
production, and agreement at the level of the numerical errors has been
found~\cite{Degrande:2014sta}.
Furthermore, the analytic expressions 
of all the $R_2$ counterterms of the MSSM model have been cross-checked 
against the results of ref.~\cite{Shao:2012ja}. In this section, we employ
\mg\ to compute total rates for several processes and specific choices
of the MSSM parameters, and compare our predictions against those obtained 
with \prospino\ (that covers the production of any pair of strongly- or 
electroweakly-interacting superpartners in the case of a degenerate squark 
mass spectrum) and \resummino\ (that supports arbitrary SUSY mass spectra for
the production of two electroweak superpartners). In the rest of this paper, 
all fermions are unambiguously four-component Dirac and Majorana ones,
so that we replace the notation $\Psi_X$ by $X$. In particular, quarks, gluino
and electroweakinos will be denoted by $q$, $\tilde g$, and $\tilde\chi$,
respectively.

\subsection{Setup of the comparison\label{sec:comp}}
Although we shall focus on superparticle-pair production here, it should be
clear that, in keeping with a general automation philosophy, \mg\ is
not restricted to simulating processes with two-body final states.
Furthermore, \mg\ lifts two other key limitations of current NLO QCD
codes: firstly, the inability to tackle QCD-mediated production 
processes (\resummino); and secondly, the inability to support without 
approximation arbitrary (non-degenerate) squark mass spectra (\prospino).
In order to highlight these differences, we have opted to present a 
comparison of the cross sections, at the 13~TeV LHC, relevant to the
following processes (where antisquarks are denoted with a star):
\begin{equation}
pp\;\longrightarrow\;\tilde{t}_1\anti{\tilde{t}_1}\,,
\;\;\;\;\;\;
pp\;\longrightarrow\;\tilde{g}\tilde{g}\,,
\;\;\;\;\;\;
pp\;\longrightarrow\;\gau[+]_1\gau[-]_1\,,
\;\;\;\;\;\;
pp\;\longrightarrow\;\tilde{e}^{+}_R\tilde{e}^{-}_R\,,
\label{eq:samplepr}
\end{equation}
in both regimes of degenerate and non-degenerate squark masses. 
We point out that we have explicitly checked that $\gau[+]_1\gau[0]_2$ 
and $\tilde{e}^{+}_L\tilde{e}^{-}_L$ production lead to the same qualitative 
conclusions as $\gau[+]_1\gau[-]_1$ and $\tilde{e}^{+}_R~\tilde{e}^{-}_R$ 
production, respectively; such final states are thus ignored in what follows.
We have not considered a direct point-wise comparison of one-loop SUSY QCD
amplitudes against those of {\sc\small FeynArts}~\cite{Fritzsche:2013fta,
Hahn:2015ghv}, as the focus of our work is on the computation of 
cross sections and observables.

\begin{table}
  \begin{center} \begin{tabular}{ll|ll|ll}
    Parameter & value & Parameter & value & Parameter & value \\\midrule\midrule
    LO PDF set & {\tt cteq6l1} & $\mu_R=\mu_F$ & 1500 & $m_t$ & 174.3 \\
    NLO PDF set & {\tt cteq66} & $m_Z$ & 91.188 & $m_b$ & 0 \\
    $\alpha_{S}(m_Z^2)$ & as per PDF set & $G_F$ & $1.16637\cdot10^{-5}$ &
       $\Gamma_{\rm all\;particles}$ & 0 \\
    $\alpha$ & $1/127.9$ & SUSY-mixing & Only between $\neu[1]$-$\neu[2]$  & 
       $(V_{\rm CKM})_{ij}$ & $\delta_{ij}$ \\\midrule
    \multicolumn{6}{c}{Degenerate SUSY mass setup} \\\midrule
    $M_{\rm prod}$ & 1500 & $M_{\rm others }$ & [100 - 1400] & $m_{\tilde{u}_L}$
       & $M_{\rm others }$ \\\midrule
    \multicolumn{6}{c}{Non-degenerate SUSY mass setup} \\\midrule
    $M_{\rm prod}$ & 1500 & $M_{\rm others}$ & [100 - 1400] & $m_{\tilde{u}_L}$
       & 1400 \\
  \end{tabular}\end{center}
  \caption{\label{tableParams}SM and SUSY parameters of the benchmark point used
  for the comparisons performed in section~\ref{sec:validate}. Dimensionful 
   quantities are
  given in GeV. $M_{\rm prod}$ denotes the mass of the produced particles in
  the processes of eq.~(\ref{eq:samplepr}), while $M_{\rm others}$ denote 
  the masses
  of \emph{all other} SUSY particles in a degenerate mass setup (this quantity
  is scanned over, hence its range value). In the non-degenerate mass setup, the
  left-handed up squark mass is set equal to a fixed value of 1.4 TeV.}
\end{table}

The model parameters corresponding to the considered SUSY benchmark point are
specified in all three codes via a similar SLHA file~\cite{Skands:2003cj},
the contents of which are summarised 
in table~\ref{tableParams}. The complicated nature of the \prospino\
inputs prompted our use of two different sets of parton distribution functions
(PDFs) for LO and NLO predictions, for which the appropriate value of
$\as(m_Z^2)$ (equal to $0.08991$ and $0.08314$, respectively) 
had to be hard-coded: by default, \prospino\ uses a fixed value of $\as(m_Z^2)$ 
independent of the PDF set. 
We have used the LO and NLO central sets 
of the CTEQ6 PDFs~\cite{Pumplin:2002vw}, and additionally turned off the 
running of $\as$. Moreover, whilst \prospino\ keeps
the exact dependence on the masses of the produced sparticles both at the LO 
and the NLO, the masses that appear in \emph{all} of the internal squark 
propagators are set equal to some averaged value when working at the NLO. 
The $K$-factor, defined as the ratio of such an ``averaged'' NLO computation 
over the corresponding LO one, is then used to multiply the exact LO result 
to get the final (and hence, approximate) NLO prediction. In order to assess 
the quality of such an approximation, we have scanned the cross sections 
obtained with all three codes in the two different mass setups defined in 
table~\ref{tableParams}.

In the case of a spectrum with degenerate SUSY masses, the mass of the
produced SUSY particle ($M_{\rm prod}$) is set equal to 1.5 TeV, while the
common mass of \emph{all other} SUSY particles ($M_{\rm others}$) is scanned
over, in the range [100, 1400]~GeV.  This insures that all of the masses
that appear in internal propagators are equal to each other (\ie\ the 
internal squarks are degenerate), with the possible exception of the 
propagators that involve the produced particles. For instance,
$\tilde{t}_1\anti{\tilde{t}_1}$ production has diagrams with internal
$\tilde t_1$ propagators: the corresponding mass is then kept equal 
to 1.5~TeV. In the case of a spectrum featuring non-degenerate 
SUSY masses, we use exactly the same configuration as for the
degenerate case, except that this time the left-handed up squark mass
($m_{\tilde{u}_L}$) is set equal 1.4~TeV. This particular choice for breaking 
the degeneracy pattern allows for an increase of the sensitivity of the 
inclusive cross section to the mass splitting $m_{\tilde{u}_L}$-$M_{\rm others}$.

We stress that it is crucial to set the mass of the produced particles 
$M_{\rm prod}$ to a value larger than all of the other masses, in order to 
insure the absence of any resonant real-emission contributions (see 
section~\ref{sec:OS}). Such contributions would in fact complicate the comparison
among the three codes, which adopt different strategies for handling them,
leading in turn to potential non-negligible differences in their predictions.

\subsection{Degenerate SUSY masses\label{sec:deg}}
\begin{figure}
  \begin{center}
    \includegraphics[width=0.487\columnwidth]{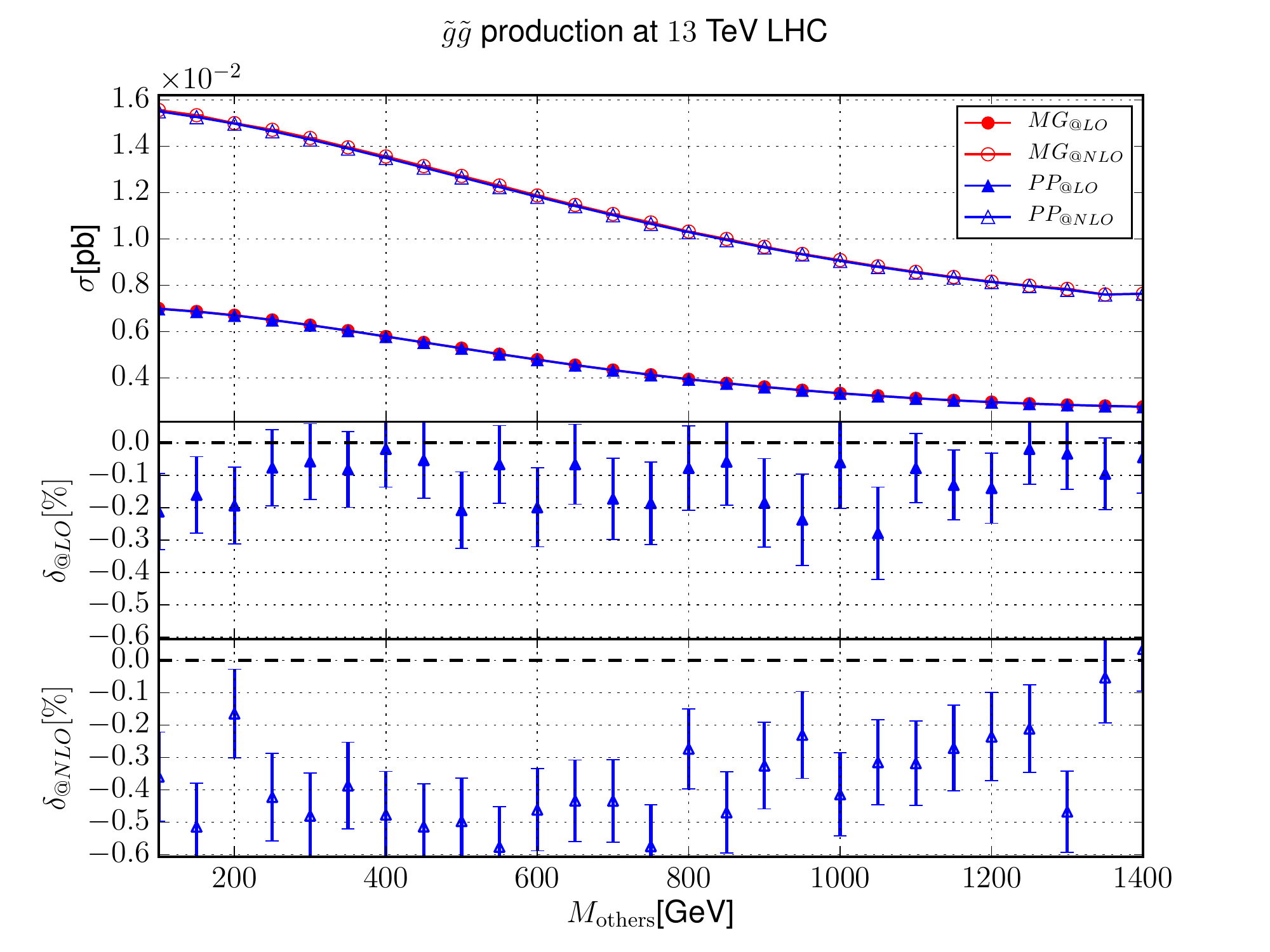}\quad
    \includegraphics[width=0.487\columnwidth]{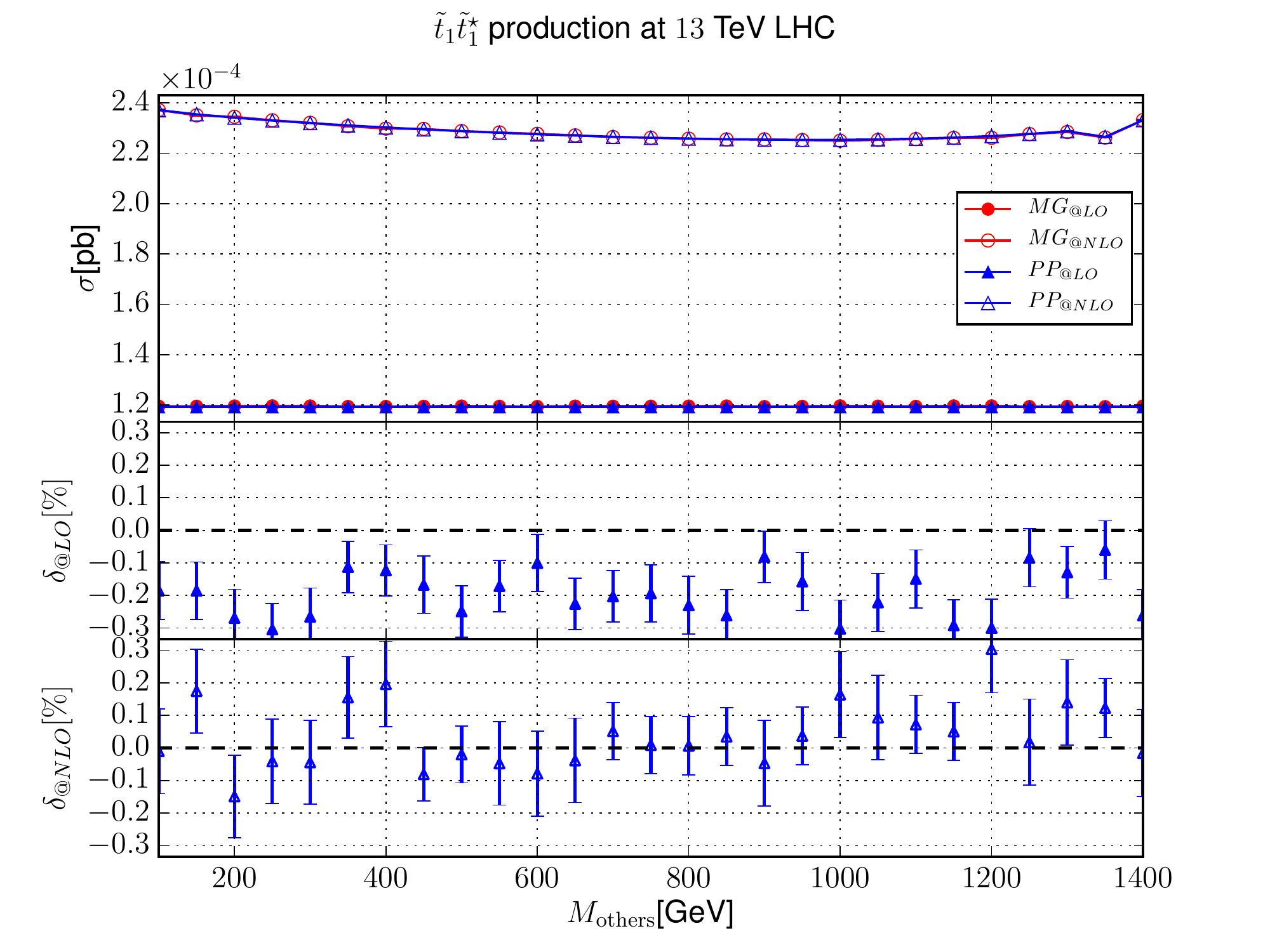}\\
    \includegraphics[width=0.487\columnwidth]{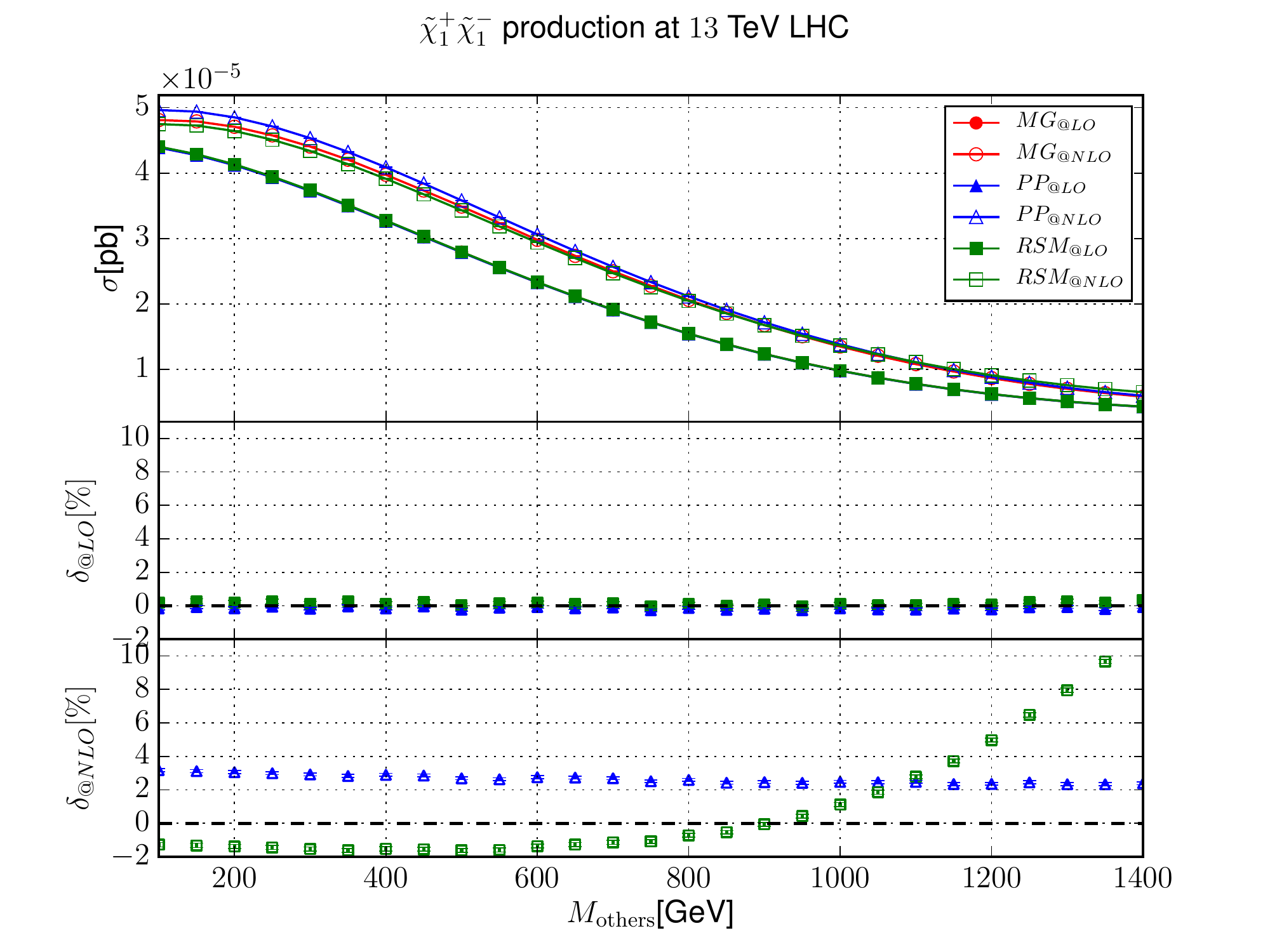}\quad
    \includegraphics[width=0.487\columnwidth]{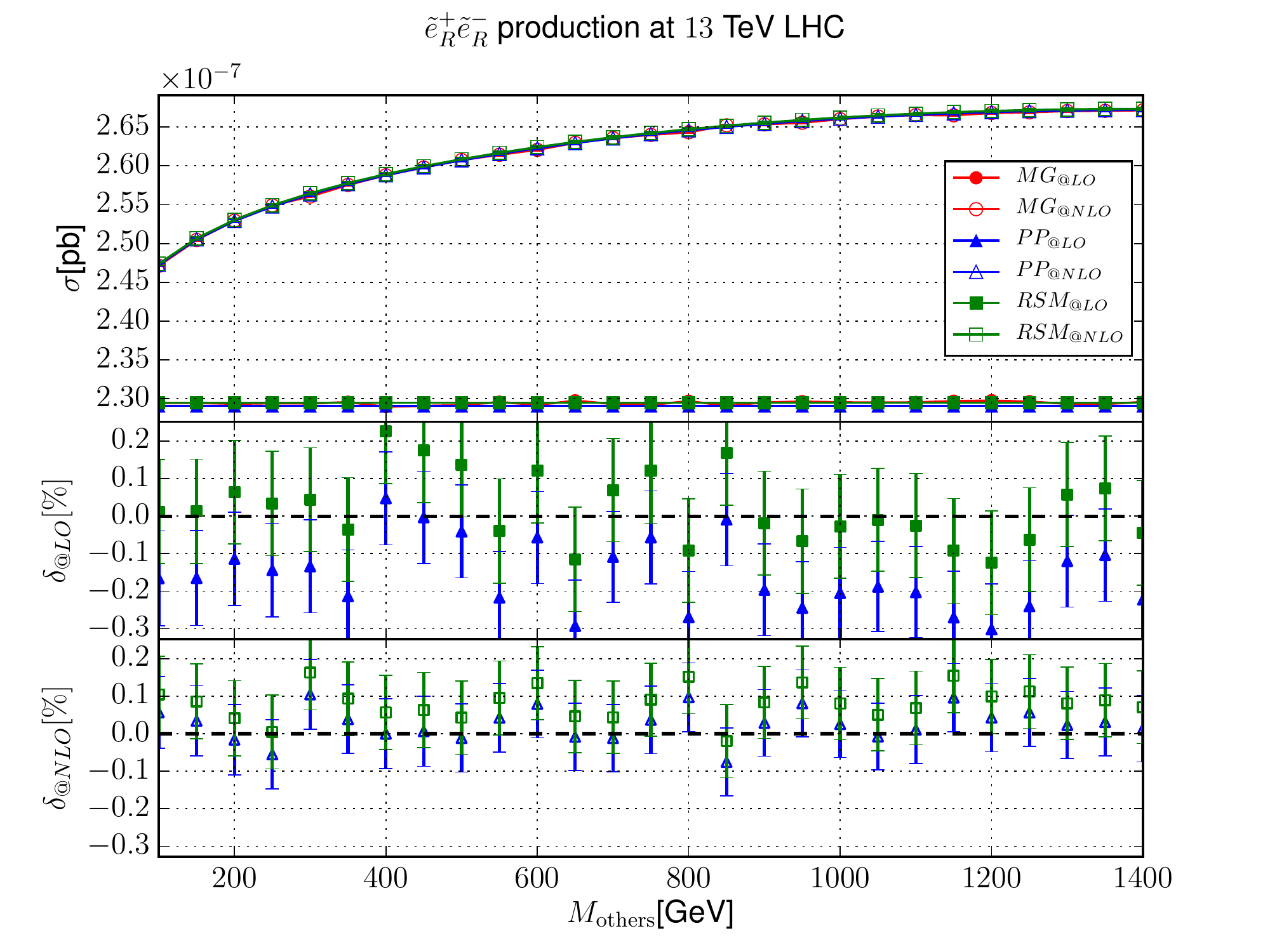}
  \end{center}
  \caption{\label{fig:degenerate_validation} 
  Comparison of inclusive LO (full symbols) and NLO (open symbols) cross
sections obtained from \mg\ (red, MG), \prospino\ (blue, PP), and \resummino\
(green, RSM) for $\tilde{g}\tilde{g}$  (top left panel),
$\gau[+]_1\gau[-]_1$ (bottom left panel), 
$\tilde{t}_1\anti{\tilde{t}_1}$ (top right panel), 
and $\tilde{e}^{+}_R\tilde{e}^{-}_R$ (bottom right panel) production at
the $\sqrt{S}=13$ TeV LHC for degenerate squark masses.  The two lower insets
of each plot show the difference $\delta$ (in percent) relative to the \mg\
predictions, at the LO and the NLO.  The width of the bands corresponds to the
Monte-Carlo uncertainty of the predictions.}
\end{figure}

We report in figure~\ref{fig:degenerate_validation} the outcome of the
comparison among the LO and NLO predictions of \mg, \prospino, and \resummino,
for a degenerate SUSY mass setup and the processes of eq.~(\ref{eq:samplepr}).
As expected, we find no dependence on $M_{\rm others}$ for inclusive stop 
(top right panel) and slepton (bottom right panel) pair production cross 
section at the LO, by virtue of the absence of internal SUSY particles of 
different flavours in the corresponding four-point tree-level amplitudes. 
This contrasts with the production of a pair of gluinos
(top left panel) and charginos (bottom left panel), which both feature 
production modes that involve $t$-channel exchanges of squarks with
different flavours.

\begin{figure}
\begin{minipage}{.48\textwidth}
  \begin{center}
    \includegraphics[width=1.\columnwidth]{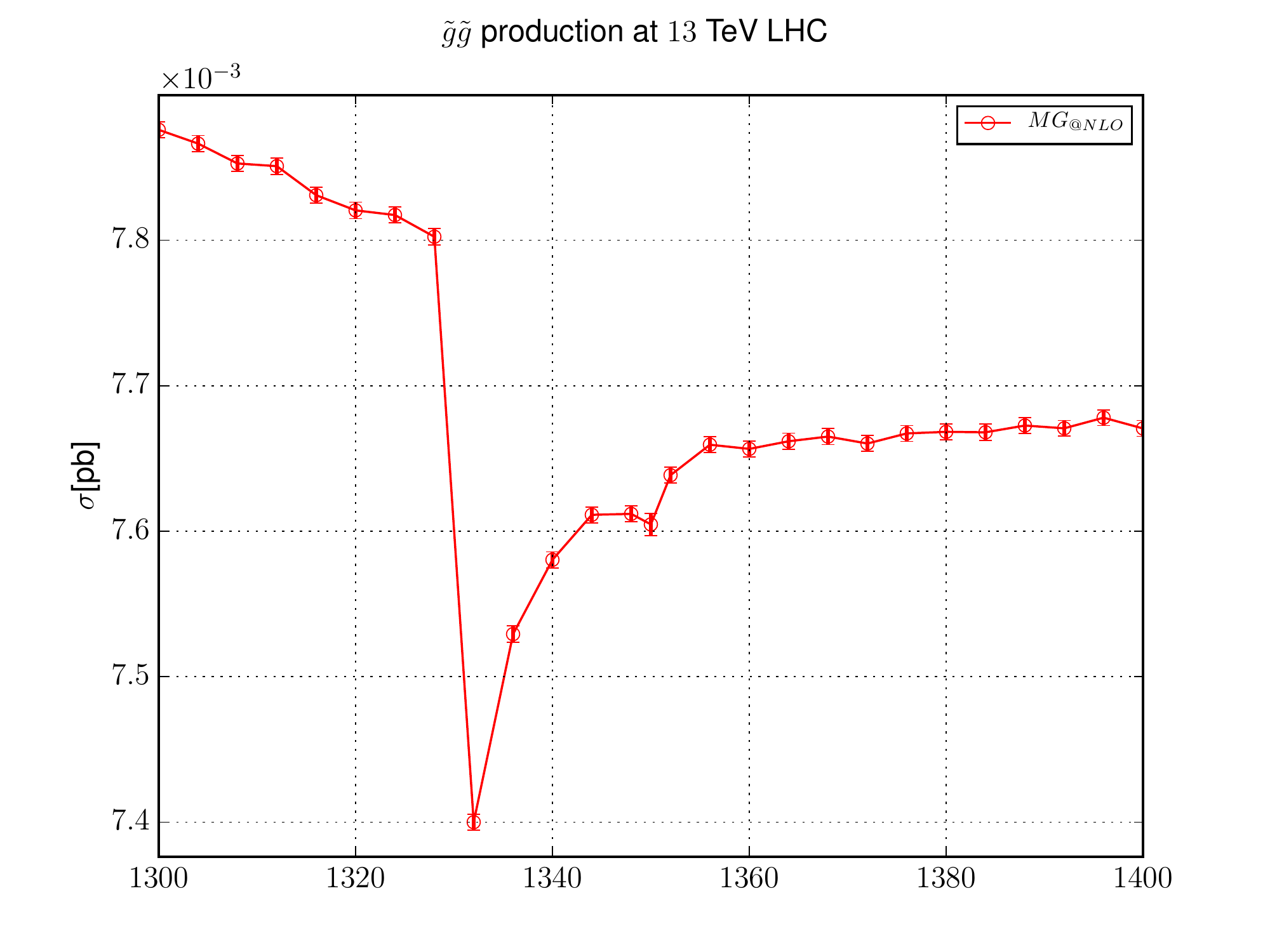}
  \end{center}
\end{minipage}
\begin{minipage}{.48\textwidth}
  \begin{center}
    \setlength{\abovedisplayskip}{6pt}
    \setlength{\belowdisplayskip}{\abovedisplayskip} 
    \setlength{\abovedisplayshortskip}{0pt}
    \setlength{\belowdisplayshortskip}{3pt} 
    \begin{align}
    \begin{split}
    \feynmandiagram[inline=(i0),horizontal=i0 to d,vertical=o2 to o1,small]{i0[particle=\(g\)] -- [gluon] d, o2[particle=\(\tilde{g}\)] --[plain,gluon] f --[charged scalar] d --[charged scalar] e -- [plain,gluon] o1[particle=\(\tilde{g}\)], f --[fermion] e};\\\\
    \feynmandiagram[inline=(i0),horizontal=i0 to d,vertical=o2 to o1,small]{i0[particle=\(g\)] -- [gluon] d, o2[particle=\(\tilde{t}_1\)] --[charged scalar] f --[fermion] d --[fermion] e -- [charged scalar] o1[particle=\(\tilde{t}_1\)], f --[plain,gluon] e};
    \end{split}\nonumber
    \end{align}
  \end{center}
\end{minipage}
  \caption{
    \label{fig:threshold_zoom} Left: \mg\ predictions for gluino pair-production
    at the 13~TeV LHC, for \mbox{$M_{\text{others}} \in [1300,1400]$~GeV}.
    Right: vertex correction diagrams featuring an anomalous threshold at $M_{\rm others}=M_{\rm prod}-m_t$.
  }
\end{figure}

The enhancement with respect to the LO results due to higher-order 
corrections is very significant for QCD-mediated processes
($\mathcal{O}(100\%)$), and considerably milder for the electroweak processes
($\mathcal{O}(10\%)$). One striking feature of the $M_{\rm others}$ dependence
of the NLO cross sections for both $\tilde{t}_1\anti{\tilde{t}_1}$ and
$\tilde{g}\tilde{g}$ production lies in the characteristic kink appearing at
$M_{\rm others}\simeq 1330$ GeV, which originates from the ``resonant'' anomalous
thresholds~\cite{Passarino:2018wix} featured by the one-loop vertex corrections
shown in figure~\ref{fig:threshold_zoom}. The latter cross a threshold at
$M_{\rm others}=M_{\rm prod}-m_t$, that is also highlighted in the zoomed-in 
figure presented in the left panel of figure~\ref{fig:threshold_zoom}.

In the degenerate mass regime, one expects to find complete agreement 
among the three codes, which is what figure~\ref{fig:degenerate_validation} 
basically shows at both the LO and the NLO, except in the case of 
chargino pair-production at the NLO, where large differences can be
seen between any two predictions. We point out that, for the other
processes, the agreement between \mg\ and \prospino\ is at the level
of 0.5\% at the worst (the latter cross sections being larger than the
former ones). This may originate from a small mismatch in the
input parameters, whose settings are especially intricate in \prospino\
as many of them are directly hard-coded in different parts of its source
code (the value of $\alpha_S(m_Z^2)$ is a prime example of this fact).  
At the NLO, the \prospino\ results for chargino
pair-production also differ by a rather flat offset of 3\% relative to
the \mg\ results. This might again be due to a mismatch in the input
parameters, which however we could not track down. Conversely, for 
chargino production the shape of the dependence on $M_{\rm others}$ 
of \resummino\ predictions is significantly different from the \mg\ one, 
and a preliminary investigation of the \resummino\ code has revealed issues 
in its SUSY-induced renormalisation of a specific vertex; this will be 
addressed in an upcoming release of the program by its authors.
Given the discrepancies found for this process, we have carried out a fully
independent analytic computation of the cut-constructible parts of the
corresponding virtual matrix elements, and found perfect point-wise agreement
with the corresponding automated numerical computations performed by \madloop.

\subsection{Non-degenerate SUSY masses\label{sec:nodeg}}

\begin{figure}
  \begin{center}
    \includegraphics[width=0.487\textwidth]{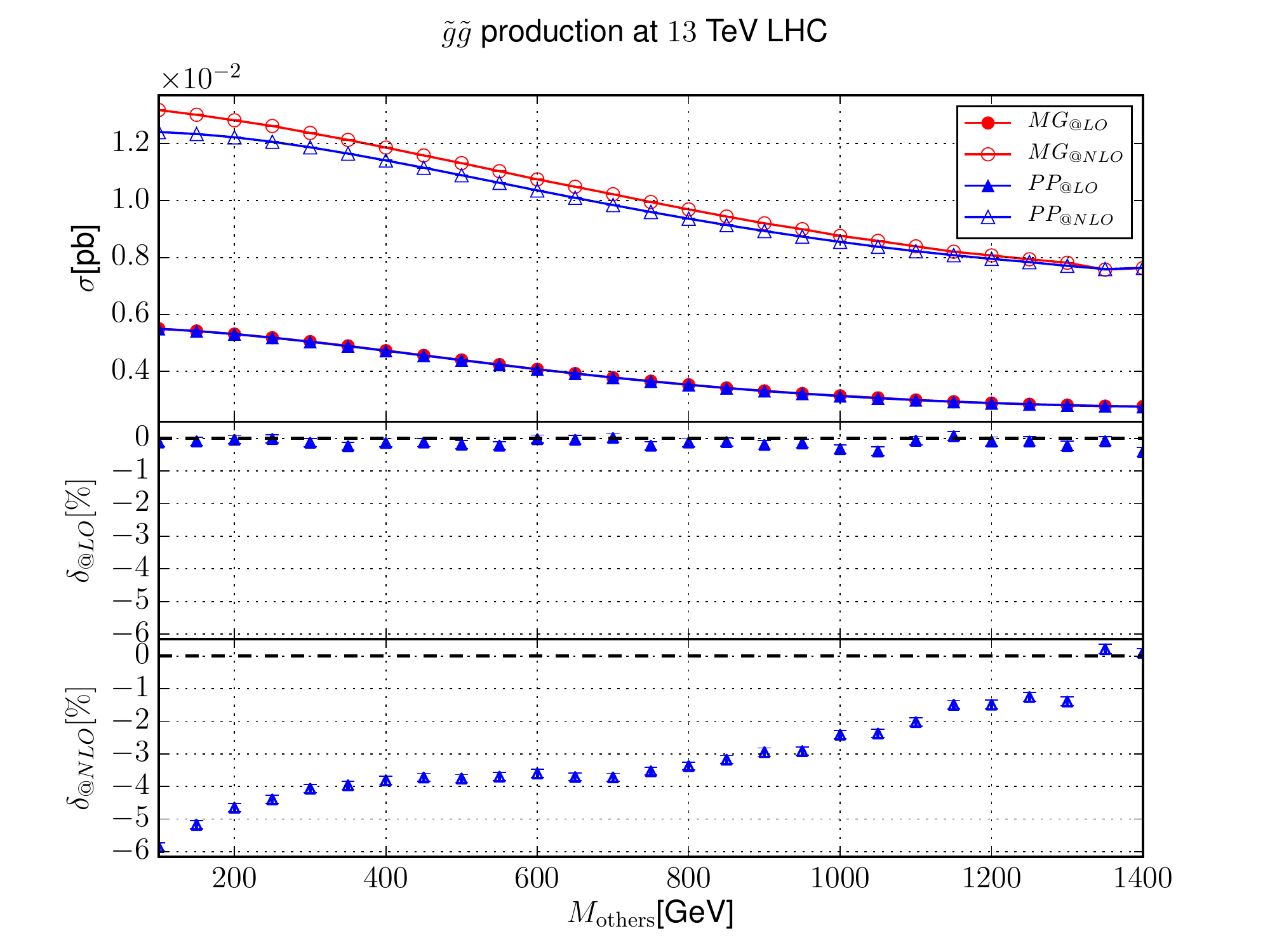}\quad
    \includegraphics[width=0.487\textwidth]{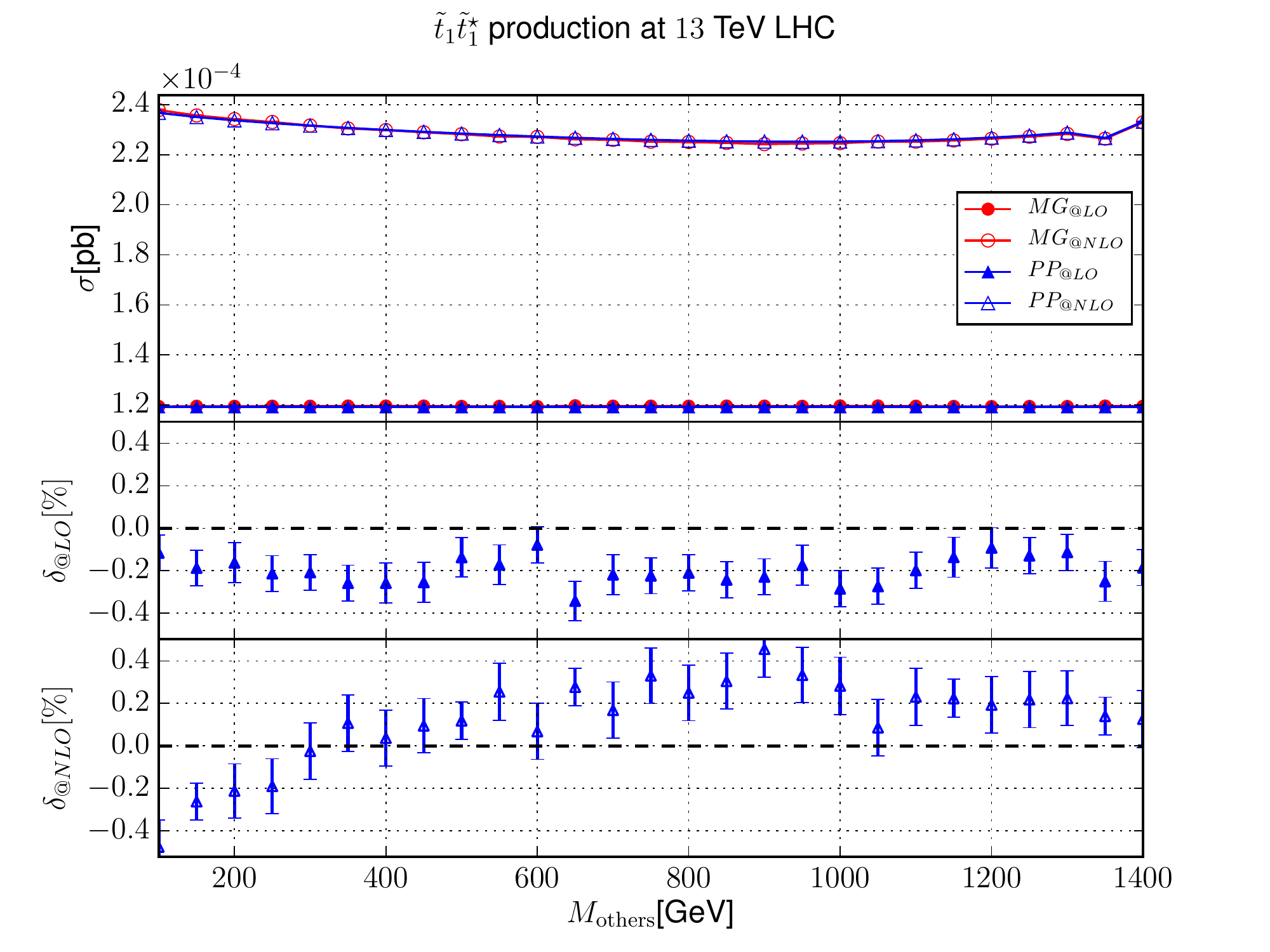}\\
    \includegraphics[width=0.487\textwidth]{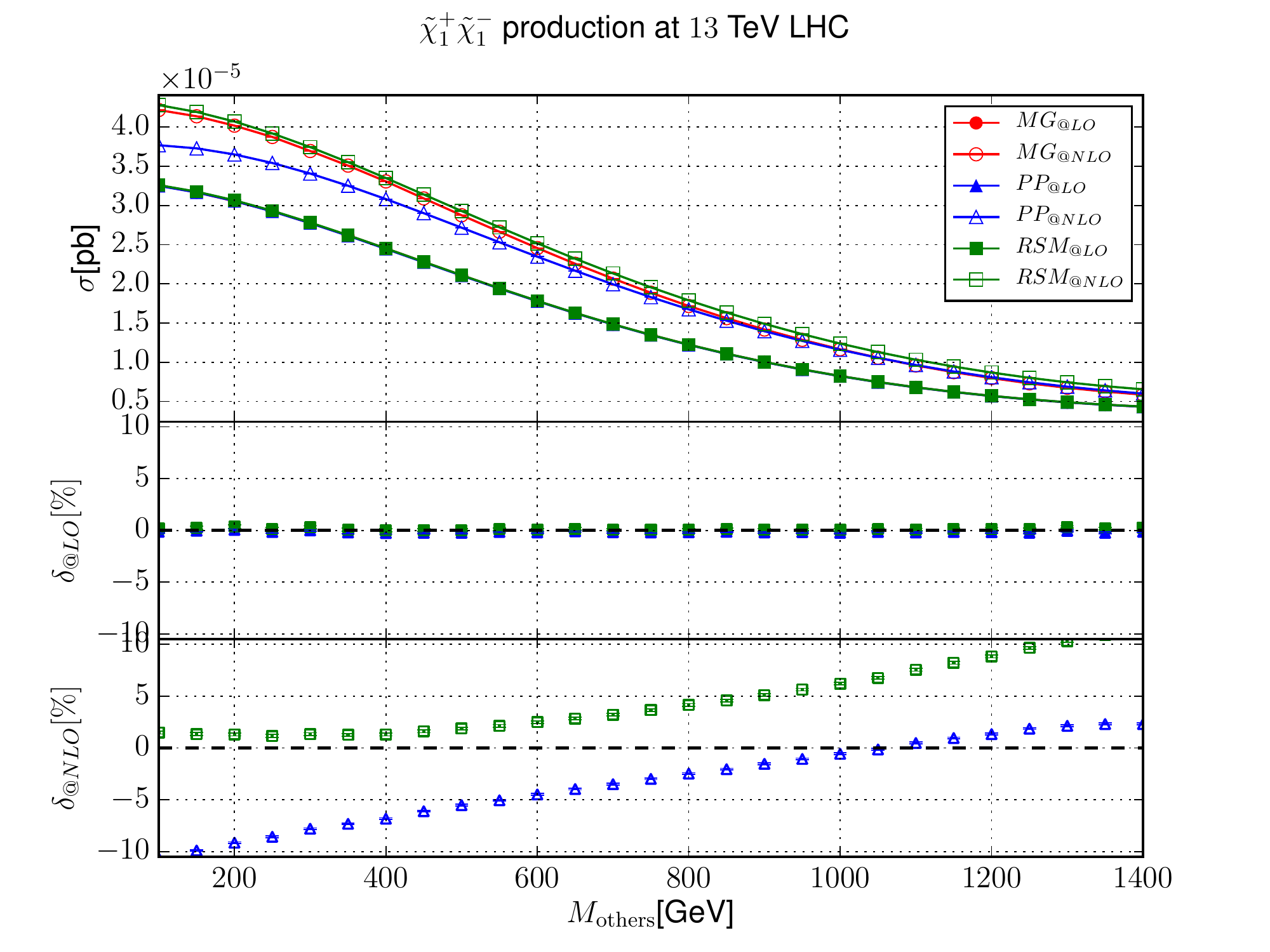}\quad
    \includegraphics[width=0.487\textwidth]{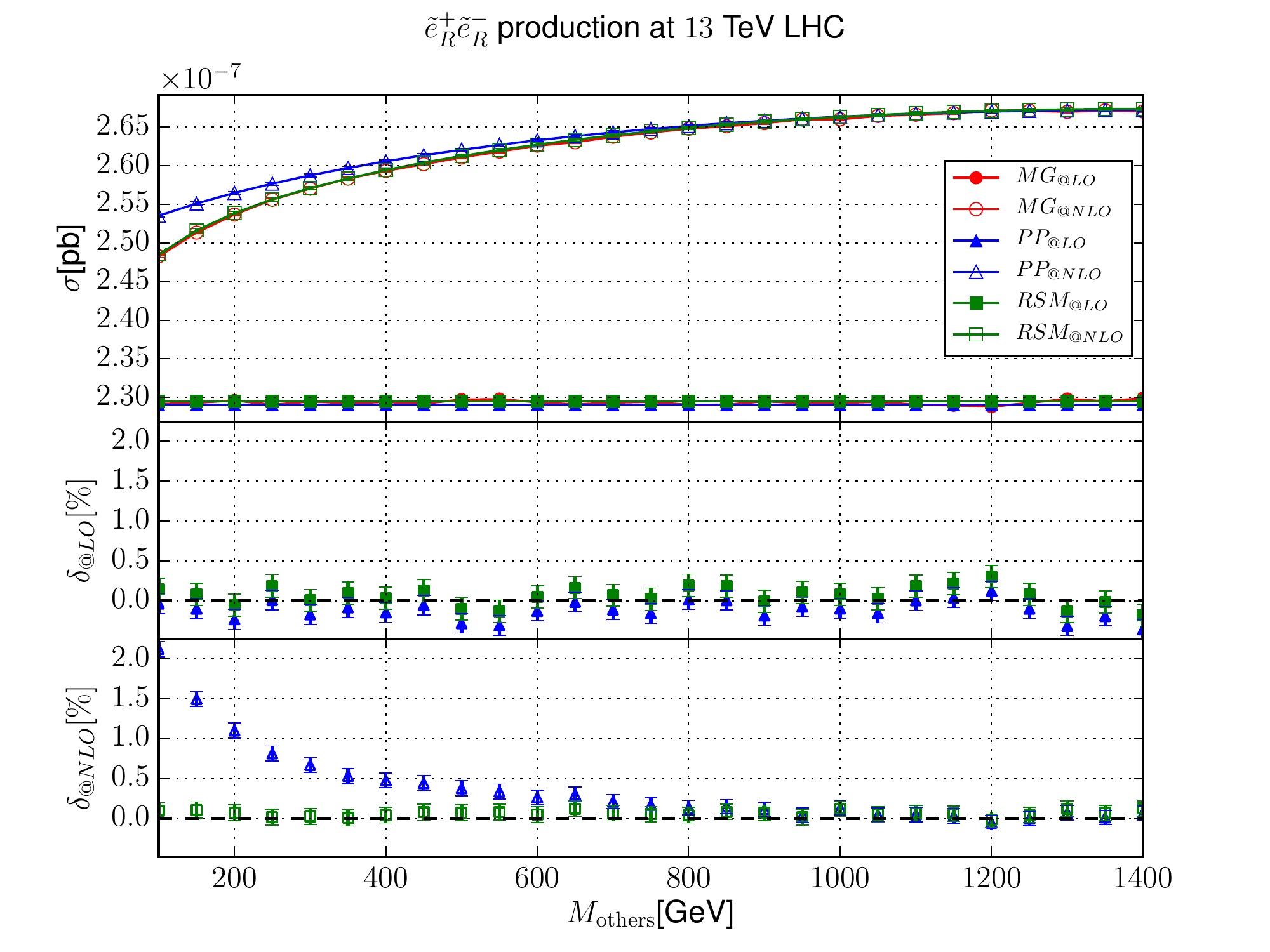}
  \end{center}
  \caption{\label{fig:nondegenerate_validation}
    Same as figure~\ref{fig:degenerate_validation}, but for non-degenerate
    squark masses.}
\end{figure}

Even in the absence of the issues we have outlined in
section~\ref{sec:deg}, in the non-degenerate squark regime we do 
not expect to find complete agreement between \mg\ and \prospino. 
Therefore, at least for the processes for which we found have agreement
in the degenerate scenario, we can assess the quality of the mass-averaging 
procedure  implemented in \prospino\ to derive the NLO cross sections 
when the spectrum is non-degenerate.
Figure~\ref{fig:nondegenerate_validation} shows that this approximation
can lead to differences with respect to the exact results (as computed by \mg) 
of several percent. The exception is stop-pair production, where the effect
of the SUSY masses being non-degenerate is small, as they appear only in 
the virtual amplitudes (and not at the level of real-emission diagrams). 
As far as \resummino\ is concerned, this program has been designed to
deal with the dependence on arbitrary SUSY mass spectra in an exact
manner. In spite of this, we do not find agreement when comparing its 
predictions with those of \mg\ for chargino-pair production, whilst the
agreement in the slepton-pair production case has only been found after a
couple of bug fixes in \resummino\ (that have been implemented in version 2.0.2;
unfortunately, these do not address the issue with charginos, which is still
under investigation.

\subsection{Summary of the comparisons}

The comparisons presented in this section and the sometimes significant
disagreement found among \mg, \prospino, and \resummino\ results, underscore
the need for a more comprehensive and robust implementation of NLO QCD
corrections for SUSY processes, that can be reliably used by collider
experiments. We believe that this is what is achieved currently by \mg, thanks
to its highly automated approach, and its history of orthogonal cross-checks
from applications and validation of the \emph{same} framework to
\emph{other} models and simulations.


\section{Total rates for supersymmetric benchmark processes in 
simplified scenarios}
\label{sec:rates}
In this section we calculate the total cross sections at the NLO in QCD for 
several supersymmetric processes in the context of simplified models,
that are typically employed for the interpretation of SUSY searches at
the LHC. In these scenarios, one assumes that only the final-state SUSY 
particles are relatively light, while all of the other superpartners are 
decoupled by their very large masses. This setup allows us to avoid to 
deal with intermediate resonances in the 
real-emission contributions, which will be extensively discussed in 
section~\ref{sec:OS}.

\subsection{Setup of the calculation} \label{sec:xsecsetup}
We consider the processes $p p \to \tilde X \tilde Y$ (where we denote by
$\tilde X$ and $\tilde Y$ two SUSY particles, which may also be identical) at 
the $\sqrt{S}=13$~TeV LHC, at its high-energy upgrade with $\sqrt{S}=27$~TeV,
and at a potential future proton-proton collider, identified as the FCC-hh, 
with $\sqrt{S}=100$~TeV. If $\tilde X$ and $\tilde Y$ are of different 
species, we enforce their masses to be equal, $m_{\tilde X}=m_{\tilde Y}$.
All of the other SUSY particles which do not appear in the final states 
are decoupled by setting their masses equal to 15~TeV (30~TeV, 110~TeV) 
when $\sqrt S=13$~TeV (27~TeV, 100~TeV, respectively), with the exception 
of the two stop states whose masses are fixed to $m_{\tilde t_1}=16$~TeV 
(32~TeV, 120~TeV) and $m_{\tilde t_2}=17$~TeV (34~TeV, 130~TeV). We refer to
appendix~\ref{app:decoupling} for details on the complexity of a numerical
implementation of the decoupling of heavy SUSY particles.
The widths of particles are taken to be equal
to zero, and the central values of the factorisation and
renormalisation scales are set as follows:
\begin{equation}
\mu_F=\mu_R\equiv\tilde{M} =\frac{m_{\tilde{X}}+m_{\tilde{Y}}}{2}\,.
\end{equation}
The theoretical uncertainty stemming from the
missing higher-order corrections is estimated by a nine-point
independent scale variation, $(\mu_F,\mu_R)=(\xi_F,\xi_R)\tilde{M}$, with
$\xi_{F/R}=1/2$, 1, 2. We use the {\tt NNPDF30\_nlo\_as\_0118}
set of parton densities~\cite{Ball:2014uwa} as provided by the \lhapdf\
interface~\cite{Buckley:2014ana}.

\subsection{Production of a pair of SUSY particles of the same species}
\label{sec:xsecSF}
\begin{figure}
  \centering
  \includegraphics[width=0.9\textwidth]{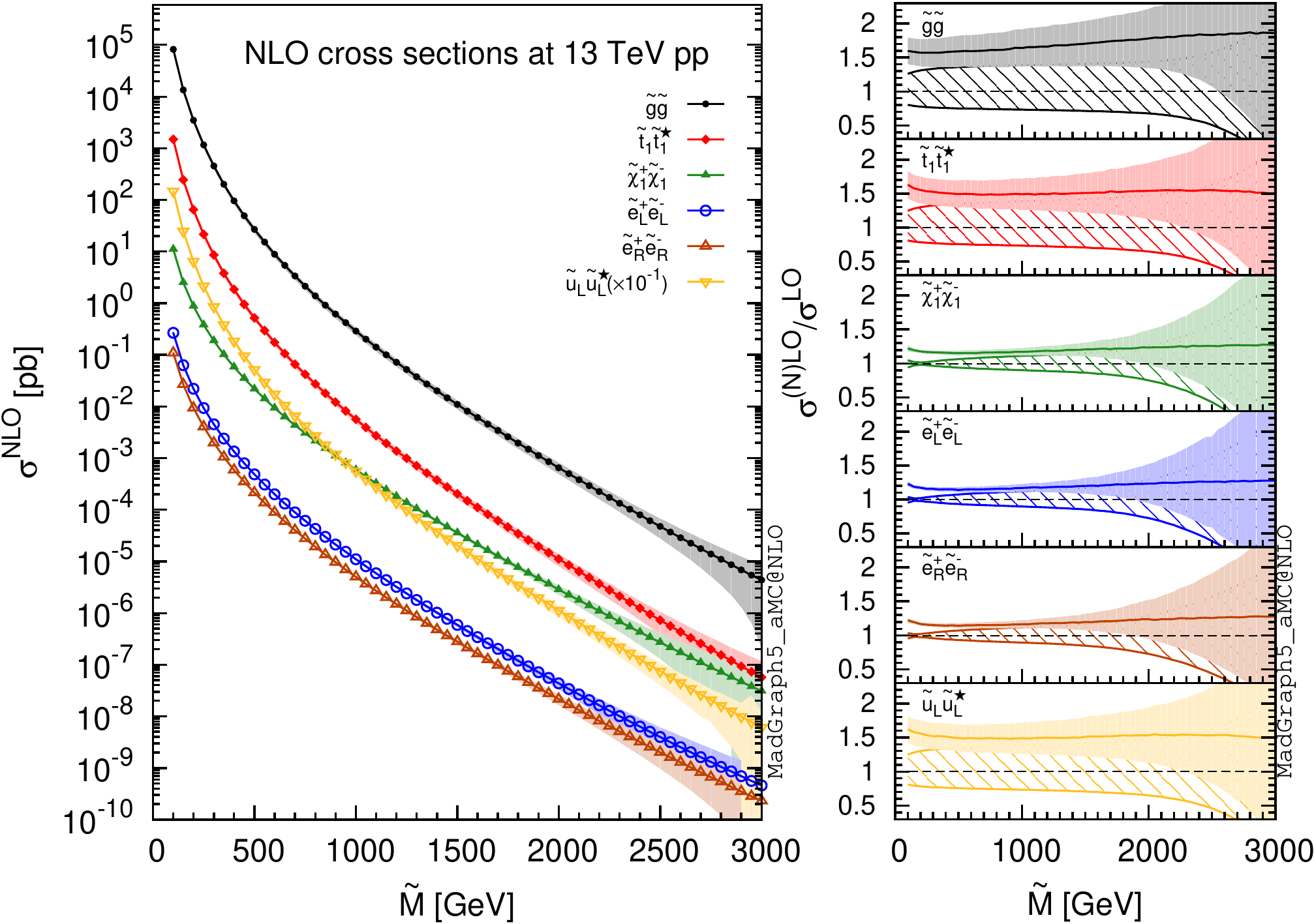}
  \caption{\label{fig:simplifiedxsect}
    Total NLO cross sections (left) and
    $K$-factors defined as the ratio of the NLO result to the corresponding LO
    one (right) for the six processes associated with the production of a pair
    of SUSY particles in the same species at $\sqrt S=13$ TeV LHC.
    The different bands correspond to the sum of the scale and PDF uncertainties
    in quadrature, the NLO ones being indicated by filled areas and the LO ones
    (shown on the right panel) by hashed areas.}
\end{figure}

\begin{figure}
  \centering
  \includegraphics[width=0.9\textwidth]{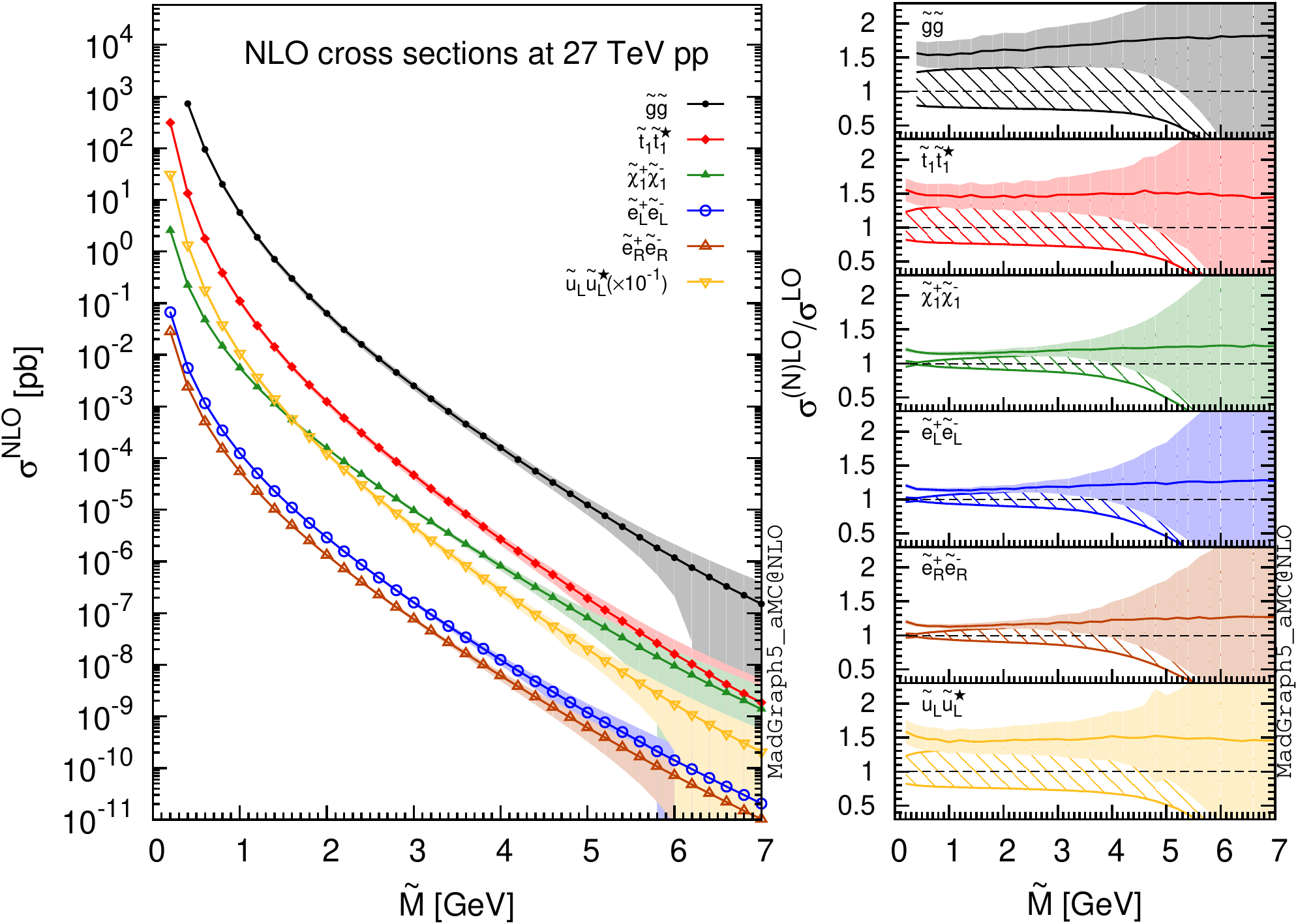}
  \caption{\label{fig:simplifiedxsect27TeV} Same as
    figure~\ref{fig:simplifiedxsect}, but for proton-proton collisions at
      $\sqrt S=27$ TeV.}\vspace*{0.5cm}
  \includegraphics[width=0.9\textwidth]{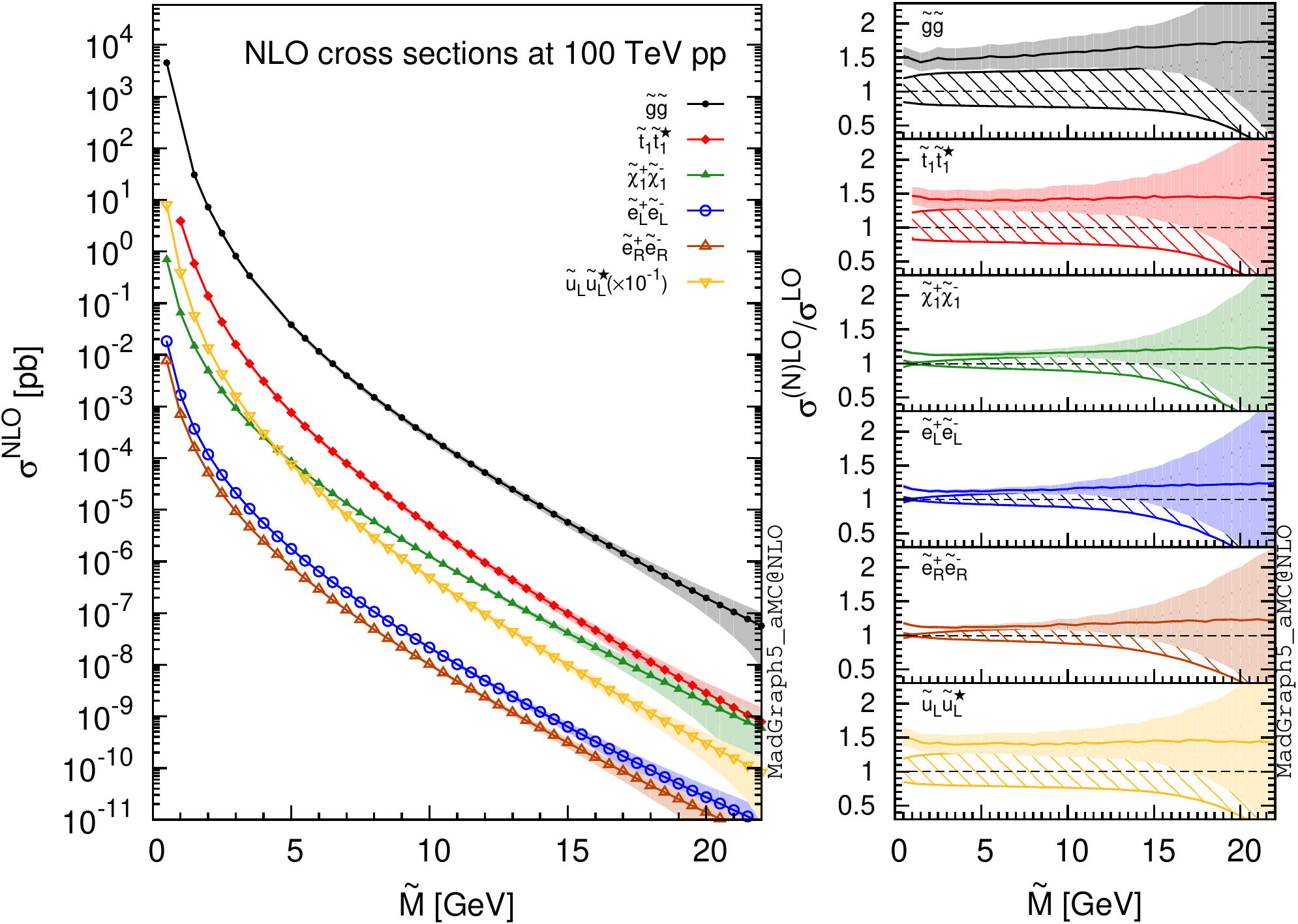}
  \caption{\label{fig:simplifiedxsect100TeV} Same as
    figure~\ref{fig:simplifiedxsect}, but for proton-proton collisions at
      $\sqrt S=100$ TeV.}
\end{figure}

In this section, we consider six pair-production processes in which
$\tilde X$ and $\tilde Y$ are of the same species. NLO total rates are 
shown in the left panel of figure~\ref{fig:simplifiedxsect}, for 
$\sqrt S=13$~TeV proton-proton collisions resulting in the production of 
a pair of gluinos ($\tilde{g}\tilde{g}$, black circles), a pair of light 
stops ($\tilde{t}_1\anti{\tilde{t}_1}$, red diamonds), a pair of left-handed 
up squarks ($\tilde{u}_L\anti{\tilde{u}_L}$, yellow triangles), a pair of 
left-handed and right-handed selectrons ($\tilde{e}_L^+\tilde{e}^-_L$ and 
$\tilde{e}^+_R\tilde{e}_R^-$, blue circles and brown triangles, respectively), 
and a pair of (opposite-sign) charginos ($\tilde{\chi}_1^+\tilde{\chi}_1^-$, 
green triangles). In order to improve the visibility of the different curves, 
we have included a rescaling factor equal to 0.1 in the case of 
$\tilde{u}_L\anti{\tilde{u}_L}$ production. Our results include the bands
associated with the theoretical uncertainties obtained from the independent
variations of the renormalisation and factorisation scales, as well as from the
PDF uncertainties; these are added in quadrature.
\definecolor{darkmagenta}{rgb}{0.55, 0.0, 0.55}

The NLO cross sections are found to span about 10 orders of magnitude when the 
SUSY mass $\tilde M$ varies from 100~GeV to 3~TeV, with the strong production 
of squark and gluino pairs larger than the electroweak production of slepton or
electroweakino pairs by orders of magnitude, for any given $\tilde M$ value.
We also show the associated $K$-factors in the right panel of the figure, where
each $K$-factor is defined as the ratio of the NLO total rate over the 
corresponding LO one evaluated at the central scale and with the central 
set of PDF. The depicted uncertainty therefore reflects the standard NLO 
cross section uncertainty, as extracted relative to the central LO 
predictions. The $K$-factors exhibit different behaviours for the
different processes. Firstly, they are larger in the cases of strong
squark and gluino production ($K\sim1.5$) than for Drell-Yan-like slepton and
chargino production ($K\sim1.2$), as is expected from the strong/electroweak 
nature of such processes. Secondly, the $K$-factor associated with the
$\tilde g\tilde g$ process shows a significant dependence on the SUSY mass
$\tilde M$, which can be traced back to the virtual amplitudes associated with
the quark-antiquark contribution to the cross section and the
large gluino colour charge~\cite{Beenakker:1996ch}.
Whilst subdominant for small SUSY masses, the quark-antiquark contribution 
becomes significant when $\tilde M$ increases (since the relative weight of
the corresponding parton luminosity increases with respect to the $gg$ one), 
and it therefore impacts the cross section to a more
significant level. Conversely, the $\tilde M$ dependence of the two
$K$-factors associated with the production of a pair of squarks is more
moderate, and almost absent in the case of the electroweak processes.

The right panel of figure~\ref{fig:simplifiedxsect} illustrates the
benefits of higher-order calculations, as it shows that theoretical
systematics are smaller at the NLO (filled areas) than at the LO
(hashed areas). However, predictions relevant to large $\tilde M$ values
are affected, both at the LO and the NLO, by significant uncertainties. 
This is because, in this region, the latter are dominated by PDF errors.
In fact, by increasing $\tilde M$, the average Bjorken $x$'s that enter 
the partonic cross sections also grow, and at large $x$'s the  PDFs are 
poorly constrained. Fortunately, one expects that a stronger constraining 
power of the searches for SUSY signals will go hand in hand with better-quality
data for SM processes at large scales, which in turn will help reduce
the PDF uncertainties.

Similar results as in figure~\ref{fig:simplifiedxsect} 
can be found when the centre-of-mass energy is set equal to
$\sqrt S=27$~TeV (see figure~\ref{fig:simplifiedxsect27TeV}) and 100~TeV (see
figure~\ref{fig:simplifiedxsect100TeV}). As is expected, the main difference
with respect to the 13~TeV case is the increase of the cross sections at any
given  $\tilde M$, due to the larger available centre-of-mass energies. By 
scaling up the SUSY mass $\tilde M$ to match the collider energy, the
behaviours of the $K$-factors are essentially identical for the 
three collider scenarios.

\subsection{Production of a pair of SUSY particles of different species}

\begin{figure}
    \centering
    \includegraphics[width=0.9\textwidth]{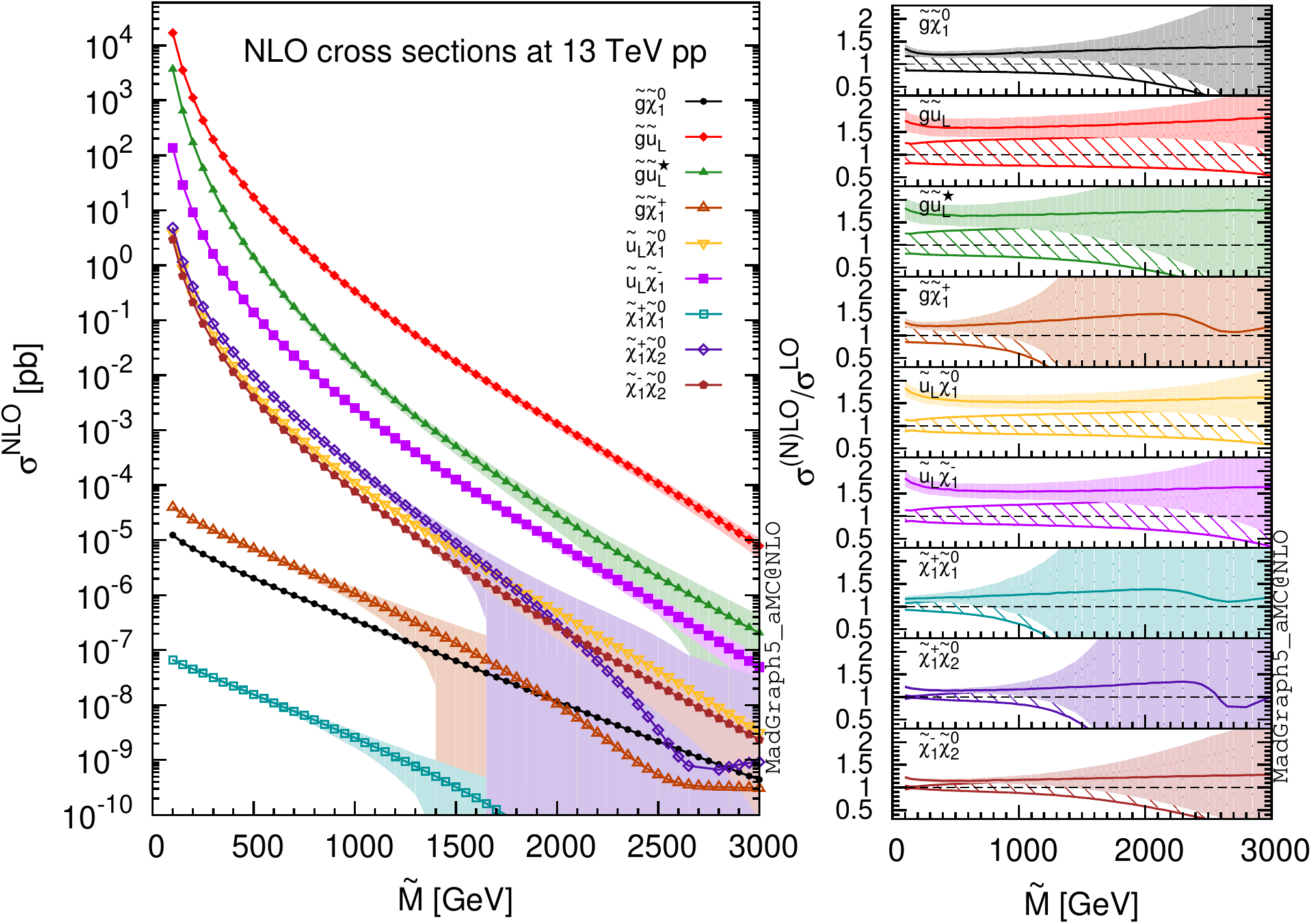}
    \caption{\label{fig:nextsimplifiedxsect} 
    Total NLO cross sections (left) and $K$-factors (right) for nine processes
    involving the production of an associated pair of different SUSY particles
    with identical mass $\tilde M$ at $\sqrt S=13$ TeV LHC.
    The different bands correspond to the quadratic sum of the scale and PDF
    uncertainties, the NLO ones being indicated by filled areas and the LO ones
    (shown on the right panel) by hashed areas.}\vspace{.5cm}
\end{figure}

We now consider the production of two SUSY particles of different species,
while still setting their masses equal to a common value $\tilde M$.
In figure~\ref{fig:nextsimplifiedxsect}, we present the dependence
of the NLO cross sections on $\tilde M$ for nine different SUSY
pair-production processes, in proton-proton collisions at $\sqrt{S}=13$~TeV. 
We focus on two strong processes in which a gluino is produced in association 
with a left-handed up squark ($\tilde{g}\tilde{u}_L$, red diamonds) or 
antisquark ($\tilde{g}\anti{\tilde{u}_L}$, green triangles), as well as four
semi-strong processes corresponding to the production of a gluino and a
neutralino ($\tilde{g}\chi_1^0$, black circles), a gluino and a chargino
($\tilde{g}\chi_1^+$, brown triangles), a left-handed up squark and a
neutralino ($\tilde{u}_L\chi_1^0$, yellow triangles) and a left-handed up squark
and a chargino ($\tilde{u}_L\chi^-_1$, magenta squares). Our results finally
also include predictions for three electroweakino pair-production processes in
which the lightest chargino is produced in association with the lightest
neutralino ($\tilde{\chi}_1^+\tilde{\chi}_1^0$, turquoise squares) or with the
next-to-lightest neutralino ($\tilde{\chi}_1^+\tilde{\chi}_2^0$ and
$\tilde{\chi}_1^-\tilde{\chi}_2^0$, purple diamonds and red pentagons).
For all of our predictions, the lightest neutralino is taken to be bino-like,
whilst the next-to-lightest neutralino and the lightest chargino are both taken
to be wino-like. Analogously to what has been done previously, we present 
the corresponding $K$-factors on the right panel of 
figure~\ref{fig:nextsimplifiedxsect}, and we include theoretical errors
estimated by summing in quadrature the uncertainties stemming from scale
variations and the PDF errors.

Gluino-squark cross sections (\ie\ with a $\tilde{g}\tilde{u}_L$ 
or a $\tilde{g}\tilde{u}_R$ final state) are identical to each other, 
these processes being driven by strong interactions that are blind with 
respect to the (s)quark chirality. Owing to the larger up-quark density in 
the proton with respect to the antiup-quark one, the corresponding conjugate
processes are 
suppressed by factors that range from a few units (for small SUSY masses) 
to almost two orders of magnitude (for large SUSY masses). This is illustrated 
in figure~\ref{fig:nextsimplifiedxsect} for $\tilde{g}\anti{\tilde{u}_L}$
production (the cross section for $\tilde{g}\anti{\tilde{u}_R}$ is identical
to the latter one, and is not shown). The $K$-factors associated with all 
these strong processes depend significantly on the SUSY mass $\tilde M$, 
and vary from about 1.5--1.6 for small $\tilde M$'s to about 1.9 for 
multi-TeV $\tilde M$. 
As was already observed in section~\ref{sec:xsecSF}, 
it is the gluino with its large associated colour charge that drives this 
dependence of the NLO $K$-factors. Theoretical systematics
follow the same pattern as those relevant to same-species production,
namely scale uncertainties are reduced at the NLO, whilst the total 
uncertainty increases for large $\tilde M$ because of the PDF behaviour.
This is especially significant in the case of the antiquark density, so that
we accordingly find a larger uncertainty for $\tilde{g}\anti{\tilde{u}_L}$
production than for $\tilde{g}\tilde{u}_L$ production.

\begin{figure}
    \centering
    \includegraphics[width=0.55\textwidth]{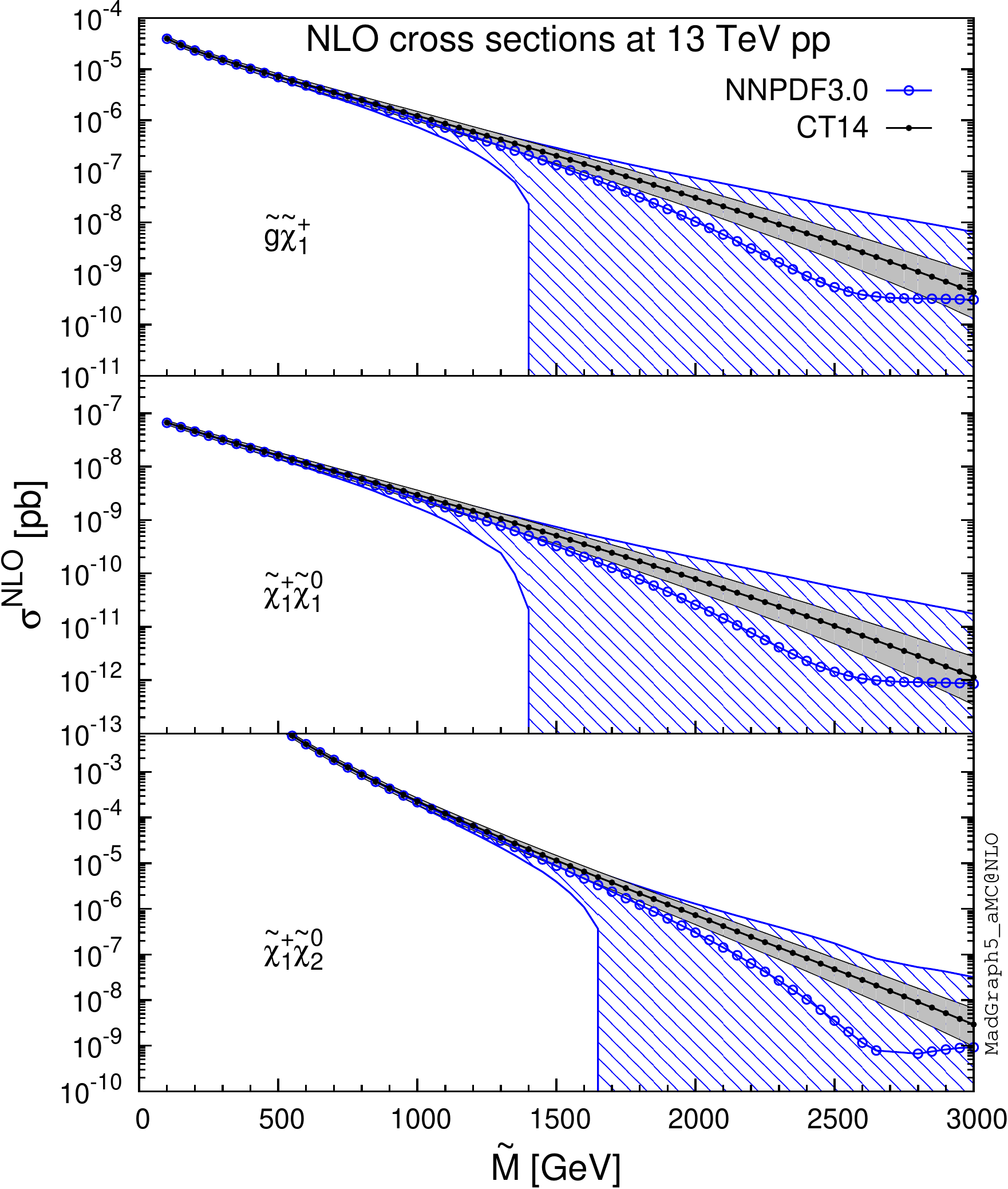}
    \caption{\label{fig:nextsimplifiedxsectPDF} Total NLO cross sections
     for $\tilde{g}\tilde{\chi}_1^+$ (upper panel), $\tilde\chi_1^+\tilde\chi_1^0$ (middle panel)
     and $\tilde{\chi}_1^+\tilde{\chi}_2^0$ (lower panel) production. We compare the results
     obtained when the NLO matrix elements are convoluted with the
     {\tt NNPDF3.0} and {\tt CT14} PDF sets.}
\end{figure}

The other processes under consideration are of either purely electroweak or 
semi strong/electroweak nature. The rates are consequently reduced by several
orders of magnitude, and the $K$-factors turn out to be smaller than for 
the purely strong processes. As is shown in the right panel of
figure~\ref{fig:nextsimplifiedxsect}, QCD corrections are in general
larger for the semi-strong processes featuring a gluino or a squark in the final
state ($K\sim 1.5$), as is expected from their sensitivity to strong 
interactions that is present already at the tree level. Conversely, the 
purely-electroweak electroweakino pair production processes exhibit smaller 
$K$-factors of about 1.2, a typical value for the Drell-Yan-like 
electroweakino pair-production that occurs when all squarks are decoupled. 
The mass dependences of the $K$-factors are moreover modest, with the 
exception of the peculiar behaviour exhibited by the 
$\tilde g\tilde{\chi}_1^+$, $\tilde{\chi}_1^+\tilde{\chi}_1^0$,
and $\tilde{\chi}_1^+\tilde{\chi}_2^0$ processes for $\tilde M\sim 2.5$~TeV. 
The dominant contribution to these three processes originates from a $u\bar d$
initial state. However, the {\tt NNPDF} densities used here are mostly unknown
at large $x$ ($x>0.1$) and are therefore associated with a large uncertainty.
Furthermore, the {\tt NNPDF} methodology (which relies on neural networks to
perform the PDF fit) yields the odd shape of the cross sections and $K$-factors
in this regime. Correspondingly, the PDF uncertainties related to these processes
grow out of control for $\tilde M > 1.5$~TeV, and the shape of the central
$K$-factor, in particular close to $\tilde M\sim 2.5$~TeV, stems from the Born
and real-emission contributions being affected differently by the
corresponding partonic luminosities.

In order to better understand these peculiar features of the
$\tilde g\tilde{\chi}_1^+$, $\tilde{\chi}_1^+\tilde{\chi}_1^0$ and
$\tilde{\chi}_1^+\tilde{\chi}_2^0$ cross sections, we show in
figure~\ref{fig:nextsimplifiedxsectPDF} the predictions obtained when the 
matrix elements are convoluted either with the {\tt NNPDF 3.0} NLO PDF set, 
or with the {\tt CT14nlo} Hessian PDF set~\cite{Dulat:2015mca}. While the 
cross sections evaluated with {\tt CT14} PDFs show a more reasonable behaviour 
at large $\tilde{M}$, this comes at the cost of introducing a 
theoretically-dominated bias on the predictions, as such a PDF set  
relies entirely, in the large-$x$ region where there is no data point to 
constrain the fit, on the extrapolation of its parameterisation at the 
initial scale.

\begin{figure}
  \centering
  \includegraphics[width=0.9\textwidth]{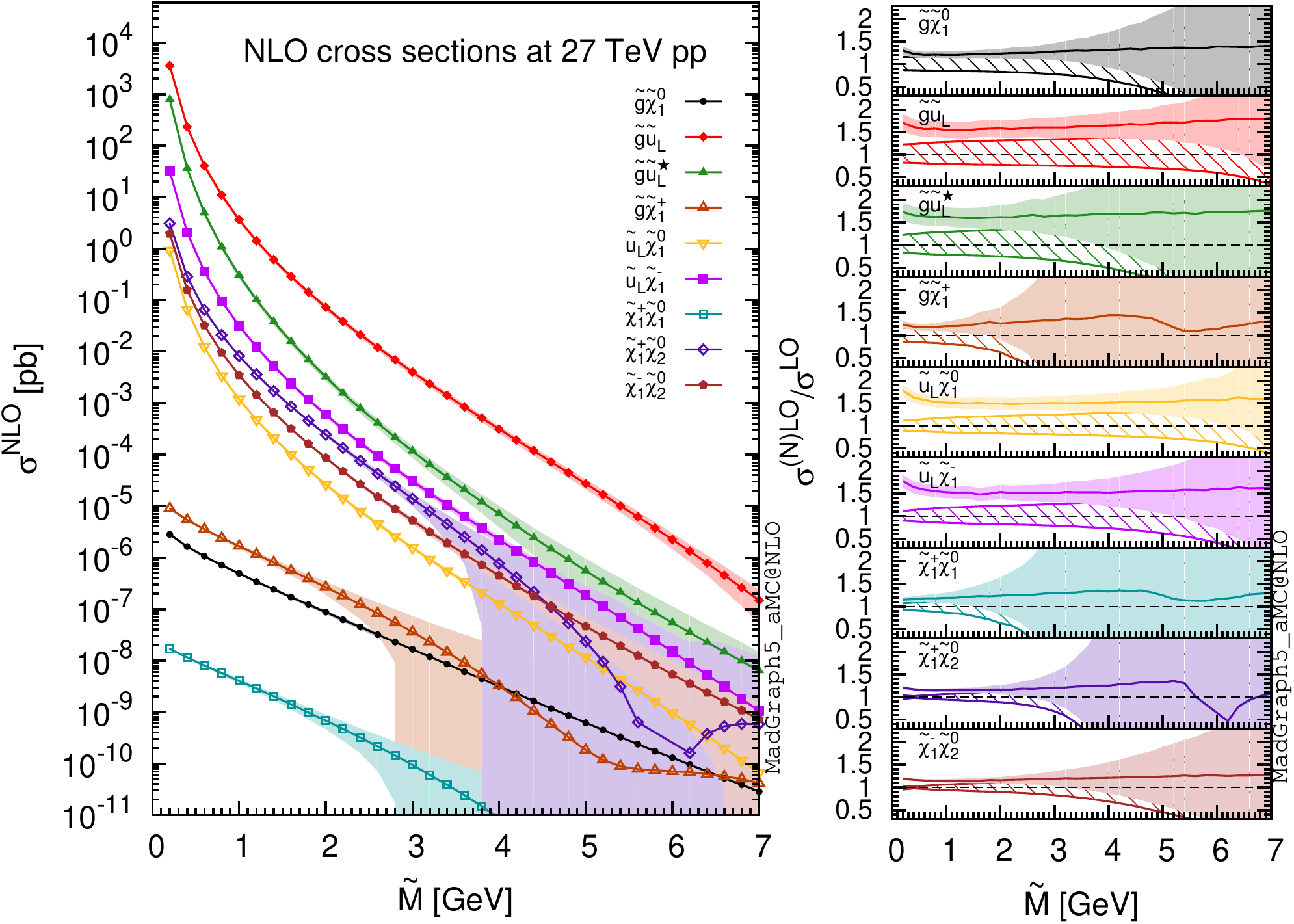}
  \caption{\label{fig:nextsimplifiedxsect27TeV} Same as
    figure~\ref{fig:nextsimplifiedxsect}, but for proton-proton collisions at
    $\sqrt S=27$ TeV.}\vspace{.5cm}
\end{figure}

\begin{figure}
  \centering
  \includegraphics[width=0.9\textwidth]{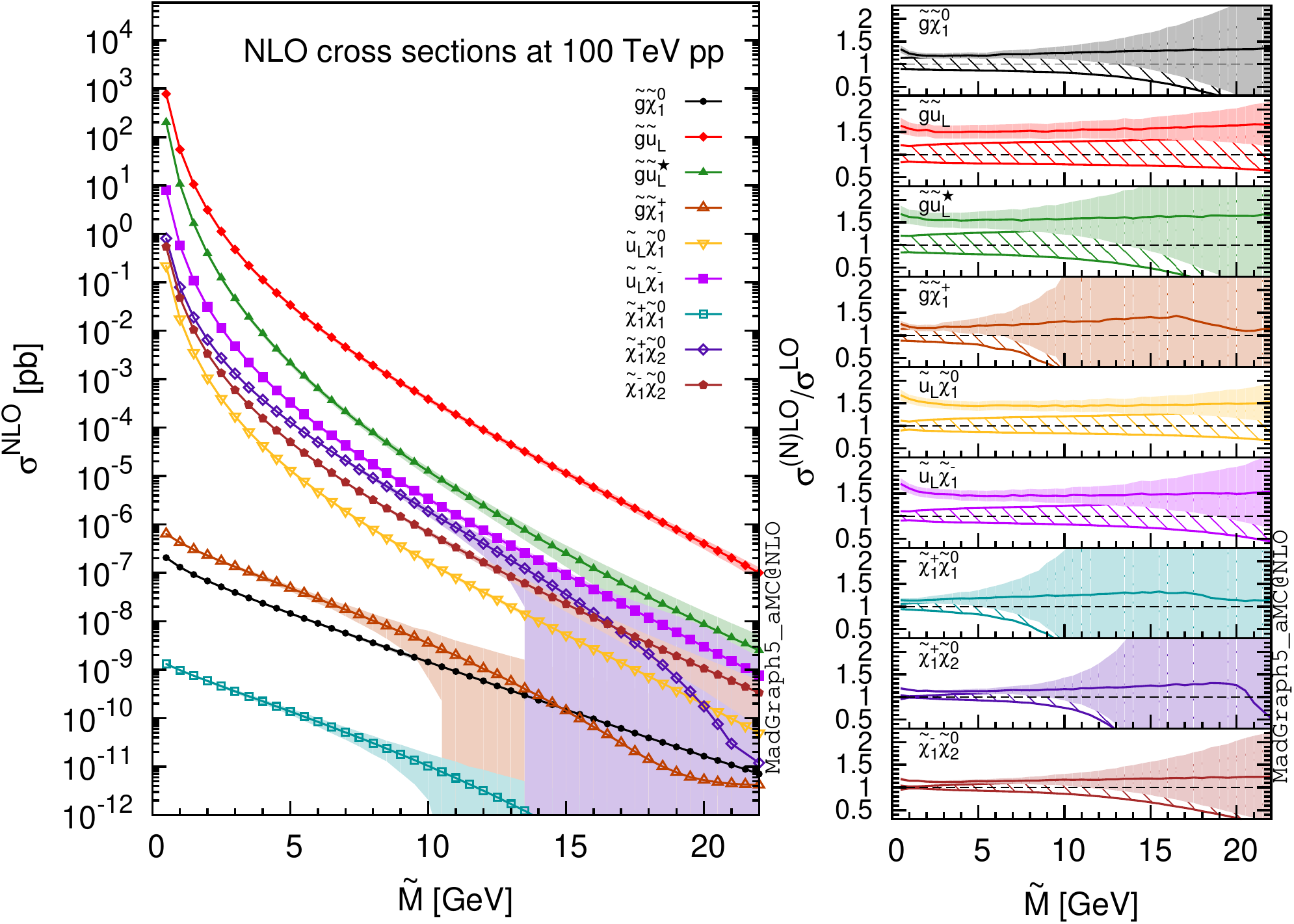}
  \caption{\label{fig:nextsimplifiedxsect100TeV} Same as
    figure~\ref{fig:nextsimplifiedxsect}, but for proton-proton collisions at
      $\sqrt S=100$ TeV.}
\end{figure}

The results obtained by increasing the centre-of-mass energy to
$\sqrt S=27$~TeV and 100~TeV are presented in 
figures~\ref{fig:nextsimplifiedxsect27TeV} 
and~\ref{fig:nextsimplifiedxsect100TeV}, respectively.
As far as the relative comparisons of these predictions with those relevant
to $\sqrt S=13$~TeV is concerned, the same observations made at the end of 
section~\ref{sec:xsecSF} for the case of same-species pair production
apply here.

We conclude this section by pointing out that tables reporting the numerical 
values that correspond to the cross sections shown in 
figures~\ref{fig:simplifiedxsect}--\ref{fig:nextsimplifiedxsect100TeV} are
provided as ancillary files on the electronic archive.

\section{Perturbative computations in the presence of resonances}
\label{sec:OS}

In this section, we discuss in general the problems posed to 
perturbative computations by the presence of narrow resonances,
and outline the strategies (called Simplified Treatments of Resonances or STR
for short) which one may employ to overcome such problems. STR include all 
of the Diagram Subtraction (DS) and Diagram Removal (DR) procedures defined 
so far in the literature, and must be seen as a systematic generalisation 
of the so-called on-shell subtractions (OSS). We also document here the 
implementation of the STR in \mg. Illustrative examples of their applications
are given in section~\ref{sec:pheno}.

\subsection{General features}
\label{sec:STRgeneral}producedures
In any theory with a sufficiently rich particle spectrum, there
is the possibility that the cross section for the production of a given 
asymptotic state $\delta$ is ill-defined in perturbation theory beyond 
the LO. Here, $\delta$ is such that its four momentum can, at least in 
principle, be reconstructed through measurements performed with a realistic 
detector, either directly or indirectly through its decay products\footnote{In 
other words, $\delta$ is not a light quark or a gluon.}. This situation can 
occur in the following case. Let
\begin{equation}
a+b\;\longrightarrow\;\delta+X
\label{delLO}
\end{equation}
be an LO contribution to the production of $\delta$; $a$ and $b$ 
denote the incoming partons that initiate the hard scattering, and $X$ a set
of final-state particles. The cross section for the process
of eq.~\eqref{delLO} may be inclusive or exclusive in $X$. At the NLO,
real-emission corrections will receive contributions from processes that 
one can write as follows:
\begin{equation}
a+b\;\longrightarrow\;\delta+\gamma+X\,,
\label{delNLO}
\end{equation}
where the nature of $\gamma$ depends on the underlying theory whose
perturbative expansion is considered. For example, in QCD $\gamma$ can
be a massless quark or a gluon, while in QED it can be a photon. Let us now
suppose that a particle $\beta$ (which must not resonantly contributes to the
process of eq.~\eqref{delLO}) exists, such that the two-body decay channel
\begin{equation}
\beta\;\longrightarrow\;\delta+\gamma
\label{betadec}
\end{equation}
is kinematically allowed and that
\begin{equation}
a+b\;\longrightarrow\;\beta+X\ ,
\label{betaLO}
\end{equation}
is a well-defined hard process, that we call the {\em underlying
resonant process}. Equations~(\ref{betadec}) and~(\ref{betaLO})
imply that, among the Feynman diagrams contributing to the process of
eq.~\eqref{delNLO}, there will be $\beta$-resonant ones, namely those
that feature a propagator associated with $\beta$. In turn, this allows
one to write the contributions of such diagrams to the
differential cross section associated with 
eq.~\eqref{delNLO} as follows:
\begin{equation}
d\sigma_{ab\to\delta\gamma X}\stackrel{m_{\delta\gamma}\to m_\beta}{\sim}
d\sigma_{ab\to\beta X}\,
\frac{1}{\left(m_{\delta\gamma}^2-m_\beta^2\right)^2}\,
d\Gamma_{\beta\to\delta\gamma}\,,
\label{OSdiv}
\end{equation}
where the first term on the r.h.s.~is the cross section for the 
production of an on-shell $\beta$. In eq.~(\ref{OSdiv}), we have denoted 
by $m_{\delta\gamma}$ and $m_{\beta}$ the invariant masses of the
$\delta\gamma$ pair and of the particle $\beta$, respectively.
Even if one Dyson-resums the propagator in eq.~(\ref{OSdiv}), thus
introducing a regularising $\Gb$ factor that prevents the
propagator from diverging at $m_{\delta\gamma}=m_\beta$, it may
still happen that
\begin{equation}
\int d\sigma_{ab\to\beta X}\,\gg\,\int d\sigma_{ab\to\delta X}\,.
\end{equation}
In this case, the NLO contribution due to eq.~\eqref{delNLO} will be
numerically (much) larger than its LO counterpart of eq.~\eqref{delLO},
thus 'spoiling' the perturbative expansion of the cross section for
$\delta$ production.

Situations of this kind are annoying because potentially relevant
for phenomenology, especially in SUSY theories where
they are ubiquitous. Examples stemming from QCD corrections include
\begin{equation}
(\delta,\gamma,\beta,X)=(W^-,\bb,\bt,t)
\label{SMex1}
\end{equation}
in the SM ($tW^-$ associated production, whose underlying resonant
process is $t\bt$ production) and
\begin{eqnarray}
(\delta,\gamma,\beta,X)&=&(\sq,q,\sg,\sq)\,,
\label{BSMex1}
\\
(\delta,\gamma,\beta,X)&=&(\sn,q,\sq,\sq)
\label{BSMex2}
\end{eqnarray}
in SUSY (squark-pair and squark-neutralino production, whose 
underlying resonant processes are squark-gluino and squark-pair
production, respectively\footnote{Squark-neutralino production
features a neutralino-gluino channel which also plays the role of
underlying resonant process.}).
A further case is that of a simplified dark matter model, achieved
\eg\ by extending the SM with a dark matter particle $\chi$ and a 
mediator $Y$, so that
\begin{equation}
(\delta,\gamma,\beta,X)=(\chi,q,Y,\chi)
\label{BSMex3}
\end{equation}
is the analogue of the processes of eq.~\eqref{SMex1}--\eqref{BSMex2}.

By adopting a commonly-used expression, which is strictly speaking incorrect 
but conveys the basic physics idea, one says that $\delta X$ production 
interferes with $\beta X$ production beyond the LO. The numerical
dominance of the latter over the former implies that the corresponding
cross section is, to a good approximation, a meaningful physical quantity
(for example, we are used to talk about measurements of the $t\bt$
cross section, which we compare with their perturbatively-computed
counterparts). Conversely, the answer to the question of whether 
it is possible (and, if so, whether it is sensible and/or convenient)
to deal with a perturbative, non-$\beta$-resonant, $\delta X$ cross
section depends on the context in which one is working.
One can introduce two conceptually different classes of applications:
\begin{enumerate}
\item Definition of the non-$\beta$-resonant $\delta X$ cross section
as a measurable quantity, for a direct comparison with experimental results.
\item Use of non-$\beta$-resonant $\delta X$ production in conjunction
with $\beta X$ production, as perturbative tools that render technically
easier the computation of the $\delta X$ cross section that includes
both resonant and non-resonant contributions.
\end{enumerate}
As a rule of thumb, applications of class 1~are mostly of interest
to SM physics, while those of class 2~are relevant to both the SM
and to SUSY (and, in general, to theories which are not confirmed
experimentally, and whose signals need to be searched for). By using
again the SM example of eq.~\eqref{SMex1}, its class 1~applications
entail the definition of the $tW^-$ cross section (see 
\eg\ refs.~\cite{Aaboud:2016lpj,Chatrchyan:2014tua} for recent
ATLAS and CMS measurements of this quantity at the LHC), for which 
the underlying $t\bt$ resonant process is considered as a background.
On the other hand, in a typical class 2~application one would exploit 
the possibility of computing {\em both} the $tW^-$ and $t\bt$ cross sections
at the NLO (\ie\ up to ${\cal O}(\aem\as^2)$ and ${\cal O}(\as^3)$,
respectively) for a phenomenologically accurate description of
$W^+W^-b(\bb)$ production which is much less demanding, from a computational
viewpoint, than the calculation of the $W^+W^-b\bb$ cross section at the
NLO.

The key point of the previous example is that the $tW^-$ cross sections
that enter the two applications are not necessarily defined in the 
same way. In general, let
\begin{equation}
\amp_{ab\to\delta\gamma X}=
\amp_{ab\to\delta\gamma X}^{(\nbeta)}+
\amp_{ab\to\delta\gamma X}^{(\beta)}
\label{ampdelX}
\end{equation}
be the amplitude associated with the process of eq.~\eqref{delNLO}.
The two quantities on the r.h.s.~of eq.~\eqref{ampdelX} denote the
non-$\beta$-resonant and the $\beta$-resonant contributions, respectively.
The matrix element will thus be proportional to:
\begin{equation}
\abs{\amp_{ab\to\delta\gamma X}}^2=
\abs{\amp_{ab\to\delta\gamma X}^{(\nbeta)}}^2+
2\Re\left(\amp_{ab\to\delta\gamma X}^{(\nbeta)}
\amp_{ab\to\delta\gamma X}^{(\beta)^\dagger}\right)+
\abs{\amp_{ab\to\delta\gamma X}^{(\beta)}}^2\,.
\label{ampdelX2}
\end{equation}
For both class 1~and class 2~applications, the contribution of
the last term on the r.h.s.~of eq.~(\ref{ampdelX2}) must be minimised.
In the context of NLO+PS simulations, this problem has been solved in
ref.~\cite{Frixione:2008yi} by introducing two different types of
procedures. In diagram removal (DR), one simply drops this
contribution, whereas in diagram subtraction (DS) the non-$\beta$-resonant
$\delta X$ cross section will feature the linear combination:
\begin{equation}
\abs{\amp_{ab\to\delta\gamma X}^{(\beta)}}^2 d\phi-
f\!\left(m_{\delta\gamma}^2\right)
\proj\left(\abs{\amp_{ab\to\delta\gamma X}^{(\beta)}}^2 d\phi\right)\,.
\label{DS}
\end{equation}
The pre-factor in the second term of eq.~\eqref{DS} is arbitrary to
a large extent, but must obey the condition
\begin{equation}
\lim_{m_{\delta\gamma}\to m_\beta}f\!\left(m_{\delta\gamma}^2\right)=1\,.
\label{fcond}
\end{equation}
The symbol $\proj$ denotes a kinematic projection that maps a generic
$\delta\gamma X$ configuration onto one that has 
\mbox{$m_{\delta\gamma}=m_\beta$}. Crucially, this map is fully local
in the phase space, so that the difference in eq.~\eqref{DS} vanishes
identically when \mbox{$m_{\delta\gamma}\to m_\beta$} (also thanks
to eq.~\eqref{fcond}); such a locality condition is essential for
the use of DS in event generators. Owing to the definition
of the $\beta$-resonant amplitude, one has
\begin{equation}
\proj\left(\abs{\amp_{ab\to\delta\gamma X}^{(\beta)}}^2\right)\propto
\abs{\amp_{ab\to\beta X}}^2
\label{nospc}
\end{equation}
by neglecting production spin correlations. Thus, eq.~\eqref{nospc}
renders it intuitively clear that the difference in eq.~\eqref{DS}
is constructed so as to avoid the double counting of the LO $\beta X$
cross section in class 2~approaches. In practice, spin correlations cannot 
be neglected, and therefore $\amp_{ab\to\beta X}$ is never used as such in 
DS procedures\footnote{It is indeed $\amp_{ab\to\delta\gamma X}^{(\beta)}$ 
(suitably projected) that is employed, in the method of 
ref.~\cite{Frixione:2007zp}, to include production spin correlations 
at the tree level in the Monte Carlo simulations of $\beta X$ production.}; this
is just as well, since it helps guarantee the local cancellation between the 
two terms in eq.~\eqref{DS}.
We finally stress that there is ample freedom in the choices of the function 
$f$ and projector $\proj$ that enter the definition of a DS cross section
through eq.~\eqref{DS}. We shall exploit this fact in the following, 
by considering several different implementations. 

As far as the second term on the r.h.s.~of eq.~(\ref{ampdelX2}) is 
concerned, DR procedures do not include it in the definition of the 
non-$\beta$-resonant $\delta X$ cross section, while DS procedures do.
Thus, for class 1~applications it is essential that both DR and DS results
be obtained, and that their difference be less than the theoretical
systematics. If that is not the case, non-$\beta$-resonant $\delta X$ 
observables are simply not physically meaningful, and both DR and DS 
predictions must be discarded. Conversely, for class 2~applications, in 
which the emphasis is on obtaining the best approximation for the full 
$\delta X$ cross section, one is interested in using DS approaches.
DR results might also be kept, provided they are statistically compatible 
with the DS ones. We also point out that in the NLO+PS simulations of
ref.~\cite{Demartin:2016axk} a third scenario has been considered
(dubbed DR2 there, and originally introduced in ref.~\cite{Hollik:2012rc}
within a fixed-order calculation), in which one keeps the second term on 
the r.h.s.~of eq.~(\ref{ampdelX2}), but does {\em not} perform the
subtraction of eq.~\eqref{DS}. As far as its usage in applications of
class 1~and 2~is concerned, this approach is quite analogous to a DS one.
However, given that no subtraction is carried out, we call it DR+I 
(for diagram removal plus interference).

In summary, the non-$\beta$-resonant $\delta\gamma X$ cross section
can be defined as follows:
\begin{eqnarray}
d\sigma_{ab\to\delta\gamma X}^{(\DR)}&\propto&
\abs{\amp_{ab\to\delta\gamma X}^{(\nbeta)}}^2 d\phi\,,
\label{sigDR}
\\
d\sigma_{ab\to\delta\gamma X}^{(\DRpI)}&\propto&
\left\{\abs{\amp_{ab\to\delta\gamma X}^{(\nbeta)}}^2+
2\Re\left(\amp_{ab\to\delta\gamma X}^{(\nbeta)}
\amp_{ab\to\delta\gamma X}^{(\beta)^\dagger}\right)\right\}d\phi\,,
\label{sigDRpI}
\\
d\sigma_{ab\to\delta\gamma X}^{(\DS)}&\propto&
\left\{\abs{\amp_{ab\to\delta\gamma X}^{(\nbeta)}}^2+
2\Re\left(\amp_{ab\to\delta\gamma X}^{(\nbeta)}
\amp_{ab\to\delta\gamma X}^{(\beta)^\dagger}\right)+
\abs{\amp_{ab\to\delta\gamma X}^{(\beta)}}^2\right\}d\phi
\nonumber
\\&-&
f\!\left(m_{\delta\gamma}^2\right)
\proj\left(\abs{\amp_{ab\to\delta\gamma X}^{(\beta)}}^2 d\phi\right)\,,
\label{sigDS}
\end{eqnarray}
in the DR, DR+I, and DS approaches, respectively. Collectively, 
the procedures implied by eqs.~\eqref{sigDR}--\eqref{sigDS} will be called
STR (that stands for \textit{Simplified Treatments of Resonances}).

STR strategies have been pursued for a long while in the 
context of fixed-order calculations and for inclusive observables in both
the SM and BSM theories (in particular in SUSY, where they are typically
called OS subtractions) -- see \eg\ refs.~\cite{Beenakker:1996ch,
Belyaev:1998dn,Tait:1999cf,Zhu:2001hw,Berger:2003sm,Campbell:2005bb,
Dao:2010nu,Hollik:2012rc}. None of these earlier procedures is apt to be
applied to exclusive event generation, and thus we believe one should
refrain from using the DR or DS tags in association with them. As far
as DR and DS procedures are concerned, either identical to or featuring
variants of those originally proposed in ref.~\cite{Frixione:2008yi},
results can be found in refs.~\cite{White:2009yt,Weydert:2009vr,
Re:2010bp,Binoth:2011xi,GoncalvesNetto:2012yt,Gavin:2013kga,Gavin:2014yga,
Demartin:2016axk}.

We now turn to giving some details about the implementation of
eqs.~\eqref{sigDR}--\eqref{sigDS} in \mg. In keeping with the general 
strategy that underpins the code, everything is fully automated and 
process- as well as model-independent\footnote{Some limitations exist,
as STR procedures cannot for instance be used within
\madspin~\cite{Alwall:2014bza} or the reweighting module of
\mg~\cite{Mattelaer:2016gcx}.}. We remind the reader that \mg\ is 
self-consistent, and thus that, in particular, it generates internally 
the Feynman diagrams and writes the corresponding 
amplitudes. This implies that the code stores the information on the
topological structure of each diagram, and therefore knows where to
find the resonances (which is essential in order to construct
$\amp_{ab\to\delta\gamma X}^{(\nbeta)}$ and 
$\amp_{ab\to\delta\gamma X}^{(\beta)}$). Furthermore, it can control the
kinematical inputs and parameter settings in a diagram-by-diagram manner if
needed. The immediate consequence of the previous observation is that the
construction of the DR and DR+I cross sections of eqs.~\eqref{sigDR}
and~\eqref{sigDRpI}, respectively, is achieved in a straightforward
(and unique) manner.

The case of the DS cross section is more involved, owing to the freedom 
in the definitions of the function $f$ and of the projector $\proj$, although
after having chosen $f$ and $\proj$, eq.~\eqref{sigDS} uniquely 
determines the corresponding DS procedure. Unfortunately, it is impossible 
to parametrise the arbitrariness in the choices of $f$ and $\proj$, and thus 
one must limit oneself to considering a finite number of physically-motivated 
options. We describe those implemented in \mg\ below, and point
out that previous results in the DS approach~\cite{Frixione:2008yi,
White:2009yt,Weydert:2009vr,Re:2010bp,Binoth:2011xi,GoncalvesNetto:2012yt,
Gavin:2013kga,Gavin:2014yga} have been obtained with a given $(f,\proj)$
pair (with the exception of ref.~\cite{Demartin:2016axk}, where two different
forms of $f$ have been compared).

\subsection{Diagram-subtraction procedures}
\label{sec:DS}
We start by pointing out that the discussion that follows is relevant
to the last term on the r.h.s.~of eq.~\eqref{sigDS}, henceforth
called the {\em DS subtraction term}. The other three
terms in the definition of the DS cross section correspond to a 
straightforward tree-level calculation, and are thus not of concern here.
We denote the kinematic of the process of eq.~\eqref{delNLO} as follows:
\begin{equation}
k_a+k_b=\kd+\kg+\sum_{i=1}^{n-1}k_i\,,
\label{kinconf}
\end{equation}
where we have assumed that the set $X$ is composed of $n-1$ particles
with momenta $k_i$. It is convenient to introduce the following
auxiliary momenta:
\begin{eqnarray}
q&=&\kb+\krec=k_a+k_b\,,
\label{qdef}
\\
\kb&=&\kd+\kg\,,
\label{kbdef}
\\
\krec&=&\sum_{i=1}^{n-1}k_i\,.
\label{krdef}
\end{eqnarray}
Although the resonance $\beta$ does not appear in the final state,
the definition of its momentum in eq.~(\ref{kbdef}) is physically
meaningful, since we are solely dealing with $\beta$-resonant diagrams.
In the centre-of-mass frame of the incoming hadrons:
\begin{equation}
k_a=x_a\frac{\sqrt{S}}{2}\left(1,0,0,1\right)
\qquad\text{and}\qquad
k_b=x_b\frac{\sqrt{S}}{2}\left(1,0,0,-1\right)\,,
\label{Bjx}
\end{equation}
with $S$ being the squared hadronic centre-of-mass energy. Its parton-level
counterpart reads thus:
\begin{equation}
s\equiv q^2=\left(k_a+k_b\right)^2=x_a x_b S\,.
\label{shat}
\end{equation}
The action of the projection $\proj$ on the partonic kinematic 
configuration is denoted as the following transformation:
\begin{equation}
\kd\;\longrightarrow\;\bkd\,,\;\;\;\;\;\;\;\;
\kg\;\longrightarrow\;\bkg\,,\;\;\;\;\;\;\;\;
k_i\;\longrightarrow\;\bk_i\;\;\;\;(1\le i\le n-1)\,.
\label{barmom}
\end{equation}
We also introduce, for consistency with eq.~\eqref{barmom}, 
the momentum of the projected resonance
\begin{equation}
\kb\;\longrightarrow\;\bkb\,,\;\;\;\;\;\;\;\;
\bkb=\bkd+\bkg\,.
\end{equation}
The DS strategies that we consider generally require the partonic incoming
momenta to be changed. This can formally be seen as also stemming from the
action of $\proj$, and thus be represented as follows:
\begin{equation}
k_a\;\longrightarrow\;\bar{k}_a\,,\;\;\;\;\;\;\;\;
k_b\;\longrightarrow\;\bar{k}_b\,.
\label{inbarmom}
\end{equation}
By taking eqs.~\eqref{Bjx} and~\eqref{shat} into account, eq.~\eqref{inbarmom}
can be equivalently written as the transformation
\begin{equation}
x_a\;\longrightarrow\;\bx_a\,,\;\;\;\;\;\;\;\;
x_b\;\longrightarrow\;\bx_b\,,\;\;\;\;\;\;\;\;
s\;\longrightarrow\;\bar{s}\equiv \bx_a \bx_b S\,.
\label{inbarmom2}
\end{equation}
While the specific form of eq.~\eqref{inbarmom2} will depend on $\proj$,
in all of our implementations we shall always choose $\bx_a$ and $\bx_b$
so that
\begin{equation}
\frac{\bx_a}{\bx_b}=\frac{x_a}{x_b}\,.
\end{equation}
This implies that the original and projected partonic centre-of-mass frames
will travel at the same speed w.r.t.~the hadronic centre-of-mass frame.
Equation~\eqref{inbarmom2} has two further implications. Firstly,
the flux factor of the DS subtraction term is equal to 
\mbox{$1/(2\bar{s})$}. Secondly, its parton-luminosity factor is given by
\begin{equation}
f_a^{(H_1)}(\bx_a)\,f_b^{(H_2)}(\bx_b)\,,
\end{equation}
with $H_1$ and $H_2$ being the incoming hadrons.

As far as the function $f$ is concerned, we shall limit ourselves
to considering the following form:
\begin{equation}
f(m^2)=\frac{{\rm BW}_\beta(m^2,x)}{{\rm BW}_\beta(\mbt,x)}\,,
\label{fchoice}
\end{equation}
for a given choice of $x$, and where
\begin{equation}
{\rm BW}_\beta(m^2,x)=\frac{\kappa}{(m^2-\mbt)^2+(x\Gb)^2}
\end{equation}
is a generalised Breit-Wigner function (the standard one being obtained
by setting $x=\mb$) in which $\kappa$ is a normalisation factor that does not
play any role. The rationale beyond eq.~\eqref{fchoice}
is that its denominator will cancel, to some extent, an analogous term
implicit in the projected matrix element of the DS subtraction term, that is
thus replaced by the numerator of eq.~\eqref{fchoice}
which supposedly models some of the off-shell-$\beta$ effects. The precise 
extent of such a cancellation depends on the interplay of several factors
(such as the choice of $x$, of the operator $\proj$, or of the PDFs),
and cannot therefore be predicted {\em a priori}. This is one of the 
reasons why in eq.~\eqref{fchoice} $x$ is treated as a free parameter.

In view of their use in the definition of $\proj$, we also introduce
the following quantities. For any four-momentum $p$, we denote the 
boost to its rest frame by:
\begin{equation}
\boost_R(p)p=\left(m,\vec{0}\right)\,.
\end{equation}
This understands that $p^2=m^2>0$; we implicitly assume that
the boost is performed along $\vec{p}$, and that it is such that:
\begin{equation}
\exp(y_{\boost_R})=\sqrt{\frac{E+|\vec{p}|}{E-|\vec{p}|}}\,\,,
\;\;\;\;\;\;\;\;
p=(E,\vec{p})\equiv (\sqrt{m^2+|\vec{p}|^2},\vec{p})\,.
\label{bstR}
\end{equation}
If in the rest frame of $p$ we impose a $1\to 2$ four-momentum
conservation,
\begin{equation}
\left(m,\vec{0}\right)=p_1+p_2\,,
\label{tmp1}
\end{equation}
then by denoting $p_i^2=m_i^2$ (with \mbox{$m>m_1+m_2$}), we have
\begin{eqnarray}
p_1&=&\Big(\varepsilon(m,m_1,m_2),\,\pi(m,m_1,m_2)\,\vec{e}\Big)\,,
\label{tmp2}
\\
p_2&=&\Big(\varepsilon(m,m_2,m_1),\,-\pi(m,m_1,m_2)\,\vec{e}\Big)\,,
\label{tmp3}
\end{eqnarray}
where $\abs{\vec{e}}=1$ and:
\begin{eqnarray}
\pi(m,m_1,m_2)&=&\frac{m}{2}\lambda(m,m_1,m_2)\,,
\label{pidef}
\\
\varepsilon(m,m_1,m_2)&=&
\sqrt{m_1^2+\pi(m,m_1,m_2)^2}
\nonumber\\*&=&
\frac{m}{2}\left(1+\frac{m_1^2-m_2^2}{m^2}\right)\,,
\label{epdef}
\end{eqnarray}
with:
\begin{equation}
\lambda(a,b,c)=\sqrt{1-\frac{(b+c)^2}{a^2}}\,\sqrt{1-\frac{(b-c)^2}{a^2}}\,.
\label{lamdef}
\end{equation}
We can now present specific details of the definition of the DS subtraction
term in the DS procedures we pursue. The reader must keep
in mind that the kinematic configuration of eq.~\eqref{kinconf} and
its associated quantities eqs.~(\ref{qdef})--(\ref{krdef}) are thought
to be given. Without loss of generality, we work in the incoming-parton
centre-of-mass frame, $\vec{q}=\vec{0}$, and the options described below are
associated in \mg\ with the function $f$ given in eq.~\eqref{fchoice} and with
either of the settings:
\begin{equation}
x=\mb\qquad\text{or}\qquad
x=m_{\delta\gamma}\,.
\label{xchoice}
\end{equation}
Other choices of $x$ would be straightforward to implement.\\

\noindent
{\bf Option A}
\vskip 0.25truecm
\noindent
This option follows the strategy first introduced in 
ref.~\cite{Frixione:2008yi}. We define:
\begin{eqnarray}
\vec{\bar{k}}_\beta&=&\vkb\,,
\label{Ab0}
\\
\bkb^0&=&\sqrt{\mbt+\vec{\bar{k}}_\beta^2}\,.
\label{Ab}
\end{eqnarray}
The momenta not associated with the resonance $\beta$ are left invariant:
\begin{equation}
\bk_i=k_i\;\;\;\;\;\;\;\;1\le i\le n-1\,.
\label{Ai}
\end{equation}
In view of eqs.~\eqref{Ab} and \eqref{Ai}, we then define:
\begin{equation}
\sqrt{\bar{s}}=\bar{q}^0=\bkb^0+\sum_{i=1}^{n-1}\bk_i^0\,.
\label{bshatA}
\end{equation}
As far as the momenta of $\delta$ and $\gamma$ are concerned,
we proceed as follows. First, we boost them in the rest frame of $\kb$:
\begin{eqnarray}
\boost_R(\kb) \kd &=& 
\Big(\varepsilon\left(\sqrt{\kbt},\md,\mga\right),\,
\pi\left(\sqrt{\kbt},\md,\mga\right)\,
\vec{e}_\delta\Big)\,,
\label{d1vec}
\\
\boost_R(\kb) \kg &=& 
\Big(\varepsilon\left(\sqrt{\kbt},\mga,\md\right),\,
-\pi\left(\sqrt{\kbt},\md,\mga\right)\,
\vec{e}_\delta\Big)\,,
\label{d2vec}
\end{eqnarray}
owing to eqs.~(\ref{tmp1})--(\ref{tmp2}). Then, by keeping the information
on $\vec{e}_\delta$ but discarding all the rest, we define
\begin{eqnarray}
\bkd&=&\boost_R^{-1}(\bkb)\,
\Big(\varepsilon\left(\mb,\md,\mga\right),\,
\pi\left(\mb,\md,\mga\right)\,
\vec{e}_\delta\Big)\,,
\label{bd1vec}
\\
\bkg&=&\boost_R^{-1}(\bkb)\,
\Big(\varepsilon\left(\mb,\md,\mga\right),\,
-\pi\left(\mb,\md,\mga\right)\,
\vec{e}_\delta\Big)\,,
\label{bd2vec}
\end{eqnarray}
which guarantee consistency with eqs.~(\ref{Ab0}) and~(\ref{Ab}), and thus 
ultimately enforce four-momentum conservation. In the case where $\kbt=\mbt$,
all of the operations above are equivalent to the identity.\\

\noindent
{\bf Option B}
\vskip 0.25truecm
\noindent
This option generalises (to an arbitrary number of final-state particles)
the strategy of ref.~\cite{Binoth:2011xi}. We define the mass of the recoil 
system $X$:
\begin{equation}
\mrect=\krec^2\,,
\end{equation}
and we keep it invariant. If the condition
\begin{equation}
\sqrt{s}\ge \mb+\mrec\,,
\label{bs0}
\end{equation}
is fulfilled, we then set
\begin{equation}
\bar{s}=s\,.
\label{bs1}
\end{equation}
Otherwise, we define\footnote{Alternatively, one can leave invariant the
partonic centre-of-mass~energy, and assign to $\kbt$ the largest value compatible
with that energy. This has the disadvantage of defining a DS subtraction
term which does not correspond to an on-shell $\beta$-resonant
cross section.}:
\begin{equation}
\sqrt{\bar{s}}=\frac{\mb+\mrec}{\sqrt{\kbt}+\mrec}\,\sqrt{s}\,.
\label{bs2}
\end{equation}
Next, we define the energies of the projected resonance and recoil system 
as follows:
\begin{eqnarray}
\bkb^0&=&\varepsilon\left(\sqrt{\bar{s}},\mb,\mrec\right)\,,
\label{tmp4}
\\
\bkrec^0&=&\varepsilon\left(\sqrt{\bar{s}},\mrec,\mb\right)\,.
\label{tmp5}
\end{eqnarray}
The corresponding three-momenta are defined by preserving the direction
of the original three-momenta, rescaling their lengths so as to impose
the mass shell conditions
\begin{eqnarray}
\vec{\bar{k}}_\beta&=&
\sqrt{(\bkb^0)^2-\mbt}\,\frac{\vkb}{|\vkb|}=
\pi\left(\sqrt{\bar{s}},\mb,\mrec\right)\,\frac{\vkb}{|\vkb|}\,,
\\
\vbkrec&=&
\sqrt{(\bkrec^0)^2-\mrect}\,\frac{\vkrec}{|\vkrec|}=
\pi\left(\sqrt{\bar{s}},\mb,\mrec\right)\,\frac{\vkrec}{|\vkrec|}\,.
\end{eqnarray}
These guarantee that \mbox{$\vec{\bar{k}}_\beta=-\vbkrec$},
since \mbox{$\vkb=-\vkrec$}. We also define
\begin{equation}
\bk_i=\boost_R^{-1}(\bkrec)\,\boost_R(\krec)\,k_i
\qquad\text{for}\qquad 1\le i\le n-1\,.
\end{equation}
Finally, $\kd$ and $\kg$ are projected using the same procedure as 
in eqs.~(\ref{d1vec})--(\ref{bd2vec}).\\

\noindent
{\bf Option C}
\vskip 0.25truecm
\noindent
This option, which is currently not implemented in \mg,
follows closely what is done for the phase-space
generation as is carried out in the module \madfks~\cite{Frederix:2009yq},
in the case relevant to a massive FKS sister. This, in turn, generalises
the massless-parton treatment of ref.~\cite{Frixione:2007vw}. The
projected partonic centre-of-mass energy is defined as follows:
\begin{equation}
\sqrt{\bar{s}}=\frac{\mb+\sum_{i=1}^{n-1}m_i}
{\min\left(\sqrt{\kbt},\,\varsigma\mb\right)+\sum_{i=1}^{n-1}m_i}\,
\sqrt{s}\,,
\label{bs3}
\end{equation}
with $\varsigma\ge 1$ being a free parameter\footnote{It is also possible
to use here the strategy outlined in eqs.~\eqref{bs0}--\eqref{bs2}, with
the formal replacement $\mrec\to\sum_{i=1}^{n-1}m_i$ there. Likewise,
eq.~\eqref{bs3}, with $\sum_{i=1}^{n-1}m_i\to\mrec$, can be used in 
option B instead of eqs.~\eqref{bs0}--\eqref{bs2}.}, typically
of ${\cal O}(1)$. One then regenerates the final-state kinematic 
configuration, using the same random numbers as for the original one,
and $\bar{s}$ instead of $s$ (the configuration thus obtained has only
a temporary role, and is denoted below in the same way as the original one).
Next, a boost $\boost$ is defined along the resonance three-momentum $\vkb$,
so that:
\begin{equation}
(\bar{q}-\boost\krec)^2=\mbt\,.
\label{Bdef}
\end{equation}
The projected momenta are then given by:
\begin{equation}
\bkb=\bar{q}-\boost\krec\qquad\text{and}\qquad
\bk_i=\boost k_i\,.
\label{basicB}
\end{equation}
Once again, the $\kd$ and $\kg$ momenta are projected using
the same procedure as in eqs.~\eqref{d1vec}--\eqref{bd2vec}.

\subsection{Using DR and DS in \mg}
\label{sec:OSinMG}
All of the STR procedures described above can be employed within \mg\ 
by downloading the \mados\
plugin\footnote{\url{https://code.launchpad.net/~maddevelopers/mg5amcnlo/MadSTRPlugin}},
and by copying the directory {\tt MadSTR} thus obtained inside the 
{\tt PLUGIN} directory, which is part of any (recent) \mg\ release. 
The current version of \mados\ is compatible with \mg\ version 2.6
 and higher; compatibility with versions 3 and higher,
that are capable of carrying out mixed-coupling perturbative 
computations~\cite{Frederix:2018nkq}, will be added in the future. 
The plugin can be activated by using the following command to start \mg:
\begin{verbatim}
  mg5_aMC --mode=MadSTR
\end{verbatim}
One can then generate NLO processes and write them to disk as usual 
(with the {\tt generate} and {\tt output} commands -- see
ref.~\cite{Alwall:2014hca} for more details). The plugin will take care
of identifying any potentially resonant contributions, of generating the
associated underlying resonant processes,
and of taking care of the extra bookkeeping in addition to that
normally performed by \mg\ in non-resonant cases. Depending on the 
mass spectrum, the contributions for which the STR is needed are 
identified at run-time. Three parameters found in {\tt run\_card.dat}, 
namely {\tt istr}, {\tt str\_include\_flux}, and {\tt str\_include\_pdf},
can be used to choose the desired STR procedure and its associated
options. More specifically, the type of STR is controlled by
{\tt istr}, which must assigned an integer value according to the
options given  in table~\ref{tab:iossubtr}. The other two parameters,
{\tt str\_include\_flux} and {\tt str\_include\_pdf}, are active only
if the STR is of DS type, and control the settings of the flux and the parton
luminosity factors, respectively, in the DS subtraction terms. In particular, 
DS procedures imply changes to the partonic centre-of-mass energy 
(see eqs.~(\ref{bshatA}) and~(\ref{bs2})). In turn, this seemingly implies 
that the flux and the luminosity factors should be changed accordingly,
when the partonic centre-of-mass energy is changed. However, this is not 
mandatory, since  it is actually part of the definition of the projection 
inherent to DS procedures and, as such, is liable to be chosen by the user.
It is in order to give one the possibility of exploring the consequence 
of this choice that the parameters {\tt str\_include\_flux} and 
{\tt str\_include\_pdf} have been introduced. By setting them equal
to {\tt True} ({\tt False}), the flux and PDF factors are (not)
re-defined. Note that the settings of the two parameters are independent
from each other, and that {\tt True} are the default values. More details
in the context of a specific example will be given in section~\ref{sec:pheno}.

\begin{table}
    \centering
    \begin{tabular}{cl}
        {\tt istr} & STR procedure \\
        \hline
        \hline
        1 & DR \\
        2 & DR+I \\
        3 & DS, option A, with $x=\mb$\\
        4 & DS, option A, with $x=m_{\delta\gamma}$\\
        5 & DS, option B, with $x=\mb$\\
        6 & DS, option B, with $x=m_{\delta\gamma}$\\
    \end{tabular}
    \caption{\label{tab:iossubtr} Possible values for the {\tt istr} 
parameter, with the corresponding STR procedure. In the case of option A
(and of option B when the condition in eq.~(\ref{bs0}) is not fulfilled),
the parameters {\tt str\_include\_flux} and {\tt str\_include\_pdf} are also
relevant (see text for details).}
\end{table}

Finally, the value of $\Gb$ (which acts as a regulator when $m_{\delta\gamma}
\to \mb$) can be controlled by changing the width of the corresponding 
particle $\beta$ in the {\tt param\_card.dat} file, for all $\beta$'s
that are potentially resonant. The code will set the widths of all 
coloured particles\footnote{This is because at the moment we restrict ourselves to the case of QCD corrections.} equal to zero everywhere except in the 
resonant real-emissions diagrams and in the corresponding DS subtraction terms,
in which the values provided in the {\tt param\_card.dat} will be employed.


\section{A case study: jets plus missing energy at the NLO+PS accuracy}
\label{sec:pheno}
We are now in the position to perform phenomenology studies in the
MSSM with a generic particle mass spectrum. As an illustrative example,
we choose the benchmark point presented in table~\ref{tab:phenobenchmark},
which is not excluded by current experimental searches at the LHC, and that
features non-trivial decay patterns. In contrast with section~\ref{sec:model}, 
the bottom squarks are taken to be non-mixing. 

In the scenario of table~\ref{tab:phenobenchmark}, the total widths and 
the relevant decay modes of the gluino and the squarks are as reported in 
table~\ref{decaytable}, together with the associated branching ratios,
these results being computed at the LO with \madwidth~\cite{Alwall:2014bza}. 
Thus, the decay widths of the squarks and gluino are sufficiently small
relatively to their masses (except for $\tilde{t}_1$, which is however never
resonant in the processes considered in this section) so that the
narrow-width approximation is sensible. Conversely, and according to
the parametrisation of long-distance effects \eg~in \pythia\ and \herwig\ 
(see refs.~\cite{Corcella:2000bw,Fairbairn:2006gg,Desai:2011su}), the
sparticle widths are sufficiently large to avoid hadronisation before
decay -- in other words, no $R$-hadrons will be present in our simulations.

\begin{table}[t]
  \begin{center}
    \begin{tabular}{cl|cl}
    Parameter & value & Parameter & value
    \\\midrule\midrule
    $m_t$                  & 172~GeV  & $n_{lf}$ & 5\\
    $m_{\tilde{g}}$        & 2000~GeV & $m_{\tilde{u}_R}$ & \texttt{$1200$} \\
    $m_{\tilde{t}_1}$      & 3000~GeV & $m_{\tilde{t}_2}$ & \texttt{$3500$}\\
    $m_{\tilde{\chi}_1^0}$ & 50~GeV   &
      $m_{\tilde{q},\tilde{q}\neq \tilde{t}_1,\tilde{t}_2,\tilde{u}_R}$ &
      2500~GeV\\
    $m_{\tilde{\chi}_i^0,i>1}$ & 5500~GeV &
      $m_{\tilde{\chi}^{\pm},\tilde{\ell}^{\mp}}$ & 5500~GeV
    \end{tabular}
    \caption{The benchmark scenario used for our phenomenological study.}
  \label{tab:phenobenchmark}
  \end{center}
\end{table}

\begin{table}
  \begin{center}
    \begin{tabular}{c|c|c|c}
     Particle & Width [GeV] & Decay mode & Branching ratio [\%]\\\hline\hline
     \multirow{2}{*}{$\tilde{g}$} &  \multirow{2}{*}{$16.6$} & $\tilde{u}_R\bar{u}$ & $50$ \\
     & & $\anti{\tilde{u}_R}u$ & $50$\\\hline
     $\tilde{u}_R$ & $2.71$ & $\tilde{\chi}_1^0u$ & $100$\\\hline
     \multirow{3}{*}{$\tilde{t}_1$} & \multirow{3}{*}{$534$} & $W^+\tilde{b}_L$ & $91.1$\\
     & & $\tilde{g}t$ & $8.8$\\
     & & $\tilde{\chi}_1^0t$ & $0.1$\\\hline
     \multirow{3}{*}{$\tilde{t}_2$} & \multirow{3}{*}{$90.3$}  & $\tilde{g}t$ & $88.7$ \\
     & & $\tilde{\chi}_1^0t$ & $8.8$\\
     & & $
     \tilde{b}_Lbt$ & $2.5$\\\hline 
     \multirow{2}{*}{$\tilde{q}_R$ with $\tilde q\neq \tilde u$} & \multirow{2}{*}{$18.5$} & $\tilde{g}q$ & $92.3$\\
     & & $\tilde{\chi}_1^0q$ & $7.7$\\\hline
     \multirow{2}{*}{$\tilde{b}_L$} & \multirow{2}{*}{$17.4$} & $\tilde{g}b$ & $98.0$\\
     & & $\tilde{\chi}_1^0b$ & $2.0$\\\hline
     \multirow{3}{*}{$\tilde{q}_L$ with $\tilde q\neq \tilde b$}
     & \multirow{3}{*}{$17.6$} & $\tilde{g}q$ & $97.0$\\
     & & $\tilde{\chi}_1^0q$ & $2.0$\\
     & & $W\tilde{g} q'$ & $1.0$
    \end{tabular}
    \caption{Decay widths and branching ratios of gluino and squarks,
in the benchmark point of table~\ref{tab:phenobenchmark}. $q'$ denotes the
down-type (up-type) quark associated with the same-generation up-type
(down-type) quark $q$.
\label{decaytable}}
  \end{center}
\end{table}
\begin{table}
  \centering
  \begin{tabular}{c||c c c}
 Process & Born signature(s) & Underlying resonant(s) & Decay(s)\\
\toprule
 $\tilde{g}\tilde{g}$ & $4j+\slashed{E}_T$ & 
 $\tilde{g}\tilde{q}_h$ & $\tilde{q}_h\to\tilde{g} q$ \\
\hline
 \multirow{2}{*}{$\tilde{g}\panti{\tilde{u}_R}$} & 
 \multirow{2}{*}{$3j+\slashed{E}_T$} & 
 $\tilde{g}\tilde{g}$ & $\tilde{g}\to \panti{\tilde{u}_R}u$ \\
  &  & 
 $\panti{\tilde{u}_R}\tilde{q}_h$ & $\tilde{q}_h\to \tilde{g}q$ \\

 $\tilde{g}\tilde{q}_h$ & $5j+\slashed{E}_T$ & 
 $\tilde{q}_h\tilde{q}_h$ & $\tilde{q}_h\to\tilde{g}q$ \\
\hline
  $\panti{\tilde{u}_R}\panti{\tilde{u}_R}$ & $2j+\slashed{E}_T$ & 
  $\tilde{g}\panti{\tilde{u}_R}$ & $\tilde{g}\to \panti{\tilde{u}_R}u$\\
  $\panti{\tilde{u}_R}\tilde{q}_h$ & $4j+\slashed{E}_T$ & 
  $\tilde{g}\tilde{q}_h$ & $\tilde{g}\to \panti{\tilde{u}_R}u$ \\
  $\tilde{q}_h\tilde{q}_h$ & $6j+\slashed{E}_T$ & 
  --- &  --- \\
  \end{tabular}
  \caption{Contributions to the signatures of eq.~(\ref{njEtm}) stemming
from the processes of eqs.~(\ref{ggprc})--(\ref{qqprc}). For each of these,
we report the Born-level signature (second column), the underlying resonant
process (third column), and the decay of the would-be resonant sparticles
(fourth column).
  \label{tab:bornsignal}}
\end{table}

\begin{figure}
  \centering
  \subfigure[$\tilde{g}\tilde{g}$]{\fbox{
    \includegraphics[width=0.31\textwidth,draft=false]{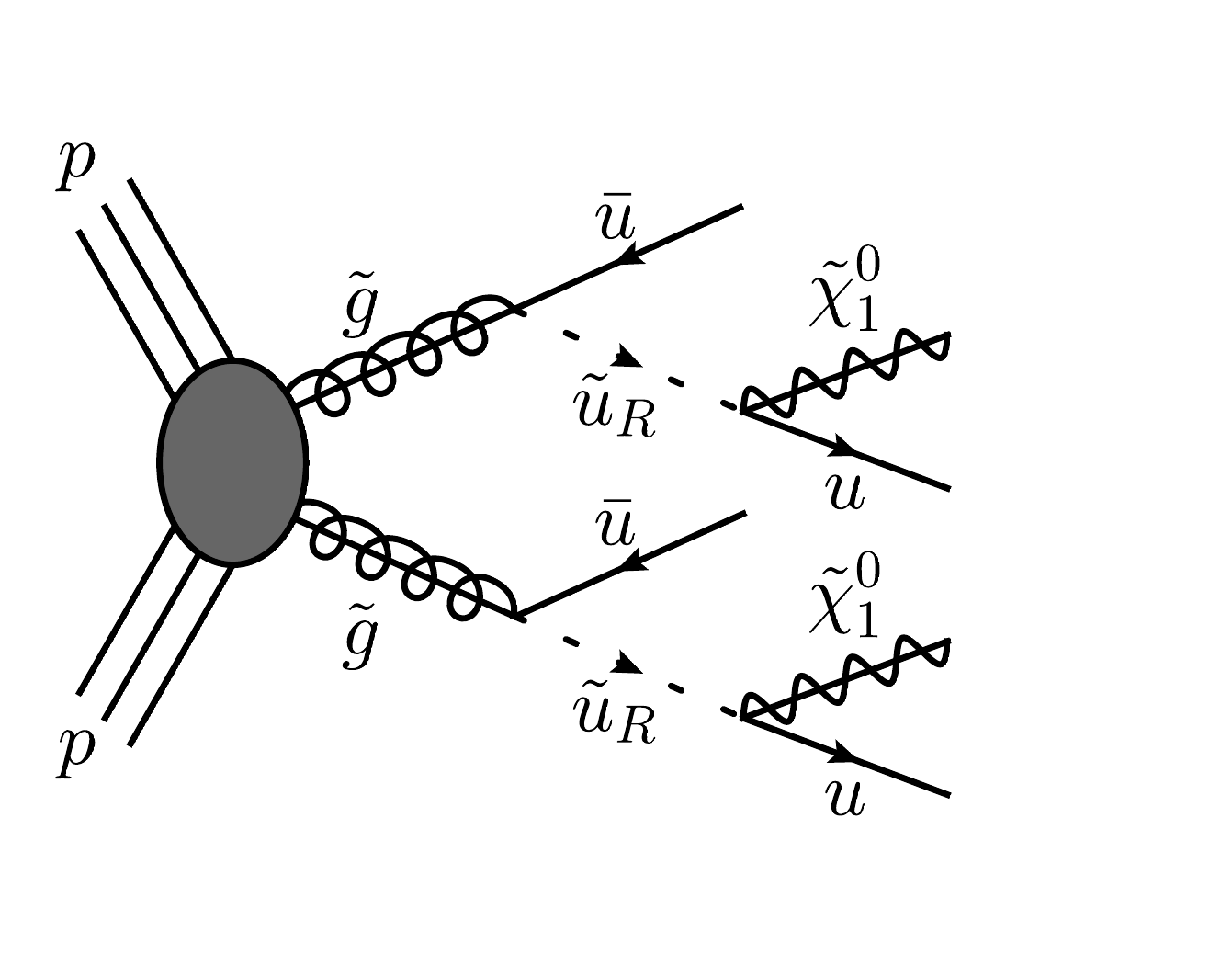}}}
  \subfigure[$\tilde{g}\tilde{q}$]{\fbox{
    \includegraphics[width=0.31\textwidth,draft=false]{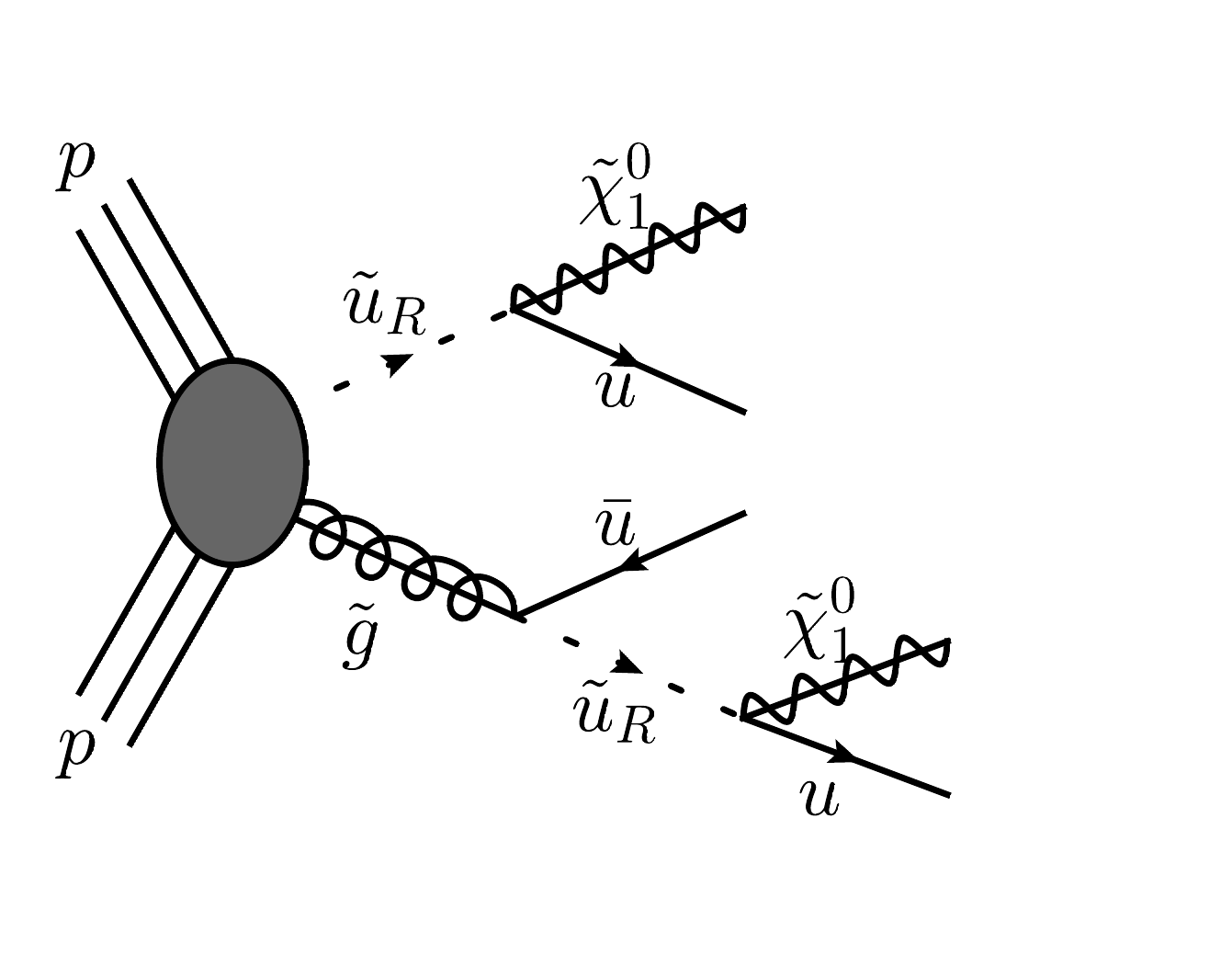}
    \includegraphics[width=0.31\textwidth,draft=false]{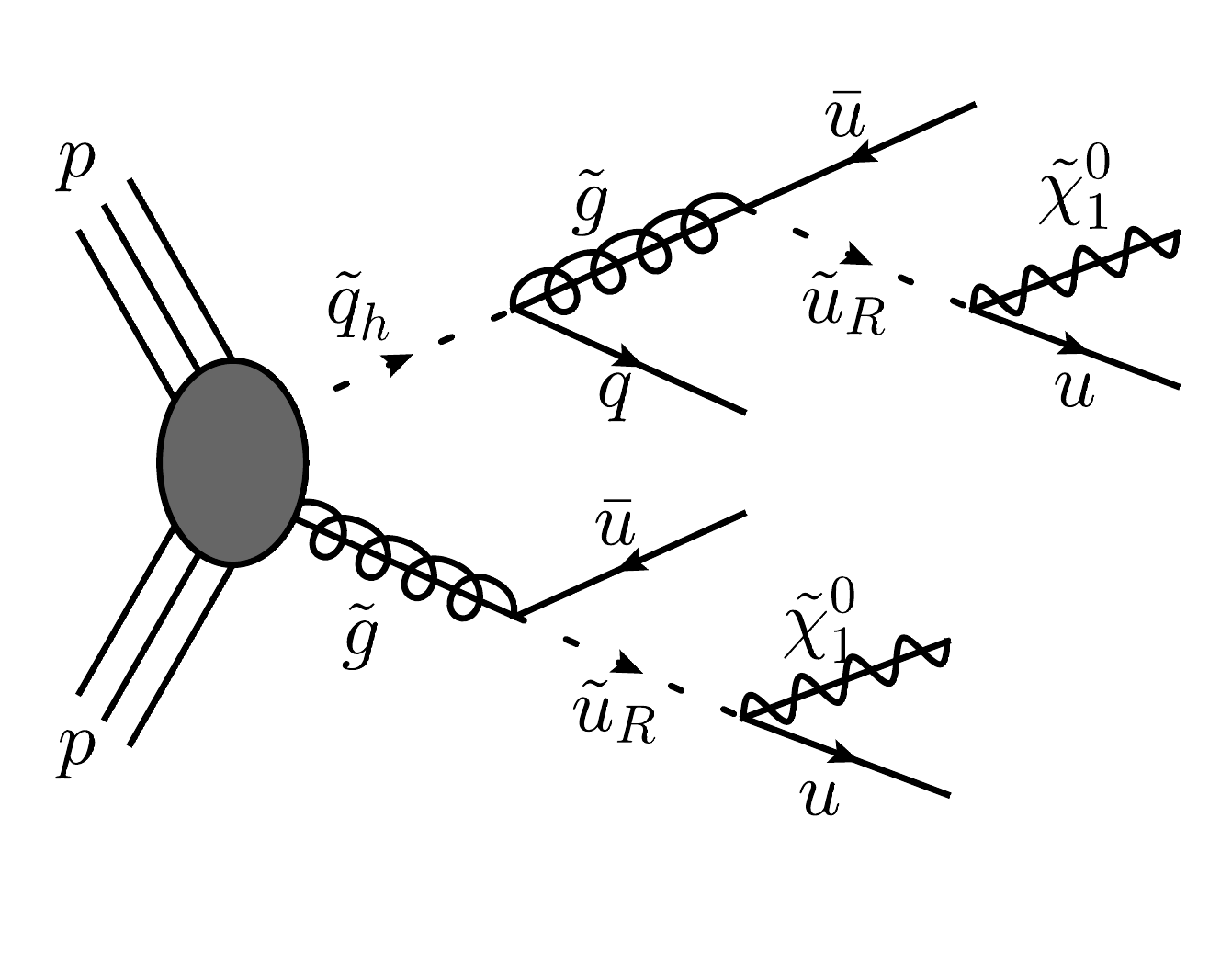}}}\\
  \subfigure[$\tilde{q}\tilde{q}$]{\fbox{
    \includegraphics[width=0.31\textwidth,draft=false]{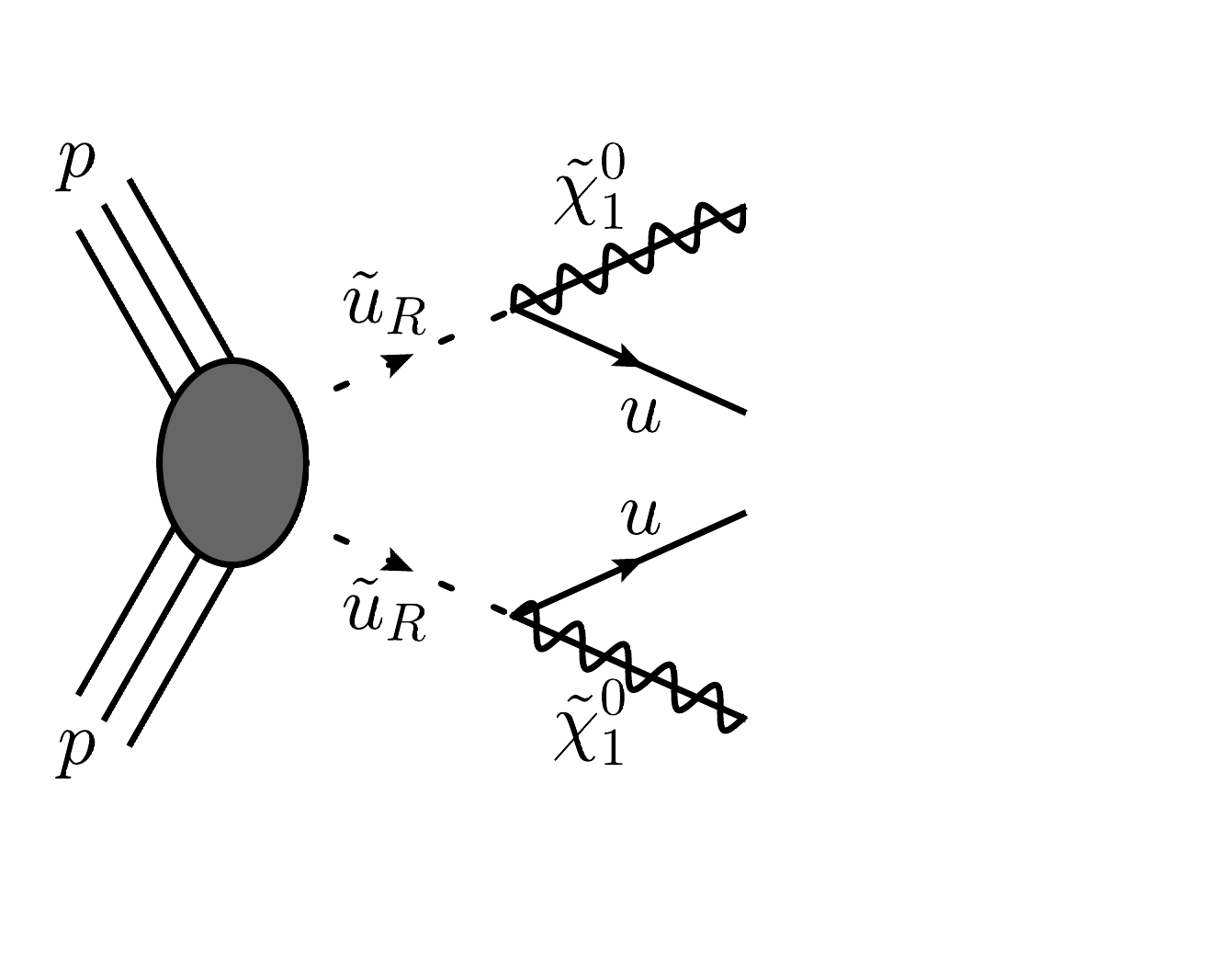}
    \includegraphics[width=0.31\textwidth,draft=false]{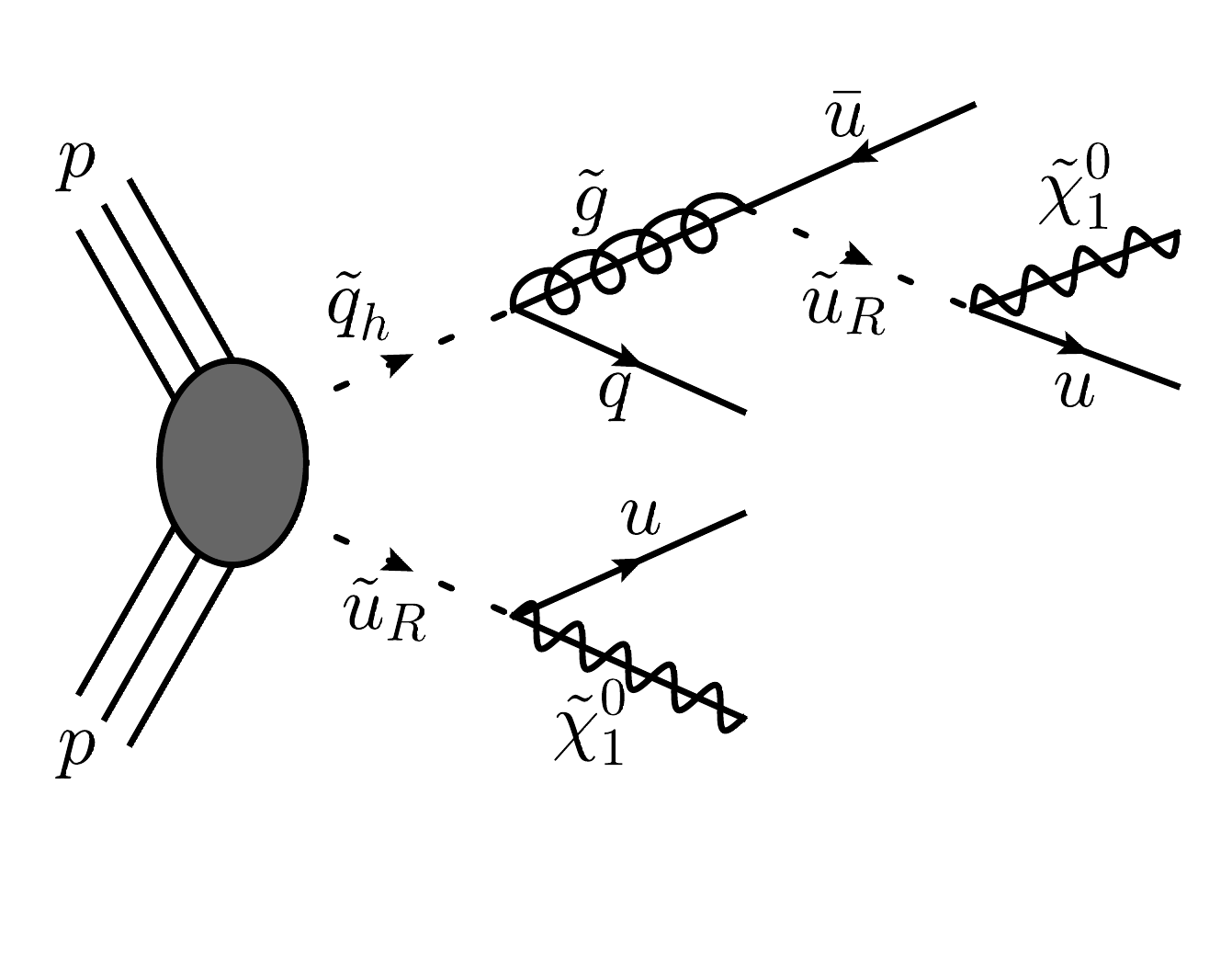}
    \includegraphics[width=0.31\textwidth,draft=false]{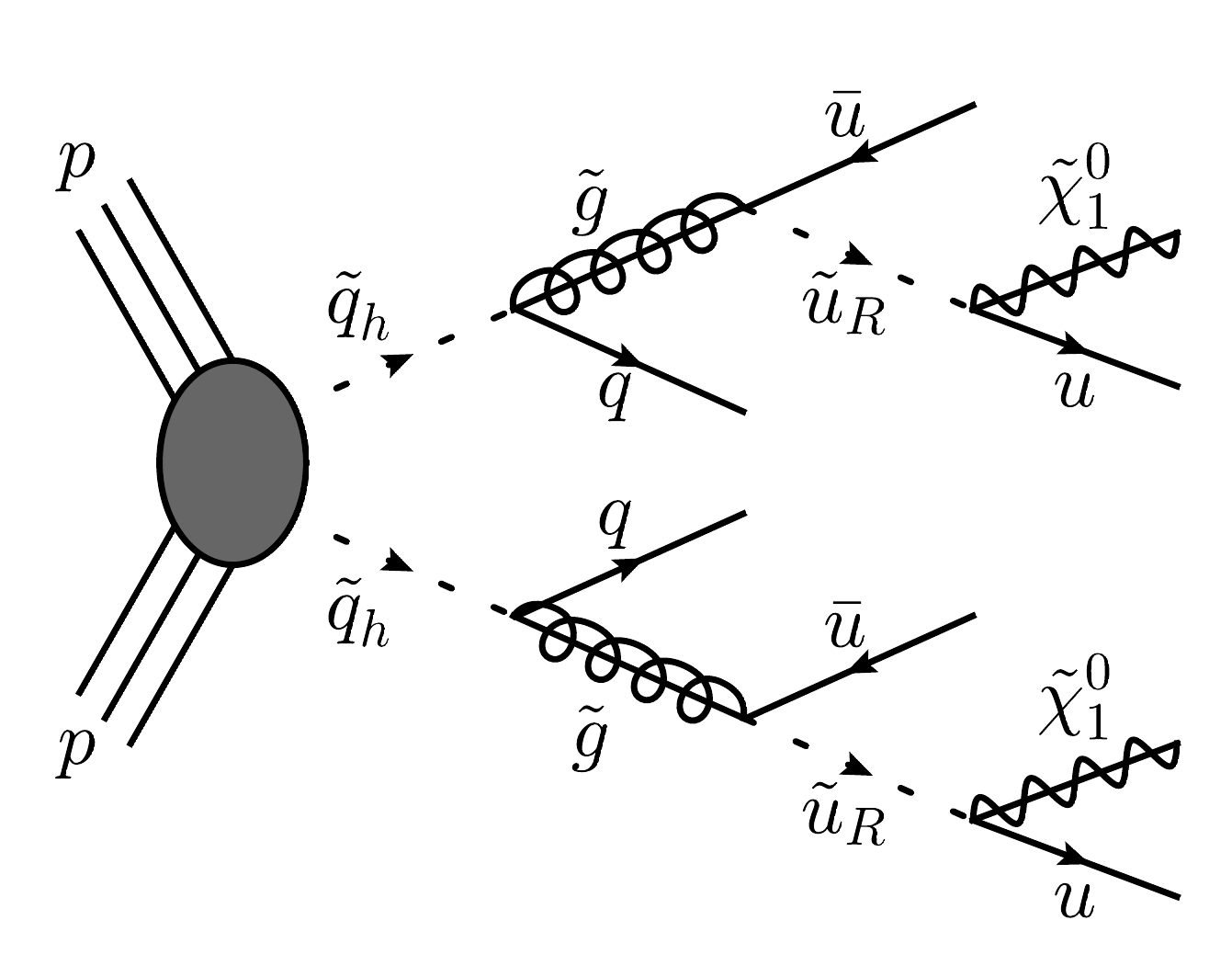}}}
  \caption{Representative Feynman diagrams for (a) $\tilde{g}\tilde{g}$,
    (b) $\tilde{g}\tilde{q}$ and (c) $\tilde{q}\tilde{q}$ production.
    All decays presented in this figure are implemented at the level of the
    parton shower programme.
  \label{fig:feynmandiag}}
\end{figure}

We shall now consider, at the 13~TeV LHC, multijet plus missing 
transverse-energy final states:
\begin{equation}
pp\;\longrightarrow\;nj+\slashed{E}_T\,,
\label{njEtm}
\end{equation}
a signature that is typical of SUSY searches at hadron colliders. 
We shall compute the contributions to eq.~(\ref{njEtm}) due to the 
following underlying processes:
\begin{eqnarray}
&&pp\;\longrightarrow\; \tilde{g}\tilde{g}+X\qquad\qquad\qquad
    \text{denoted by }\tilde{g}\tilde{g}\ ,
\label{ggprc}
\\
&&pp\;\longrightarrow\; \tilde{g}\panti{\tilde{q}}+X\qquad\qquad\quad
    \text{denoted by }\tilde{g}\tilde{q}\ ,
\label{qgprc}
\\
&&pp\;\longrightarrow\; \panti{\tilde{q}_i}\panti{\tilde{q}_j}+X\qquad\qquad 
    \text{denoted by }\tilde{q}\tilde{q}\ .
\label{qqprc}
\end{eqnarray}
The subsequent hadronic decays of the final-state sparticles are carried
out by the parton shower programme according to the results of
table~\ref{decaytable}; sample diagrams,
corresponding to the signatures of eq.~(\ref{njEtm}) being induced by 
the processes of eqs.~(\ref{ggprc})--(\ref{qqprc}) plus sparticle decays,
are depicted in figure~\ref{fig:feynmandiag}. In order to ensure
that squarks only decay into light-flavoured jets, we restrict our
simulations by solely considering the first two generations of squarks. 
We also distinguish, in the rest of this section, the light (anti-)squark
$\panti{\tilde{u}_R}$ from all of the other heavier (anti-)squarks of the 
first two generations, that we denote by $\tilde{q}_h$. 

The situation is summarised in table~\ref{tab:bornsignal}. For each
of the processes of eqs.~(\ref{ggprc})--(\ref{qqprc}), which we also
distinguish according to whether a light squark is present in the
primary (\ie\ before decay) process, we report the Born-level signature
(second column), the underlying resonant process, defined according to 
eq.~(\ref{betaLO}) (third column), and the relevant decays of the primary
sparticles, according to eq.~(\ref{betadec}) (fourth column). The experimental
signature of eq.~\eqref{njEtm} is obtained by considering the parton-level
process definitions of eqs.~\eqref{ggprc}--\eqref{qqprc} whose Born
contributions do not feature any resonance\footnote{This implies
that the SUSY particle decays shown in figure~\ref{fig:feynmandiag} must be
handled
at the level of the parton shower programme.}. Their corresponding real-emission
contributions therefore include at most one SUSY resonance which is then
necessarily subject to an STR procedure. In this way, our simulation setup is
guaranteed to remain within the scope of STR applicability as described in
section~\ref{sec:STRgeneral}.

All of our simulations are NLO+PS accurate, whereby NLO matrix elements
are matched with the \pythia~\cite{Sjostrand:2014zea} parton showers 
according to the MC@NLO formalism~\cite{Frixione:2002ik}, automated
in \mg. The resulting hadron-level events are clustered by making use of 
the anti-$k_T$ algorithm~\cite{Cacciari:2008gp} with jet radius $R=0.4$, as
implemented in \fastjet~\cite{Cacciari:2011ma}.

For our phenomenological analysis, we implement an event selection similar to
the one of the CMS SUSY search of ref.~\cite{Sirunyan:2017cwe}. Firstly, 
jets are required to have transverse momentum larger than 30 GeV, and 
pseudorapidity \mbox{$|\eta|<2.4$}. Events are kept if they
feature at least $N_{\rm jet} \ge 2$ jets. Secondly, the total hadronic 
activity $H_T$, defined as the scalar sum of the transverse momenta of all 
reconstructed jets, must be larger than 300~GeV. Thirdly, the missing 
transverse hadronic energy $H_T^{\rm miss}=|\overrightarrow{H}_T^{\rm miss}|$,
with $\overrightarrow{H}_T^{\rm miss}$ 
the negative of the vector sum of the transverse momenta of all 
reconstructed jets with a pseudorapidity $|\eta|<5$, must be larger than 
300~GeV. Finally, the two leading jets and $\overrightarrow{H}_T^{\rm miss}$
are imposed to be well separated in azimuth,
$\Delta\phi(H_T^{\rm miss},j_{1,2})>0.5$. When a third and a fourth jet 
are within the acceptance defined before, we additionally impose
$\Delta \phi(H_T^{\rm miss},j_{3,4}) > 0.4$.

\renewcommand{\arraystretch}{1.4}
\begin{table}
  \begin{tabular}{c||c|c|c|c|c|c|c|c}
  & \multirow{2}{*}{[fb]} & \multicolumn{6}{c|}{{\tt istr}} & \multirow{2}{*}{LO} \\\rule{0pt}{3ex}
  & & 1 & 2 & 3 & 4 & 5 & 6 & \\[1mm]
  \hline\hline \rule{0pt}{3ex}
  \multirow{2}{*}{$\tilde{g}\tilde{g}$} & $\sigma_{\rm inclusive}$ & $0.331$ & $0.330^{+19\%}_{-18\%}\pm28
  \%$ & $0.327$ & $0.322$ & $0.330$ & $0.330$ & $0.187^{+44\%}_{-29\%}\pm 27\%$ \\\cline{2-9}\rule{0pt}{3ex}
  & $\sigma_{\rm fiducial}$ & $0.228$ & $0.227^{+19\%}_{-18\%}\pm 28\%$ & $0.225$ & $0.222$ & $0.228$ & $0.227$ & $0.128^{+44\%}_{-29\%}\pm 27\%$ \\\hline\rule{0pt}{3ex}
  \multirow{2}{*}{$\tilde{g}\tilde{q}$} &  $\sigma_{\rm inclusive}$ & $8.42$ & $8.39^{+12\%}_{-14\%}\pm 6.9\%$ & $8.38$ & $8.35$ & $8.41$ & $8.40$ & $5.49^{+38\%}_{-25\%}\pm 7.0\%$ \\\cline{2-9}\rule{0pt}{3ex}
  & $\sigma_{\rm fiducial}$ & $5.93$ & $5.91^{+12\%}_{-14\%}\pm 6.9\%$ & $5.90$ & $5.87$ & $5.93$ & $5.92$ & $3.86^{+38\%}_{-26\%}\pm 7.0\%$ \\\hline\rule{0pt}{3ex}
  \multirow{2}{*}{$\tilde{q}\tilde{q}$} &  $\sigma_{\rm inclusive}$ & $20.4$ & $20.4^{+7.8\%}_{-10\%}\pm 2.2\%$ & $20.4$ & $20.4$ & $20.4$ & $20.4$ & $14.9^{+30\%}_{-22\%}\pm 2.2\%$ \\\cline{2-9}\rule{0pt}{3ex}
  & $\sigma_{\rm fiducial}$ & $14.8$ & $14.8^{+7.8\%}_{-9.9\%}\pm 2.2\%$ & $14.8$ & $14.7$ &  $14.8$ & $14.8$ & $10.8^{+30\%}_{-21\%}\pm 2.2\%$
  \end{tabular}
  \caption{Total inclusive and fiducial cross sections (in fb) at the
    $\sqrt{S}=13$ TeV LHC. The leftmost errors stem from scale variations,
    the rightmost ones from PDF uncertainties. We have set 
    {\tt str\_include\_pdf=True} and 
    {\tt str\_include\_flux=True} (see section~\ref{sec:OSinMG}).
\label{tab:xslhc}}
\end{table}

Total cross sections with ($\sigma_{\rm fiducial}$) and without
($\sigma_{\rm inclusive}$) the above cuts at the 13 TeV LHC are presented in
table~\ref{tab:xslhc}, for each of the six STR procedures listed
in table~\ref{tab:iossubtr}.
Gluino-pair production leads to the smallest cross
sections, as a result of the large gluino mass and the correspondingly small
gluon PDF at large Bjorken $x$'s. The NLO inclusive (fiducial) cross section 
turns out to be equal to about 0.33~fb (0.23~fb), with a large $K$-factor 
of about 1.8 both for the inclusive and fiducial cases. This large $K$-factor 
originates from the large colour charge associated with the gluino, and the 
purely strong nature of the Born process, as was already discussed in
section~\ref{sec:rates}. Scale uncertainties are reduced
by a factor of two with respect to the LO ones and, as a consequence of the 
typically large Bjorken $x$ values associated with the production of 
a pair of 2~TeV gluinos, the theoretical error is dominated by
the PDF uncertainties. Both scale and PDF errors are essentially
independent of the STR procedure adopted, which is the reason why we
only report them for the {\tt istr=2} case. As far as the STR dependence
itself is concerned, it is about 3\%, and thus much smaller than the 
other theoretical uncertainties. As is well known, this is an 
observable-dependent statement, and we shall show later that at the
differential level things are more involved. Finally, we remark that 
the Monte Carlo integration errors are equal to about 0.2\%, and have
therefore been ignored.

Because of the smaller $\tilde{u}_R$ mass, the cross sections of gluino-squark
associated production and of squark pairs are 25 to 60 times larger than the
gluino-pair ones. This behaviour is driven by that of the parton luminosities:
valence quarks contribute significantly, since one is in an $x$-region where 
their PDFs are large. Correspondingly, the PDF uncertainties are much smaller 
than for gluino-pair production, as deep-inelastic scattering, fixed-target 
experiment data, LHC Drell-Yan, forward $W$-boson, and $Z$-boson data allow 
one to strongly constrain the fit of the valence quark densities at
large $x$'s. This implies that for these processes, at variance with the
case of gluino pairs, the PDF errors are smaller than the scale ones,
in spite of the fact that the latter are a factor of three smaller at 
the NLO than at the LO (as opposed to a factor of two for gluino-pair
production). As far as the STR-option dependence is concerned, it is
below 1\% and thus, as in the case of the gluinos, largely subdominant
with respect to the other uncertainties.

\begin{figure}
  \centering
  \foreach \page in {4,8,2,12,14,16}{
  \includegraphics[page=\numexpr \page\relax, width=0.45\textwidth,draft=false]{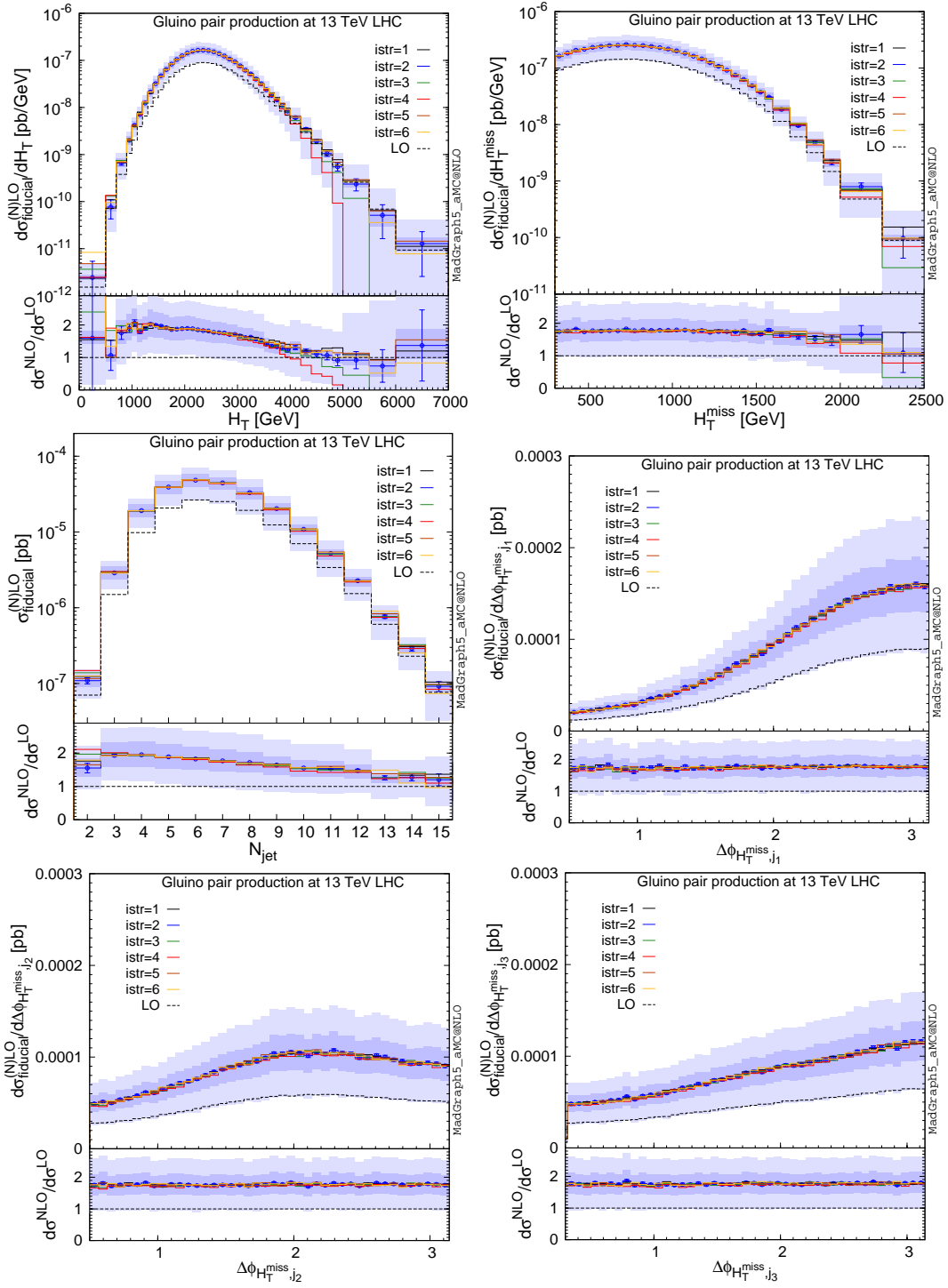}
}
  \caption{
    Representative differential distributions for gluino-pair production
    at $\sqrt{S}=13$ TeV LHC, relevant to the CMS analysis of
    ref.~\cite{Sirunyan:2017cwe}. We consider the $H_T$ (upper left),
    $H_T^{\rm miss}$ (upper right) and jet multiplicity (centre left) spectra,
    as well as the azimuthal separation between the $H_T^{\rm miss}$ vector and
    the three hardest jets (centre right and bottom panels). The error bars 
    represent the Monte Carlo integration errors in the default {\tt istr=2} 
    choice. The NLO results are obtained with {\tt str\_include\_pdf=True} 
    and {\tt str\_include\_flux=True}.
  }
  \label{fig:gogo}
\end{figure}

\begin{figure}
  \centering
  \foreach \page in {4,8,2,12,14,16}{
  \includegraphics[page=\numexpr \page\relax, width=0.45\textwidth,draft=false]{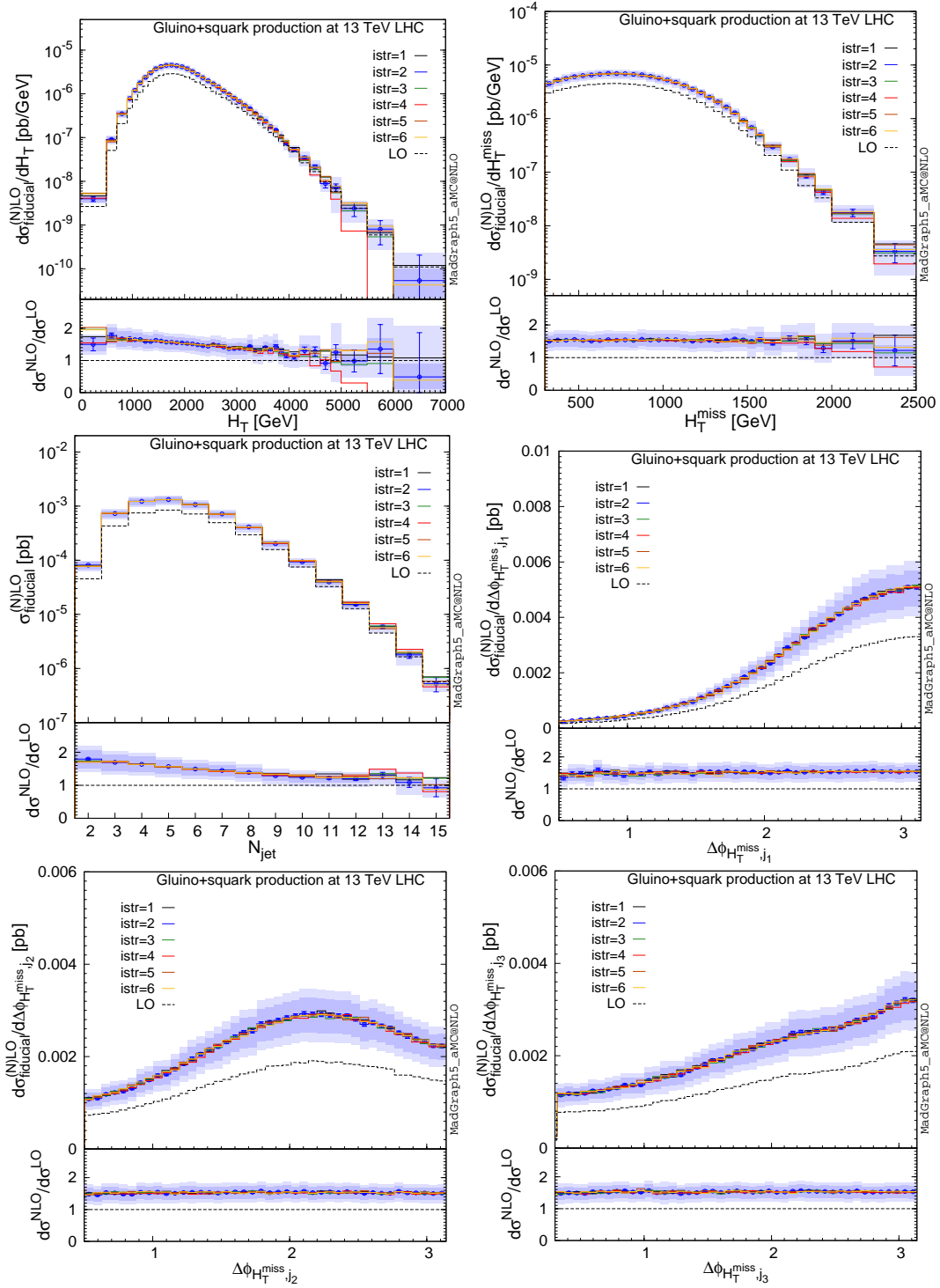}
  }
  \caption{Same as in figure~\ref{fig:gogo} but for gluino-squark production.}
  \label{fig:gosq}
\end{figure}

\begin{figure}
  \centering
  \foreach \page in {4,8,2,12,14,16}{
  \includegraphics[page=\numexpr \page\relax, width=0.45\textwidth,draft=false]{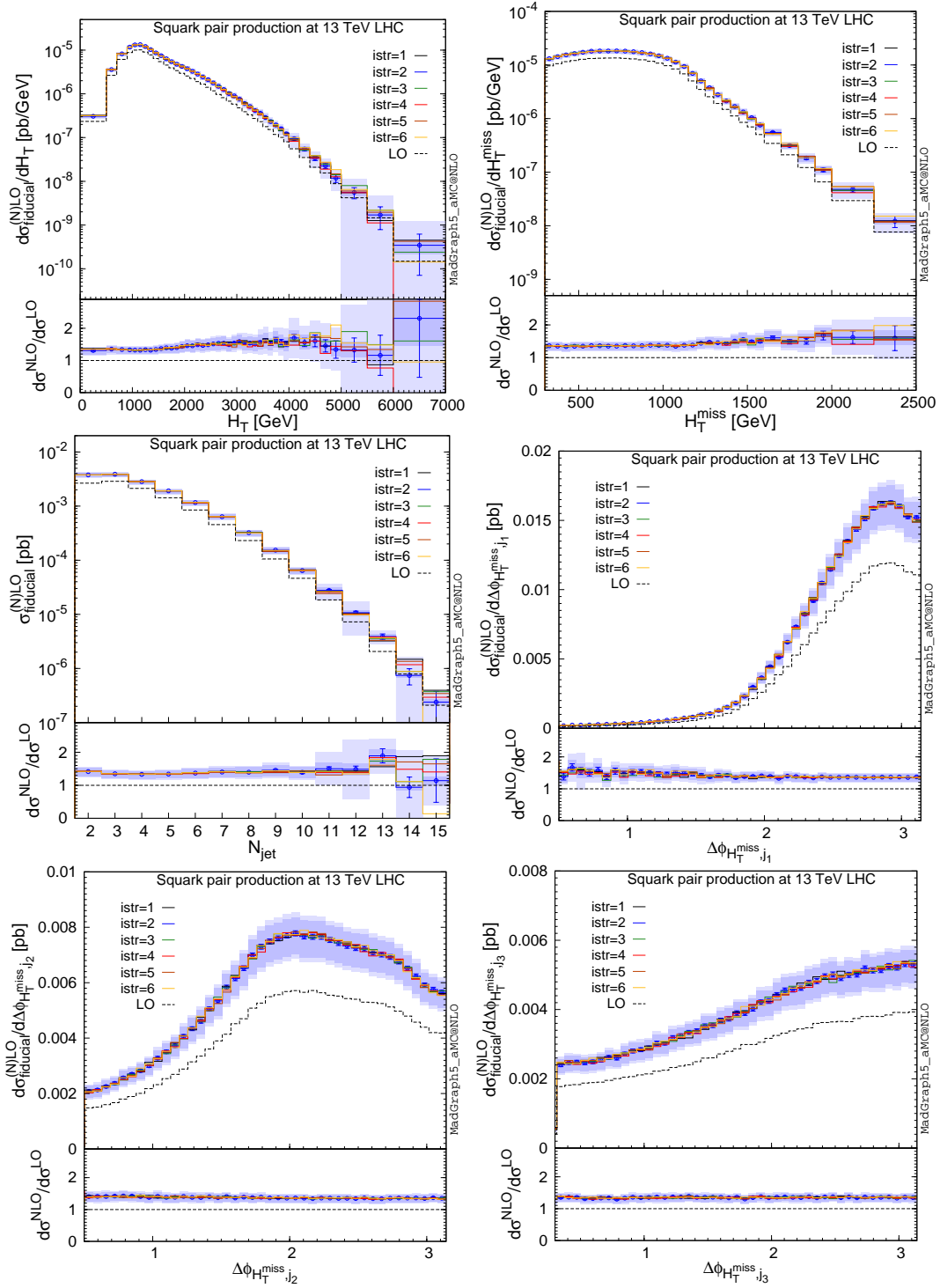}
  }
  \caption{Same as in figure~\ref{fig:gogo} but for squark-squark production.}
  \label{fig:sqsq}
\end{figure}

In figures~\ref{fig:gogo}, \ref{fig:gosq}, and \ref{fig:sqsq} we
present six differential distributions for gluino-pair, gluino-squark, and 
squark-pair production, respectively. Such observables are directly relevant 
to the CMS SUSY search of ref.~\cite{Sirunyan:2017cwe}, and thus all 
results are obtained by applying the fiducial-volume cuts defined before. 
Each panel of each figure has the same layout, namely: the main frame
presents the differential distributions in pb/GeV or pb, at the LO (dashed
black histogram) and at the NLO (solid coloured histograms; there are
six of these, corresponding to the six STR options we have considered).
The statistical errors are shown as error bars, while the scale uncertainties 
are displayed as dark-blue bands; finally, the linear sums of the scale and
PDF errors are represented as light-blue bands.
The lower inset presents the ratios of the six NLO predictions over the
LO one, the latter evaluated with the central scale choice and central PDF 
(\ie, these are the standard $K$ factors).

Mirroring what has been found for total rates, PDF uncertainties are quite 
substantial for gluino-pair production, as one is always in a kinematic
regime where large Bjorken $x$'s are relevant. For the other two processes, 
such uncertainties are reduced, except in those phase-space corners 
where again one is sensitive to large Bjorken $x$'s values, \eg\ for
large $H_T$ or $H_T^{\rm miss}$. It is therefore obvious that, despite
the progress made in the computations of the short-distance cross sections,
precise simulations at large scales can only be achieved by constraining
much more strongly the PDFs at $x\to 1$. As far as the $K$-factors are
concerned, they are found to be relatively flat for most observables,
and close to $2$ for $\tilde{g}\tilde{g}$ production, to $1.5$ for 
$\tilde{g}\tilde{q}$ production, and to $1.4$ for $\tilde{q}\tilde{q}$ 
production, respectively. Distinctly non-flat $K$-factors are obtained 
in particular in the cases of the $H_T$ and jet-multiplicity distributions.
This helps stress the fact that, in general, the re-scaling of LO+PS
predictions by an overall constant factor is a dangerous procedure that
may lead to unreliable results, which underscores one of the main motivations
of the present paper.

It is interesting to observe that the dependence on the STR choice is
generally very mild also at the differential level, which gives one 
confidence on the description of multijet SUSY-induced signatures through 
the production-and-decay picture of eqs.~(\ref{ggprc})--(\ref{qqprc}).
A notable exception is the large-$H_T$ region (for all of the three
processes, although it is particularly prominent in the case of
gluino-pair production), where the STR dependence becomes 
extremely large, and appears to be pathological. We shall argue
that, in fact, this behaviour allows the method to self-diagnose that it
is being applied in regions where its founding assumptions are dubious 
at best. Let us start with a technical explanation. The STR procedures
that differ the most from the median are those associated with 
{\tt istr=3,4} -- these are DS procedures, option~A (see 
table~\ref{tab:iossubtr}). In these cases, through a reshuffling
operation the momenta of the incoming partons are changed, thereby
implying that the corresponding Bjorken $x$'s are changed too.
This affects the value of both the flux factor and the PDFs.
In the notation of section~\ref{sec:DS}:
\begin{equation}
f\!\left(m_{\delta\gamma}^2\right)
\proj\left(\abs{\amp_{ab\to\delta\gamma X}^{(\beta)}}^2 {\rm d}\phi\right) \sim 
\frac{x_a x_b}{\bx_a \bx_b} \; 
\frac{f_a^{(H_1)}(\bx_a)\,f_b^{(H_2)}(\bx_b)}{f_a^{(H_1)}(x_a)\,f_b^{(H_2)}(x_b)}\,,
\label{eq:DScnt}
\end{equation}
which shows the effect on the flux and parton-luminosity factors (first and 
second  terms on the r.h.s.~of eq.~(\ref{eq:DScnt}), respectively) at the 
level of the subtraction cross section (last term on the r.h.s.~of
eq.~(\ref{sigDS})). If one generates a far off-shell kinematic configuration, 
$m_{\delta\gamma} \gg m_\beta$, then $x_a x_b \gg \bx_a \bx_b$. Therefore,
since eventually all PDFs decrease with increasing Bjorken $x$'s, 
both terms on the right-hand side of eq.~\eqref{eq:DScnt} are large,
which implies a strong suppression of the physical cross section of
eq.~(\ref{sigDS}) in this kinematic region (since the subtraction term
is large); so strong, in fact, that it may become negative. Owing to the 
structure of the function $f(m^2)$ given  in eq.~\eqref{fchoice}, this 
feature is more pronounced for {\tt istr=4} than for {\tt istr=3}, which 
clearly shows in the figures. 

In fact, although the mechanism we have just described is responsible
for the behaviour of the cross sections in the large-$H_T$ region, 
its effects are particularly dramatic in gluino-pair production owing 
to $x_{a,b}\simeq 1$ there
(in other words, if the gluino had a smaller mass, the subtraction
cross section would be significantly smaller). For such Bjorken $x$'s,
the central values of the PDFs are very suppressed and poorly constrained, 
and thus affected by very large residual uncertainties: indeed, we see
from figures~\ref{fig:gogo}--\ref{fig:sqsq} that as soon as the STR-choice
dependence becomes very large, so does the PDF uncertainty, the effect being
larger when parton luminosities involving the gluon density are relevant.
We have verified that, by ``removing'' the luminosity factor from
eq.~(\ref{eq:DScnt}) by setting {\tt str\_include\_pdf} equal to {\tt False}
(see section~\ref{sec:OSinMG}), the results obtained with {\tt istr=3,4} 
are much closer to the others, thus showing that the large STR-choice
dependence is driven by the PDFs.

We point out that sizable differences due to STR options should be
expected also for $m_{\delta\gamma} \ll m_\beta$. However, while the 
Breit-Wigner distribution is symmetric around $m_\beta$, the flux and 
PDF ratios are not. This implies that the explanation given above 
may not apply in this kinematic regime. However, regardless of which
mechanism is responsible for an enhanced STR-choice dependence, the
take-home message is the following: such a dependence is the signal
that describing the process of interest by means of a cross section
dominated by resonant production of (s)particles which subsequently
decay is simply not adequate, and a full (unfortunately, more complicated)
computation is required. Ultimately, the decision of where to stop
trusting a simplified computation rests with the user, and it depends
on many factors, in particular on whether one is interested in a 
class.~1 or class.~2 approach (see the itemised list below eq.~(\ref{BSMex3})
in section~\ref{sec:OS}). It is because of this that \mg\ does not set to
zero the subtraction term in eq.~(\ref{sigDS}) when far off-shell:
if needed, such a cut must be implemented at the analysis level.


\section{Conclusions}\label{sec:conclusion}
With the steady increase of the statistics accumulated by the LHC 
experiments, and the absence of positive results in the searches for
new physics, it becomes necessary to improve the accuracy of the 
simulations of BSM signals, thus matching that of their SM backgrounds.
By far and large, this currently means matrix elements computed at the
NLO in QCD matched to parton showers (NLO+PS). Such a necessity has spurred
some recent theoretical activity, whereby authors have addressed the
needs of specific search strategies.

The goal of this paper has been that of rendering such an improvement
systematic (\ie\ achievable for arbitrary processes and for a vast class
of renormalisable theories). This is feasible thanks to the powerful and
flexible environment constituted by the automated program \mglong\ and 
the physics models it can use for simulations. In particular, for the
sake of the present work, two major limitations have been lifted.
Firstly, at the level of the construction of models, we have 
overhauled the way in which \fr\ and \nloct\ deal with on-shell 
renormalisation schemes, so that a much larger flexibility is achieved,
that helps deal with the difficult cases especially relevant to supersymmetric
theories. Secondly, at the level of the \mglong\ code proper, we have
automated a variety of solutions to the problem posed by the presence
of partonic channels that appear beyond the leading order in perturbative
computations, and that feature narrow resonances; this problem is 
particularly acute in theories with a complex mass spectrum.
Such solutions, that we have dubbed Simplified Treatment of Resonances (STR),
generalise the so-called on-shell subtractions, and encompass the Diagram 
Removal (DR) and Diagram Subtraction (DS) strategies introduced in the last 
few years in the context of NLO+PS simulations. Technically, these two
pieces of work have been implemented, respectively, in a plugin for
\fr, called \bfl, and in a plugin for \mg, called \mados.

As a proof-of-concept, we have generated an NLO UFO model for the MSSM with 
a widely-used renormalisation scheme, and we have studied processes that
feature intermediate resonances at the LHC, at the NLO+PS accuracy and
with a realistic set of final-state cuts. We have performed thorough
self-consistency checks of our implementation, and compared some loop
matrix elements generated by the code with those resulting from 
analytical computations. It is important to bear in mind that the core
structure of the \mglong\ code, which has been only minimally affected by the present 
work, has by now been very extensively validated in countless simulations.
As a further a posteriori validation, we have compared total-rate
leading- and next-to-leading order results for benchmark $2\to 2$ processes 
with those of the public codes \prospino\ and \resummino, with the
restrictions that these two programs enforce. Such comparisons, whose 
details can be found in section~\ref{sec:validate}, in some cases show 
disagreements (of different origins) among the various predictions,
and further underscore the advantages of a general, process-independent,
and automated implementation.

In conclusion, with the present work we have achieved, for the first time, 
the complete automation of NLO+PS simulations for supersymmetric-particle 
production at hadron colliders in the framework of the MSSM with a generic
particle spectrum, and set up the tools for dealing with similarly involved 
theories by means of a user-driven framework. 
We point out, however, that we have not yet implemented and tested
the general solution, introduced in ref.~\cite{Frederix:2018nkq}, to the 
problem posed by unstable resonances in the context of the complex-mass 
scheme~\cite{Denner:1999gp,Denner:2005fg}.
Such a solution requires further developments in \fr\ and \nloct,
so that NLO UFO models contain the necessary routines for
dynamically and automatically selecting, according to the particle
spectrum, the appropriate Riemann sheets for the calculation
of the UV counterterms. This is left to future work.

All of the computer programs 
relevant to this paper are publicly available -- on top of \mglong, the MSSM 
model, together with the \bfl\ plugin and ready-to-be-used \mthmtc\ notebooks, 
can be found at\\
$~~~~$\url{http://feynrules.irmp.ucl.ac.be/wiki/MSSMatNLO},\ \\
while the  \mados\ plugin can be downloaded from\\
$~~~$\url{https://code.launchpad.net/~maddevelopers/mg5amcnlo/MadSTRPlugin}.\\
This paper is accompanied by ancillary files, stored on the electronic
archive, that collect NLO QCD results for total rates of pair-production
supersymmetric processes, which we have refrained from including here
for reasons of space.


\section*{Acknowledgements}

We would like to thank Dorival Goncalves and David Lopez-Val for collaboration
at early stages of the development of \mados\ and for discussions on related
topics. We also thank Davide Pagani for discussions on these topics.
BF and HSS are supported by the LABEX ILP (ANR-11-IDEX-0004-02, ANR-10-LABX-63).
KM is supported by JSPS KAKENHI Grant No.~18K03648.
MPAS is supported by the BMBF under contract 05H18PMCC1 and the DFG through the 
Research Training Network 2149 "Strong and weak interactions - from hadrons to 
dark matter".
MZ is supported by the Netherlands National Organisation for 
Scientific Research (NWO).
SF is grateful to the CERN TH division for the hospitality during the
course of this work.
VH is supported by the European Research Council 
(ERC) grant No 694712 (PertQCD), and by the Swiss National Science Foundation 
(SNSF) grant No 179016.

\clearpage

\appendix
\section{Conventions for one-point and two-point functions}\label{app:PV}
In the analytical formulas presented in this paper, all $A$ and $B$
loop-integrals have been normalised as
\be
  A_0(m^2) = \int {{\rm d}^D q \over i \pi^2}
     \frac{ (2\pi \mu_R)^{2 \epsilon}}{\big[q^2 - m^2\big]}
  \ \ \text{and}\ \
  B_{\{0,\mu\}}(p^2; m_1^2,m_2^2) = \int {{\rm d}^D q \over i \pi^2}
     \frac{ (2\pi \mu_R)^{2 \epsilon}\ \{1,q_\mu\}}{\big[q^2 - m_1^2\big]
    \big[(q + p)^2 - m_2^2\big]} \ ,
\ee
where we recall that $D=4-2\epsilon$ is the number of spacetime dimensions and
$\mu_R$ is the regularisation scale (taken equal to the renormalisation scale).
The $B_\mu$ vectorial integral has been further reduced to a scalar integral
using Lorentz covariance,
\be
  B_\mu(p^2; m_1^2,m_2^2)= p_\mu B_1(p^2; m_1^2,m_2^2) \ ,
\ee
and the $B_1$ integral is connected to several $B_0$ integrals as
\be
    B_1(p^2,m_1^2,m_2^2) = -\frac12 B_0(p^2;m_1^2,m_2^2)
    + \frac{m_2^2 -m_1^2}{2p^2} \Big[ B_0(p^2;m_1^2,m_2^2) -
      B_0(0;m_1^2,m_2^2)\Big] \ ,
\ee
the $p^2\to 0$ limiting case involving a derivative of the $B_0$ function with
respect to the $p^2$ variable instead of the squared bracket. Explicitly, one gets
\be\bsp
  & A_0(m^2) = m^2 \Big[\frac{1}{\epsbar}+1+\log{\frac{\mu_R^2}{m^2}}\Big]\ ,\\
  & B_0(p^2;m_1^2,m_2^2)= \frac{1}{\epsbar} + 2 - \log\frac{p^2}{\mu_R^2}
    +\sum^2_{i=1}
      \Big[\gamma_i \log\frac{\gamma_i-1}{\gamma_i} - \log(\gamma_i-1) \Big]\ ,
\esp\ee
where
\be
  \gamma_{1,2} = \frac{p^2-m_2^2+m_1^2\pm\sqrt{(p^2-m_2^2+m_1^2)^2-4p^2m_1^2}}
    {2p^2}\ ,
\ee
and with the ultraviolet-divergent parts of the integrals being written in terms
of the number of spacetime dimensions and the Euler-Mascheroni constant $\gamma_E$,
$\frac{1}{\epsbar}=\frac{1}{\epsilon} - \gamma_E + \log{4\pi}$.
Several special limits for the $B_0$ function and its $B_0^\prime$ derivative
are useful,
\be\bsp
  B_0(0,m^2,m^2)=&\ \frac{1}{\bar\epsilon} + \log{\frac{\mu_R^2}{m^2}}\ ,\\
  B_0(0,m_1^2,m_2^2) = &\ \frac{1}{\bar\epsilon} + 1 +
     \frac{m_1^2\log{\frac{\mu_R^2}{m_1^2}}-m_2^2\log{\frac{\mu_R^2}{m_2^2}}}
      {m_1^2-m_2^2}\ ,\\
  B_0(m^2,m^2,0)=&\ \frac{1}{\bar\epsilon}+2+\log{\frac{\mu_R^2}{m^2}}\ ,\\
  B_0^\prime(0,m^2,m^2)=&\ \frac{1}{6m^2}\ ,\\
  B_0^\prime(0,m_1^2,m_2^2)=&\ \frac{m_1^2+m_2^2}{2(m_1^2-m_2^2)^2}
    +\frac{m_1^2m_2^2}{(m_1^2-m_2^2)^3}\log{\frac{m_2^2}{m_1^2}}\ , \\
  B_0^\prime(m^2,m^2,0)=&\ -\frac{1}{2m^2} \Big[\frac{1}{\epsbar} + 2 +
     \log{\frac{\mu_R^2}{m^2}}\Big]\ .
\esp\ee
We stress that since this paper does not consider the complex-mass scheme, the
renormalisation counterterms are defined using \emph{only} the real part of the
two-point functions.

\section{The \bfl\ package}\label{app:bfl}
In order to circumvent the lack of flexibility concerning the choice of the
renormalisation scheme in the current \fr\ release, we have developed a plugin,
called \bfl, that is fully flexible in the way a bare
Lagrangian can be renormalised. This package can be downloaded from the wikipage
collecting details about the MSSM at NLO implementation in \fr\ introduced in
this paper, \url{http://feynrules.irmp.ucl.ac.be/wiki/MSSMatNLO}.

In practice, the user starts by loading \fr\ and any given model implementation
in the \mthmtc\ session. The \bfl\ plugin can then be imported as any \mthmtc\
package, by typing,
\begin{verbatim} Begin["MoGRe`"]; << MoGReoop.m; End[]; \end{verbatim}
where one assumes that the current directory is the one containing the
plugin.

\subsection{The main method {\tt MoGRe\$Renormalize} and its options}
\label{sec:options}
The main function of the \bfl\ plugin is
called {\tt MoGRe\$Renormalize} and takes a Lagrangian as an input, as for instance
in the following example
\begin{lstlisting}
  MoGRe$\$$Renormalize[LMSSM]
\end{lstlisting}
where \verb+LMSSM+ stands for the MSSM Lagrangian. The user is
allowed to specify three options that modify the behaviour of the method and that
respectively address the treatment of the four-scalar vertices, loop-induced
field mixing and the nature of the interaction in which the loop-corrections
are evaluated. These options can be set following a standard \mthmtc\ syntax, as
for instance through the command
\begin{lstlisting}
  SetOptions[MoGRe$\$$Renormalize,
     Exclude4Scalars -> True,
     FlavorMixing    -> {{st1, st2}, {sb1, sb2}},
     CouplingOrders -> {QCD}
  ];\end{lstlisting}
This first indicates, through the {\tt Exclude4Scalars} option, that all model
four-scalar interactions have to be ignored in the renormalisation procedure, so
that the corresponding counterterms are not evaluated. The default choice for
this option (set to {\tt True} in our example) is {\tt False}. While strictly
speaking ignoring the renormalisation of the four-scalar vertices is incorrect,
these vertices rarely appear at tree-level so that the associated counterterms
are often not necessary. Avoiding their calculation and their inclusion in the
final UFO model therefore allows for an increase of the efficiency of the
computations, both at the \nloct\ and \mg\ level.

Secondly, we provide information on the different sets of fields that mix at the
one-loop level through the {\tt FlavorMixing} option of the
{\tt MoGRe\$Renormalize} method. If set to {\tt
True} (the default choice), all fields carrying the same quantum numbers and
lying in the same spin and colour representations are assumed to mix. In
contrast, all loop-level mixings are forbidden if this option is set to {\tt
False}. The user has also the possibility, like in our example, to provide a
list with the different sets of fields that mix at the one-loop level. Any
physical field can be used in such a list, and the code further checks whether
the input is compatible with the representation of the involved fields under the
model gauge groups. In our
example, we have forbidden any loop-level mixing, except the one of the
two stop-eigenstates (denoted {\tt st1} and {\tt st2} in the \fr\ model) and the
one of the two sbottom-eigenstates (denoted {\tt sb1} and {\tt sb2} in the \fr\
model), as given by eq.~\eqref{eq:stopsbotmix}.

Finally, the last option
indicates which type of interaction should be renormalised, among all the
interactions declared in the {\tt MR\$InteractionOrderHierarchy} option of the
\fr\ model implementation. In the MSSM implementation, two types of interactions are
available, namely the \verb+QCD+ and \verb+QED+ ones. In the above example, that
matches the physics goals of this paper (NLO QCD corrections for the MSSM), we
have solely selected the QCD interaction type \verb+QCD+.

Before describing how the bare Lagrangian is technically renormalised, we detail
in the next subsections various methods that can be used to simplify the model
and modify the way in which {\tt MoGRe\$Renormalize} works.

\subsection{Simplifying the procedure}\label{app:BFL_simp}

In general, all external parameters have to be renormalised, which yields a
heavy renormalisation procedure for complex models like the MSSM. However, some
parameters may not need to be renormalised, like the electromagnetic coupling
constant that does not receive any correction at one loop in QCD. Whilst this
type of information can be useful to speed up the renormalisation procedure, the
programme cannot guess it at this stage where no calculation has been done yet.
For this reason, the user is allowed to declare the parameters that should not
be renormalised with the \verb+MoGRe`DefineUnrenormalizedParameters+ method that
takes, as arguments the corresponding symbols as implemented in the \fr\ model.
The arguments can also be provided as a unique list. Going back to the
considered example where only QCD corrections matter, the command
\begin{verbatim} MoGRe`DefineUnrenormalizedParameters[
  Gf, aEWM1, MZ, MUH, alp, tb,
  Mx1, Mx2, mHu2, mHd2, meL, mmuL, mtauL, meR, mmuR, mtauR,
  Sequence @@ Flatten[Table[{au[i, i], ad[i, i]}, {i, 1, 2}]],
  Sequence @@ Flatten[Table[{ae[i, i]}, {i, 1, 3}]],
  Sequence @@ Flatten[Table[{VV[i, j], UU[i, j]}, {i, 1, 2}, {j, 1, 2}]],
  Sequence @@ Flatten[Table[{NN[i, j]}, {i, 1, 4}, {j, 1, 4}]
 ]
\end{verbatim}
leads to declaring all parameters connected to the electroweak sector (namely
the electroweak inputs, the Higgs sector parameters, the electroweak gaugino and
scalar soft masses as well as the chargino and neutralino mixing matrices) to be
insensitive to QCD corrections at one loop. The renormalisation of all
relevant internal parameters is accordingly and automatically simplified through
their functional dependence on the above parameters.

Similarly, the code assumes by default that all fields get renormalised,
although this may not be the case in practice. For instance, the weak boson
two-point functions are insensitive to QCD corrections at one loop. This type of
information can be passed to the code by means of the
\verb+MoGRe`DeclareUnrenormalizedFields+ method that takes as arguments all fields
that should not be renormalised. The arguments can be provided either
sequentially or under the form of a list. For instance in the MSSM
implementation worked out in this paper, we could use
\begin{verbatim} MoGRe`DeclareUnrenormalizedFields[
  seL, seR, smuL, smuR, stau1, stau2, sne, snm, snt, A, W, Z
 ];
\end{verbatim}
although the sleptons are in principle not necessary as they do not appear in any
QCD vertex (but we keep them here for illustrative purposes).
This prevents all charged sleptons, sneutrinos and electroweak bosons from being
renormalised. All remaining fields will be renormalised, the associated
wave-function renormalisation constants being taken complex by default. Reality
conditions can be enforced through the usage of the
\verb+MoGRe`RealFieldRenormalisation+ method, that takes as an
argument the symbols associated with the concerned fields. All relevant symbols
can be given again either
under the form of a list or of a sequence. In the case where no argument is
provided, all wave-function renormalisation constants are taken real, as with
the following example,
\begin{verbatim} MoGRe`RealFieldRenormalisation[ ]\end{verbatim}
For mixing fields for which matrix renormalisation is in order, the
method acts on all the elements of the renormalisation matrix.

\subsection{Restrictions}\label{sec:restrictions}
The \fr\ model implementation may contain (external or internal) parameters that
are numerically vanishing when default values are accounted for. While it is in
general safer to keep these parameters all along the renormalisation procedure,
so that they are renormalised and get associated with potentially non-vanishing
renormalisation constants, this is often not necessary and only
leads to heavier subsequent calculations.
The \bfl\ package by default takes care of the removal of these zero
parameters both from the tree-level Lagrangian and from the rules dedicated to
the replacement of the bare quantities by the renormalised ones. In this way, those
parameters will not be renormalised as not present in the bare Lagrangian
anymore, and the code will make sure that they do not re-appear through the
renormalisation of other quantities. For instance, the CKM matrix is often taken
diagonal, so that \bfl\ by default removes all its off-diagonal elements. This
behaviour can be turned off by issuing the command
\begin{verbatim} EnforceZeros = False; \end{verbatim}

\subsection{Specifying the renormalisation scheme}\label{sec:scheme}
As detailed in section~\ref{sec:susy-onshell}, there is no unique way to define
an OS renormalisation scheme in the MSSM and in SUSY in general. This
consisted in the main reason that has led to the development of the \bfl\
package. Scheme-specific renormalisation conditions can be added by the user by
making use of a dedicated method named \verb+AddRenormalizationCondition+. The
latter takes two arguments, a renormalisation constant (associated with either a
parameter or a field) and a function of different parameters, fields and other
renormalisation constants. As a result, the first renormalisation constant will
be considered equal to the function given as the second argument. The way in
which field renormalisation constants should be input follows the \fr\ syntax,
wave-function renormalisation constants being provided as
\begin{verbatim}
  FR$deltaZ[{fld1,fld2}, {{}}]]
  FR$deltaZ[{fld1,fld2},{{"L"}}]]    FR$deltaZ[{fld1,fld2},{{"R"}}]]
\end{verbatim}
for non-fermionic, left-handed fermionic and right-handed fermionic fields
respectively. In the diagonal case, the two field symbols \verb+fld1+ and
\verb+fld2+ are equal. In a non-diagonal field mixing case, they can be
different. For a parameter \verb+prm+, the corresponding syntax reads
\begin{verbatim}
  FR$delta[{prm}, {}]]
\end{verbatim}
For instance, the stop mixing angle conditions of eq.~\eqref{eq:renomix} could
be implemented as
\begin{verbatim}
 AddRenormalizationCondition[FR$delta[{Rtop[1,1]}, {}], 1/4 Rtop[2,1] *
    ( FR$deltaZ[{st1,st2},{{}}] -  Conjugate[FR$deltaZ[{st2,st1},{{}}]] )];
\end{verbatim}
where the \verb+Rtop+ symbol represents the stop mixing matrix $S_{\tilde t}$ of
eq.~\eqref{eq:3rdgenmix} in our \fr\ implementation. This is equivalent to
indicating to the \bfl\ package that
\be
  (\delta S_{\tilde t})_{11} = -s_{\tilde t}\ \delta\theta_{\tilde t}
     = \frac14\ (S_{\tilde t})_{21}\
       \Big[\delta Z_{\tilde t,12}-\delta Z_{\tilde t,21}^\ast\Big] \ .
\ee

\subsection{Clearing a renormalisation scheme}
All the options detailed above can be reset by calling the
\verb+MoGRe`ClearRenormalizationScheme[]+ method.

\subsection{Technical details on the functioning of the method}
\subsubsection{Initialization}\label{sec:bfl:initmain}
The {\tt MoGRe\$Renormalize} method begins with a check that all
parameters passed as options are meaningful. Appropriate error messages are
printed to the screen if necessary.

In a second step, still prior to any computation,
simplifications are performed and the Lagrangian is put under an internal format
allowing for a more efficient run. More precisely, the Lagrangian is
truncated from its constant terms, and all parameters that are
vanishing are removed except if the \verb+EnforceZeros+ flag has been set to
{\tt False} (see section~\ref{sec:restrictions}). The Lagrangian is then expanded so
that all (unphysical) gauge eigenstates are replaced by physical mass
eigenstates. In the case where four-scalar interactions are requested
not to be
renormalised (see section~\ref{sec:options}), they are removed from the
Lagrangian. They will however be reintroduced at the very end of the
renormalisation procedure, as those interactions could appear into given loop
diagrams. Moreover, \bfl\ requires that all tree-level bilinear terms are
canonically normalised and that all kinetic and mass mixings have been
appropriately rotated away by the user. In practice, the code ignores all
bilinear terms provided by the user and reintroduces the canonical ones directly
on the basis of the model field content.

As a last initialisation step, \bfl\ makes use of the \fr\ model information on
the external and internal parameters that must be exchanged during the
renormalisation procedure (the \fr\ \verb+FR$LoopSwitches+
option~\cite{Alloul:2013bka}). In the \fr\ syntax, \verb+FR$LoopSwitches+
consists in a list
of 2-tuples of parameters,
\begin{lstlisting}
  FR$\$$LoopSwitches = { {prm1, prm2} , .... }
\end{lstlisting}
in which \verb+prm1+ is external and \verb+prm2+ is internal. Prior to the
renormalisation of the model, \verb+prm2+ is made external and \verb+prm1+
internal, the dependence of this last parameters on the other parameters being
derived by the code. For instance, in many publicly available \fr\ models, the
$W$-boson mass $m_W$ is derived from the other electroweak inputs. OS
renormalisation however requires $m_W$ to be external. The
\verb+FR$LoopSwitches+ is then used to trade it, for example, with the Fermi
constant $G_F$. We refer to the \fr\ manual~\cite{Alloul:2013bka} for more
information.

Some parameters can moreover be doubly defined in \fr\ models, like the Yukawa
couplings and the fermion masses in the SM that are taken as
different input parameters, although being numerically equal. The idea behind
this trick consists in allowing for massless fermions and non-zero Yukawa
couplings at tree-level. However, this makes the theory ill-defined when
renormalisation is at stake, as all these parameters must be enforced to be
equal when counterterms are evaluated. This can be achieved by means of the
\verb+FR$RmDblExt+ \fr\ parameter, that consists in a list of \mthmtc\
replacement rules mapping one of the doubly-defined parameters to the other. For
instance,
\begin{lstlisting}
  FR$\$$RmDblExt = { ymb -> MB, ymc -> MC, ymdo -> MD, yme -> Me,
   ymm -> MMU, yms -> MS, ymt -> MT, ymtau -> MTA, ymup -> MU};
\end{lstlisting}
replaces every single Yukawa coupling (normalised to be exactly equal to the
associated mass parameter) by the corresponding fermion mass. Such a replacement
is also enforced in \bfl\ in the case where the \verb+FR$RmDblExt+ parameter
exists. Concerning the model introduced in this work, all parameters are
uniquely defined so that this is irrelevant.

Along these lines, we recall that we rely on \nloct~\cite{Degrande:2014vpa} for
the analytical
computation of the various counterterms. This implies that care must be taken
with any coupling depending on particle masses, like the Yukawa couplings or
the trilinear scalar interaction strengths of eq.~\eqref{eq:TtoA}. By default,
\nloct\ renormalises them in the $\msbar$ scheme regardless of any finite pieces
that are relevant in the OS scheme. A correct treatment therefore requires
to replace them by their analytical expression (that involves the fermion
masses), prior to the call to \nloct. This can be achieved straightforwardly
with the \verb+RemovingInternalCst+ method of \bfl\ that re-expresses a given
parameter in terms of the others. Concretely, this method removes a given
parameter and the associated renormalisation constant from the model and
replaces them by the corresponding analytical expressions. Whilst the
replacement rule of the parameter itself is taken from the model file, the one
of the renormalisation constant is derived on the fly.

In the context of the MSSM implementation presented in this paper, we have
implemented
\begin{verbatim} MoGRe`RemovingInternalCst/@{
  gs, Sequence@@Flatten[Table[{yu[i,i], yd[i,i], tu[i,i], td[i,i]}, {i,1,3}]]
 };
\end{verbatim}
so that all Yukawa and trilinear scalar interactions have been replaced
according to their dependence on the external quark masses. We have also made
use of this method to define the renormalisation of the strong coupling through
$\alpha_S$ and not $g_s$.

\subsubsection{Field renormalisation}
In order to get the list of fields that should be renormalised, the programme
starts by extracting from the Lagrangian all relevant interaction terms on the
basis of the information passed as the value of the \verb+CouplingOrders+ option
of the {\tt MoGRe\$Renormalize} method (see section~\ref{sec:options}). The
corresponding field content is subsequently extracted and the relation between
the bare and renormalised quantities are derived following
eq.~\eqref{eq:renofield}. Matrix renormalisation is by default considered for
what concern fields lying under the same representation of the gauge and
Poincar\'e groups. An expansion over all flavour indices is then performed and
the field mixing restrictions passed as the \verb+FlavorMixing+ option of the
{\tt MoGRe\$Renormalize} method (see section~\ref{sec:options}) are finally
enforced.

The wave-function renormalisation constants associated with the left-handed
and right-handed chiralities of a Majorana fermion being equal, the code
simplifies the resulting expression by mapping the right-handed one onto the
left-handed one,
\be
  \Psi \to \Big[ 1 + \frac12 \delta Z_\Psi \Big]\Psi\ .
\ee
This has the advantage to prevent \nloct\ from calculating twice the same
quantity
and to yield more compact expressions for the counterterms.

\subsubsection{Parameter renormalisation}
The renormalisation of the model parameters is accounted for in three steps.
External parameters (including internal parameters that have been made external
with \verb+FR$LoopSwitches+, see section~\ref{sec:bfl:initmain}), particle
masses and internal parameters (including external parameters that have been
made internal with \verb+FR$LoopSwitches+) are handled separately. All
parameters and masses are directly renormalised according to
eq.~\eqref{eq:renoprm}, after removing all doubly-declared parameters defined
through the \verb+FR$RmDblExt+ variable (see section~\ref{sec:bfl:initmain}).
The code additionally takes care of deriving all relations connecting the
renormalisation constants of the internal parameters to those of other (external
or internal) parameters. Those relations are truncated at the one-loop level.
For instance, the renormalisation constant associated with the third generation
trilinear coupling strength of eq.~\eqref{eq:TtoA},
$\delta({\bf\hat T^u})_{33}$, can be written in terms of the top-quark mass and
$({\bf A^u})_{33}$ renormalisation constants,
\be
  \delta ({\bf \hat T^u})_{33} = \bigg[
    \frac{\delta m_t}{m_t} + \delta({\bf A^u})_{33}
   \bigg]({\bf \hat T^u})_{33} \ .
\ee
The obtained set of relations do not include any dependence on the
renormalisation constants that would be associated with unrenormalised
parameters (see section~\ref{app:BFL_simp}) and vanishing parameters have been
removed (see section~\ref{sec:restrictions}).

As a last step, the code applies all renormalisation conditions that have been
provided by the user via the \verb+AddRenormalizationCondition+ method. The
relations between bare parameters and the corresponding renormalised ones are
modified so that the dependent renormalisation constants are replaced by their
functional form. This also concerns the rules relating the renormalisation
constant of an internal parameter to other parameters and their renormalisation
constants.

\subsubsection{Renormalisation of the Lagrangian}
All the previously derived parameter and field redefinitions are finally applied
to the Lagrangian. A truncation at the one-loop level is performed by the code,
so that each Lagrangian term is at most linear in the renormalisation constants.
The Lagrangian is then formatted so that \nloct\ can be called to derive
the UV and $R_2$ counterterms, following the techniques
detailed in ref.~\cite{Degrande:2014vpa}. As a consequence, while \bfl\ lifts
some of the limitations inherent to a joint use of \fr\ and \nloct, it naturally
inherits all limitations that are strictly bounded to \nloct. For instance,
couplings independent of the particle masses are renormalised in the $\msbar$
scheme and $\as$ has to be renormalised either in the $\msbar$ scheme or as
described in section~\ref{sec:renoas}. This will be addressed in future work.

\section{Decoupling of heavy SUSY particles in \mg}\label{app:decoupling}
Arbitrary SUSY mass spectra often feature SUSY particles with masses much larger than the collision energy scale, and which are therefore expected to decouple in the corresponding cross-section computations. 

In the tree-level matrix elements, this decoupling property applies individually to each tree-level Feynman diagram featuring a propagator of the decoupling heavy SUSY particle. In the loop matrix elements, however, this decoupling is realised in a more complicated way involving cancellations among several loop diagrams featuring the decoupling particle(s) running in the loop. These cancellations become more severe as one approaches the decoupling limit, and numerical inaccuracies in the loop computations will eventually spoil them, yielding incorrect predictions.

\begin{figure}
 \center\includegraphics[width=0.75\textwidth]{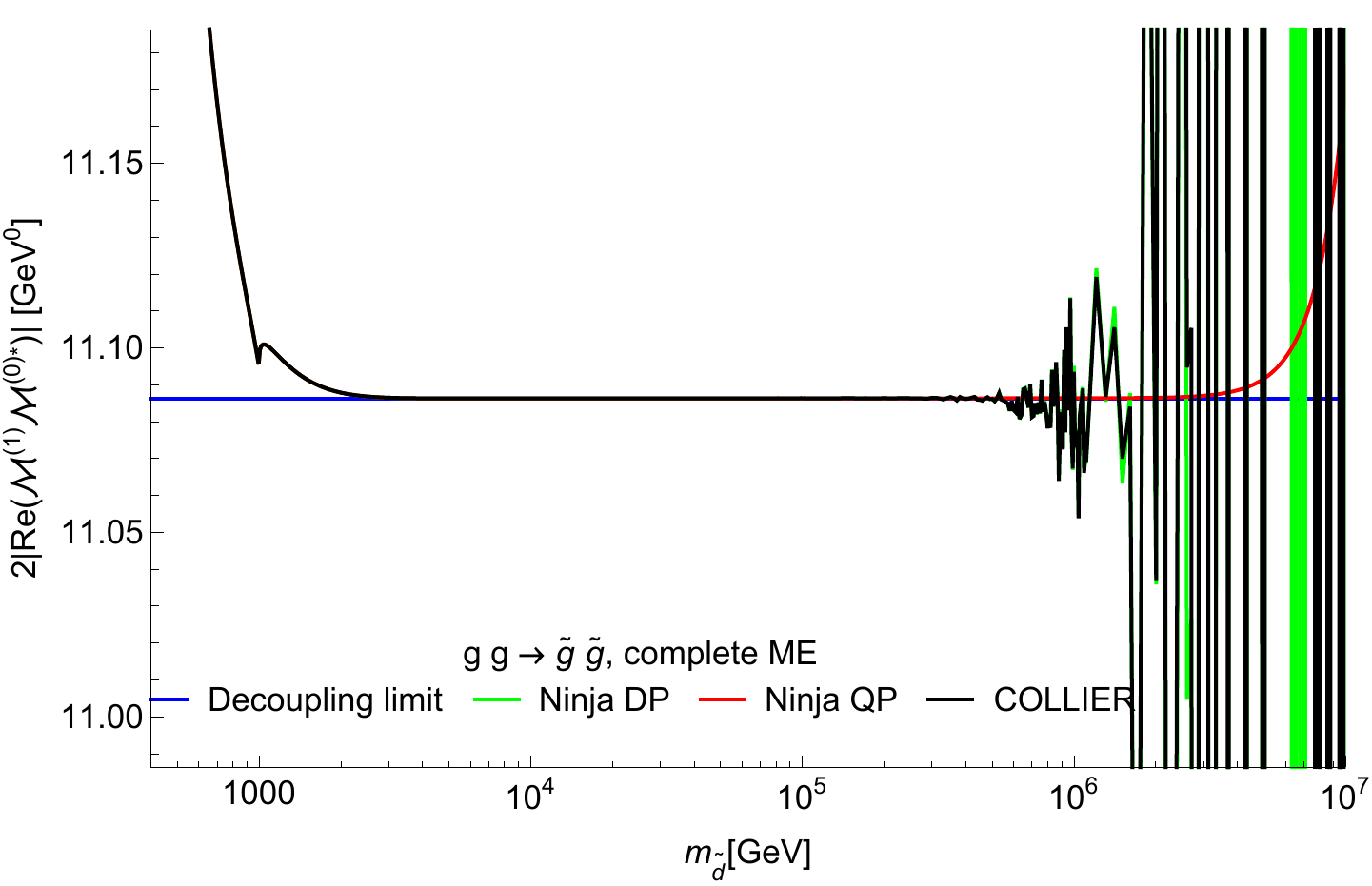}
 \caption{Numerical stability of the one-loop matrix element (ME) associated with the
   process $g g \rightarrow \tilde{g} \tilde{g}$ as a function of the mass of
   the down squark $\tilde{d} \equiv \tilde{d}_L, \tilde{d}_R$. The considered kinematic configuration satisfies $\sqrt{s}=4 m_{\tilde{g}}=2$ TeV. The constant line in blue indicates the decoupling limit, as obtained from computing the one-loop matrix element in the absence of down squarks. The other lines denote a numerical evaluation using various one-loop reduction algorithms and implementations of scalar master integrals: \collier\ (black) and \ninja\ in double (DP, green) and
in quadruple (QP, red) precision.\label{fig:LoopStabilty}}
\end{figure}

The solution to this problem simply amounts to completely removing the problematic heavy modes from the process definition,  effectively enforcing the decoupling property of the resulting matrix elements. 
It is however rather impractical having to settle for different process definitions depending on the masses of the heaviest particles in the spectra, and it is therefore desirable to determine quantitatively when this explicit removal procedure really becomes mandatory from a numerical standpoint.
To this end, we consider the one-loop matrix element for the pair production of
gluinos from initial-state gluons (\ie\ $g g \rightarrow \tilde{g} \tilde{g}$),
for the reason that this particular process features the worst numerical behaviour in
the decoupling limit amongst all $2\rightarrow2$ one-loop SUSY matrix elements.
We fix the kinematic configuration to $\sqrt{s}=4 m_{\tilde{g}}=2$ TeV, and report in figure~\ref{fig:LoopStabilty} the numerical evaluation of this matrix element as a function of the mass of a decoupling down squark.
We observe that the numerical evaluation starts to depart away from the decoupling limit for down-squark masses around 500~TeV in double precision and 5000~TeV in quadruple precision. However, these numerical instabilities will start to
significantly impact the accuracy of the final result only when reaching even larger masses of about $10^3$ TeV and $10^4$ TeV respectively.
In view of these results, we conclude that a conservative recommendation is to
manually enforce the decoupling of certain heavy particles by removing them explicitly from the process definition when their masses are larger than about \emph{a thousand times} the characteristic energy probed by the observable.
We stress however that in any case we tested that the associated numerical instabilities are correctly detected
by \madloop\ and will therefore adequately be reported to the integrator, which will in turn set the corresponding weight to zero if the accuracy is too poor. If such exceptional configurations occur too frequently, a clear warning is issued, hence preventing the user from inadvertently using incorrect results.

\bibliographystyle{JHEP}
\bibliography{library}

\end{document}